\documentclass[12pt]{article}
\makeatletter
\gdef\@fpheader{ }
\usepackage{jheppub}

\usepackage{amsmath}
\usepackage{amsfonts}
\usepackage{amssymb}
\usepackage{dsfont}
\usepackage{mathrsfs}
\usepackage{bm,slashed}
\usepackage{color}
\usepackage{hyperref}
\usepackage{MnSymbol}
\hypersetup{
    colorlinks=true, 
    linktoc=page,    
    linkcolor=blue,  
    urlcolor=black
}

\usepackage{subcaption}
\usepackage{color,psfrag}
\usepackage{graphicx,tikz}

\usetikzlibrary{calc}
\usetikzlibrary{decorations.pathmorphing}

\def\centerarc[#1](#2)(#3:#4:#5)
    { \draw[#1] ($(#2)+({#5*cos(#3)},{#5*sin(#3)})$) arc (#3:#4:#5); }
    
\def\centerarcnodes[#1](#2)(#3:#4:#5)(#6,#7)
    {\coordinate(#6) at ($(#2)+({#5*cos(#3)},{#5*sin(#3)})$);
    \coordinate(#7) at ($(#2)+({#5*cos(#4)},{#5*sin(#4)})$);
    \draw[#1] ($(#2)+({#5*cos(#3)},{#5*sin(#3)})$) arc (#3:#4:#5); }

\def\angcircle(#1)(#2)(#3:#4) {\coordinate(#1) at ($(#2)+({#4*cos(#3)},{#4*sin(#3)})$); }

\DeclareFontFamily{U}{rsfs}{}         
\DeclareFontShape{U}{rsfs}{m}{n}{<5> rsfs5 <6><7> rsfs7          %
  <8><9><10><10.95><12><14.4><17.28><20.74><24.88> rsfs10}{}     %
\DeclareMathAlphabet{\mathfs}{U}{rsfs}{m}{n}                     %
\newcommand{\be}{\nopagebreak[3]\begin{equation}}
\newcommand{\ee}{\end{equation}}

\newcommand{\bee}{\nopagebreak[3]\begin{equation*}}
\newcommand{\eee}{\end{equation*}}

\newcommand{\ba}{\nopagebreak[3]\begin{eqnarray}}
\newcommand{\ea}{\end{eqnarray}}

\newcommand{\beq}{\begin{eqnarray}}
\newcommand{\eeq}{\end{eqnarray}}

\newcommand{\baa}{\nopagebreak[3]\begin{eqnarray*}}
\newcommand{\eaa}{\end{eqnarray*}}

\newcommand{\la}{\label}
\newcommand{\n}{\nonumber}

\newcommand{\C}{\mathbb{C}}
\newcommand{\N}{\mathbb{N}}
\newcommand{\R}{\mathbb{R}}
\newcommand{\Z}{\mathbb{Z}}

\newcommand{\vp}{\vec{p}}

\def\pa{\partial}
\def\rd{\mathrm{d}}
\newcommand{\va}{\scriptscriptstyle}

\newcommand\myeq[1]{\stackrel{{#1}}{=}}

\def\eps{\varepsilon}

\def\label{\langle}
\def\ra{\rangle}

\def\f{\frac}

\def\SU{\text{SU}}
\def\ISU{\text{ISU}}
\def\SO{\text{SO}}

\def\SL{\text{SL}}
\def\su{\mathfrak{su}}
\def\so{\mathfrak{so}}
\def\sll{\mathfrak{sl}}

\def\km{\alpha}
\def\vA{\vec{A}}
\def\vX{\vec{X}}
\def\vP{\vec{P}}
\def\vS{\vec{S}}

\def\vx{\vec{x}}

\def\bra{\langle}
\def\ket{\rangle}

\def\dag{{}^\dagger}
\def\next{\,,\quad}
\def\w{\wedge}

\def\nn{\nonumber}
\def\pp{\partial}
\def\vSigma{\vec{\Sigma}}

\def\cJ{{\cal J}}
\def\cM{{\cal M}}

\def\cG{{\cal G}}

\def\cB{{\cal B}}

\def\tl{\tilde{l}}
\def\tm{\tilde{m}}

\title{Gravitational edge modes: From Kac-Moody charges to Poincar\'e networks
}
\author[a]{Laurent Freidel,}
\author[a,b]{Etera Livine,}
\author[a]{Daniele Pranzetti}

\affiliation[a]{Perimeter Institute for Theoretical Physics,  Waterloo, Ontario N2L 2Y5, Canada}
\affiliation[b]{Univ Lyon, ENS de Lyon, Univ Claude Bernard, CNRS, LPENSL, 69007 Lyon, France}

\emailAdd{lfreidel@perimeterinstitute.ca}
\emailAdd{etera.livine@ens-lyon.fr}
\emailAdd{dpranzetti@perimeterinstitute.ca}

\abstract{
We revisit the canonical framework for general relativity in its connection-vierbein formulation,
recasting the Gauss law, the Bianchi identity and the space diffeomorphism bulk constraints as conservation laws for boundary surface charges, respectively electric, magnetic and momentum charges.
Partitioning the space manifold into 3D regions glued together through their interfaces, we focus on a single domain and its punctured 2D boundary. The punctures carry a ladder of Kac-Moody edge modes, whose 0-modes represent the electric and momentum charges while the higher modes describe the stringy vibration modes of the 1D-boundary around each puncture.
In particular, this allows to identify missing observables in the discretization scheme used in loop quantum gravity and leads to an enhanced theory upgrading spin networks to tube networks carrying Virasoro representations. In the limit where the tubes are contracted to 1D links and the string modes neglected, we do not just recover loop quantum gravity but obtain a more general structure:  Poincar\'e charge networks, which carry a representation of the 3D diffeomorphism boundary charges on top of the $\SU(2)$ fluxes and gauge transformations.

}

\begin{document} 
\maketitle
\flushbottom

\date{\today}


\section{Introduction}

In our quest for quantum gravity, an essential task is to reach a proper understanding of the degrees of freedom and of the symmetries of gravity associated with local subregions. 

There are three interlaced insights around this fundamental question.
On the one hand, we have the insight about the local holographic behavior of general relativity  \cite{tHooft:1993dmi, Susskind:1994vu,Bekenstein:1994bc, Bousso:2002ju}, which stems from the idea that black holes carry a geometrical entropy proportional to their area.
On the other hand, there is the deep experience rooted in canonical gravity that general relativity is a constraint system
\cite{De-Witt, Ashetkar,Henneaux:1992ig}.
This means that the total energy associated to any space like region, for an arbitrary time flow, is entirely encoded in its boundary \cite{Brown:1992br,Hayward:1993ph,Hawking:1995fd,Iyer:1994ys,Hawking:1996ww,Freidel:2013jfa}. A simple and profound fact that distinguishes gravity from any 
other physical systems.
The third key ingredient is more recent. It is the understanding that we can assign gravitational edge modes to any boundary surface in space \cite{Freidel:2015gpa, Donnelly:2016auv}. This requires to identify a boundary symmetry algebra associated with any surface and to understand the gravitational edge modes as its representation states.

Interlacing these three insights leads naturally to the picture that the relevant geometrical degrees of freedom live on boundaries, that their dynamics and  the fabric of quantum  space itself is  encoded 
 into their entanglement. The central point is that this entanglement  is derived  from the fusion properties of the gravitational boundary symmetry algebra. This is the new perspective that we are proposing and developing here. 

The idea of gravitational edge modes has a long subtle history, most of it tied up with trying to understand 
the nature of black hole thermodynamics and/or infinity.
 It can be traced back to Regge--Teiltelboim \cite{Regge:1974zd}   and to Carlip \cite{Carlip:1996yb, Carlip:1998wz} who gave  them  the colorful name ``would-be-gauge-degrees-of-freedom''
and to Balachandran \cite{Balachandran:2018lcf} in the  canonical formulation of gravity.
In these works, the differentiability of the Hamiltonian constraints is the main reason behind the need for new boundary degrees of freedom.
This is the  same reason that led to the understanding of Kac--Moody symmetry as an edge mode symmetry in 3D Chern--Simons
theory \cite{Elitzur:1989nr,Banados:1994tn,Geiller:2017xad}. 
It can also be recognized, in the covariant formalism, in the work of Brown and York \cite{Brown:1992br} and is implicit in the ``membrane paradigm'' representation of Black-holes \cite{Thorne:1986iy}.
It also resurfaced, in quantum gravity, in the study of black hole thermodynamics: First in work by   Ashtekar-Baez-Corichi-Krasnov \cite{Ashtekar:1997yu} following an insight by Smolin \cite{Smolin:1995vq} 
and later further developed in \cite{Smolin:1998qp,Ashtekar:2000eq,Engle:2010kt,Pranzetti:2014tla}. In this approach, a particular boundary condition adapted to the black hole horizon reveals a set of boundary Chern--Simons edge modes as the states that play an essential role in black hole counting. 

The idea that  the boundary edge modes are representation states for the boundary symmetry algebra   not only to black hole horizons but also to  
  arbitrary finite boundaries \cite{ Freidel:2015gpa, Donnelly:2016auv,Freidel:2016bxd} 
  has led to a renewed interest in
  understanding  the nature and dynamics of gravitational edge modes for finite boundaries 
  \cite{Speranza:2017gxd,Geiller:2017xad,Geiller:2017whh,Wieland:2017zkf,Camps:2018wjf,Wieland:2019hkz}. 
  It is also important to appreciate that
  this new line of investigation is also deeply connected with the renewed understanding of the meaning and importance of the  asymptotic symmetry group and the corresponding soft modes  in the study of the S-matrix
  \cite{ Balachandran:2018lcf, Compere:2018ylh, Haco:2018ske} 
 (see especially \cite{Strominger:2017zoo} and references therein  for a detailed and comprehensive  review of this exciting subject). 

From the perspective of a canonical framework for general relativity, whose goal is to describe the propagation and evolution in time of the geometry, the edge modes live on corners, that is the 2D boundary surfaces of 3D regions of space-like hypersurfaces. In the context of holography, they are supposed to reflect and represent the degrees of freedom of the 3D geometry. 
Since classical spacetime is thought as a manifold, which can be described as an atlas of charts or in physical terms a union of bounded sets, space can also be modeled as the union of 3D regions glued together through their interfaces. One can then envision gravity to be written as a theory of edge modes living on the boundary surfaces of these patches of 3D geometry \cite{Freidel:2015gpa,Donnelly:2016auv}. This translates into a picture of space as a network of  ``bubbles'' as in \cite{Freidel:2018pvm}, which should eventually lead to quantum gravity as dynamical networks of quantum edge modes. This picture is naturally compatible with local-holography and it  is designed to offer a perfect setting to study the coarse-graining of gravity both at the classical and the quantum levels \cite{Livine:2013gna, Dittrich:2014ala, Dittrich:2014mxa, Dittrich:2016tys}.
The main reason one expect good  coarse-graining property is that we propose to label the states by boundary symmetry charges. Coarse-graining the charges of symmetry is then a more robust selection process
that can be implemented as a fusion of the symmetry algebra (see   \cite{Delcamp:2016lux,Delcamp:2016yix,Bodendorfer:2018csn}  for an instantiation of this process).
Here we present such a reformulation at the kinematical level, while we postpone the inclusion of dynamics (action of the time diffeomorphisms) to future work.

There are two main ingredients to our framework.
First, we rely on a reformulation of the (kinematical) Hamiltonian constraints of general relativity, generating the local $\SU(2)$ gauge invariance and 3D diffeomorphisms in the bulk, as conservation laws for boundary charges on 2D boundary surfaces---the corners. 
The key idea is to understand the constraint as a conservation law for the boundary charges. 
%
Second, we introduce distributional sources of curvature and torsion on the boundary surfaces and carefully describe the structure and algebra of edge modes that they induce. This leads to a full picture of the gravitational degrees of freedom carried by 2D boundary surfaces, as a starting point for the quantum theory.
This line of research,  inspired by \cite{Freidel:2015gpa}, was started  in \cite{Freidel:2016bxd}, where it was recognized that the boundary degrees of freedom can be reorganized into a ``gravity string''.

\begin{figure}[h!]
\begin{subfigure}[t]{47mm}
\includegraphics[height=35mm]{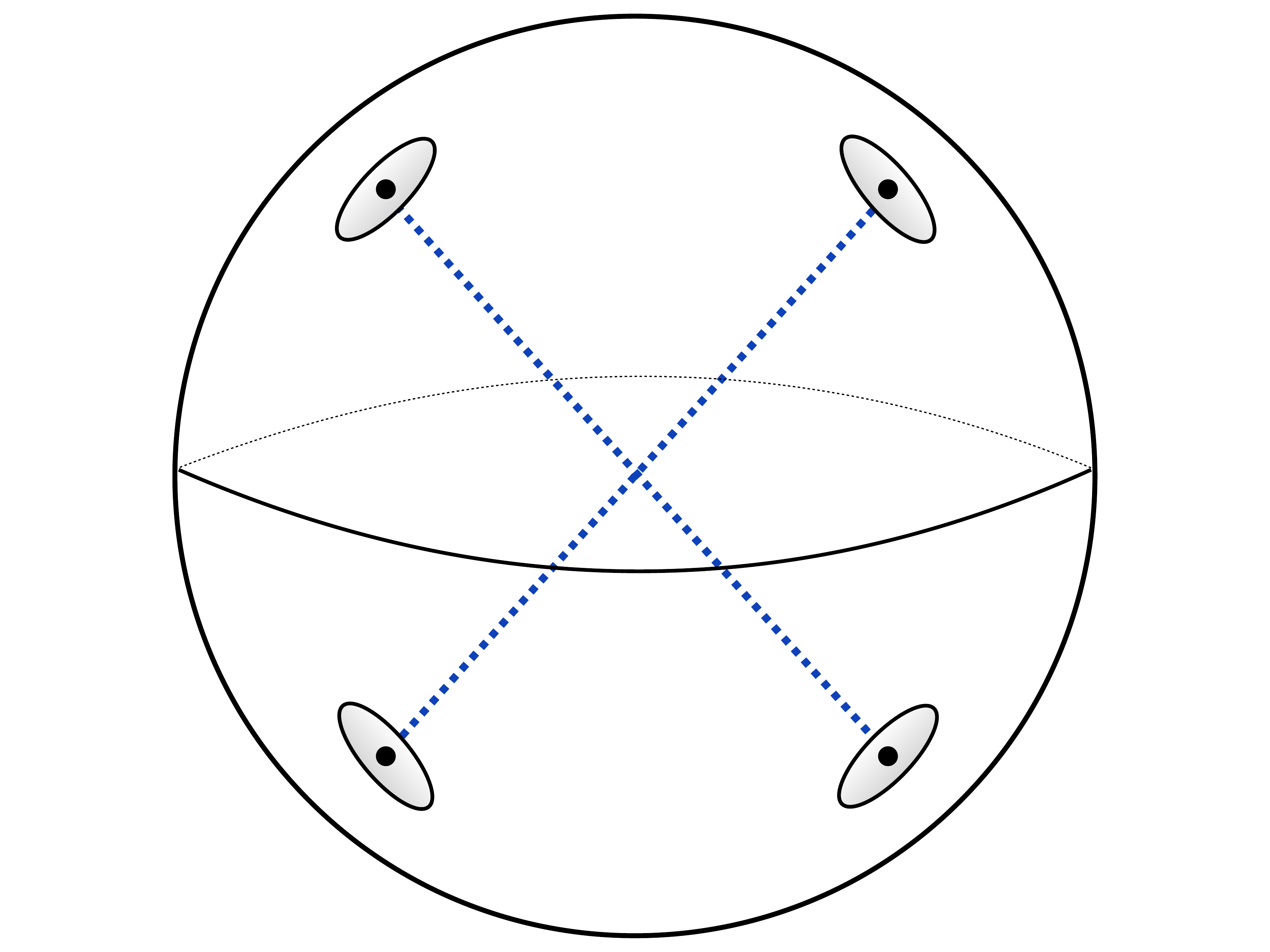}
\caption{A 3D region with 3-ball topology and its boundary surface punctured  by  disks representing its interfaces.}
\label{fig:fullsphere}
\end{subfigure}
\hspace*{3mm}
\begin{subfigure}[t]{47mm}
\includegraphics[height=35mm]{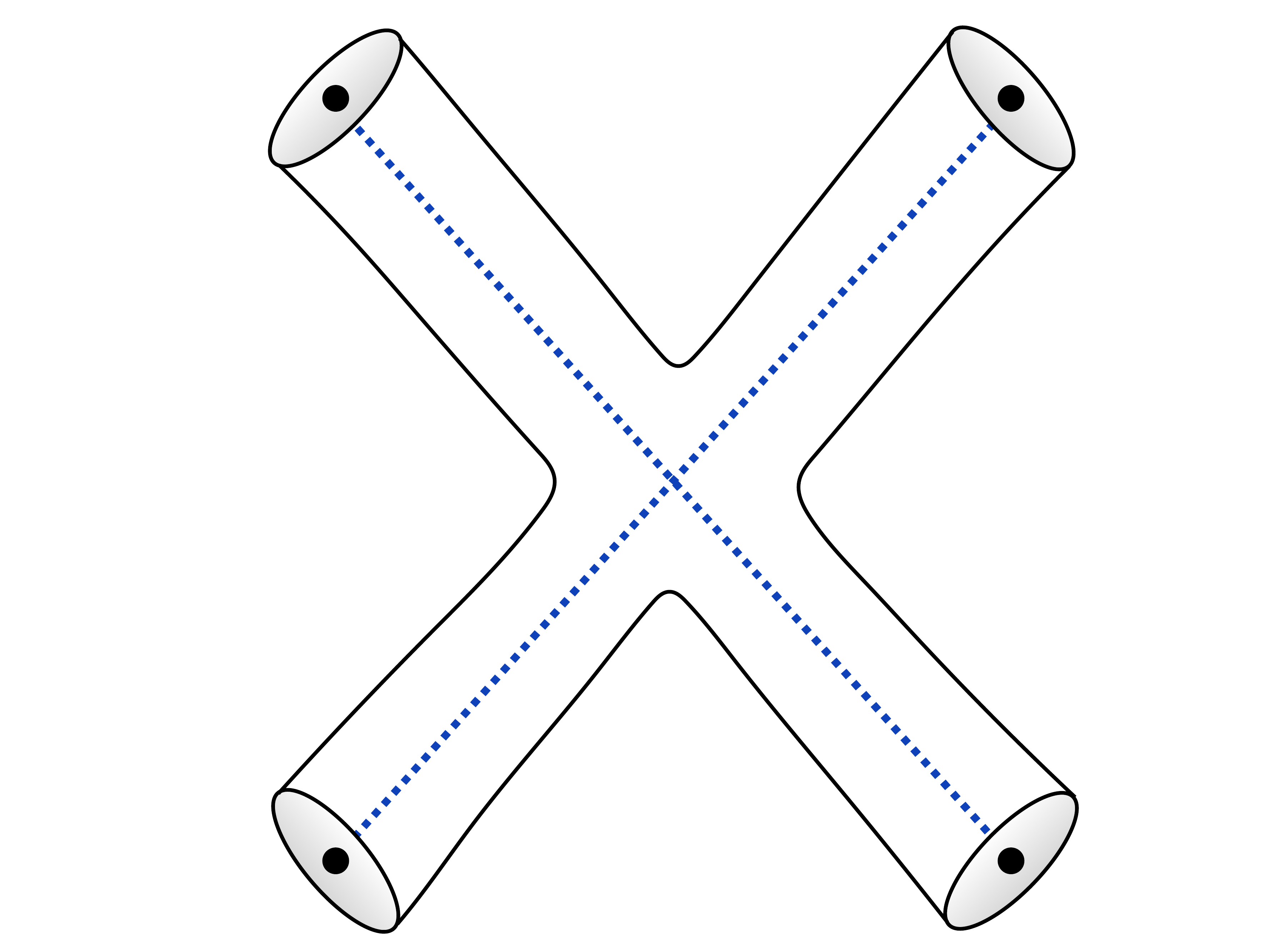}
\caption{The punctures are sources of curvature and torsion on the boundary, which propagate to tubes in the 3D bulk.}
\label{fig:slimtube}
\end{subfigure}
\hspace*{3mm}
\begin{subfigure}[t]{47mm}
\includegraphics[height=35mm]{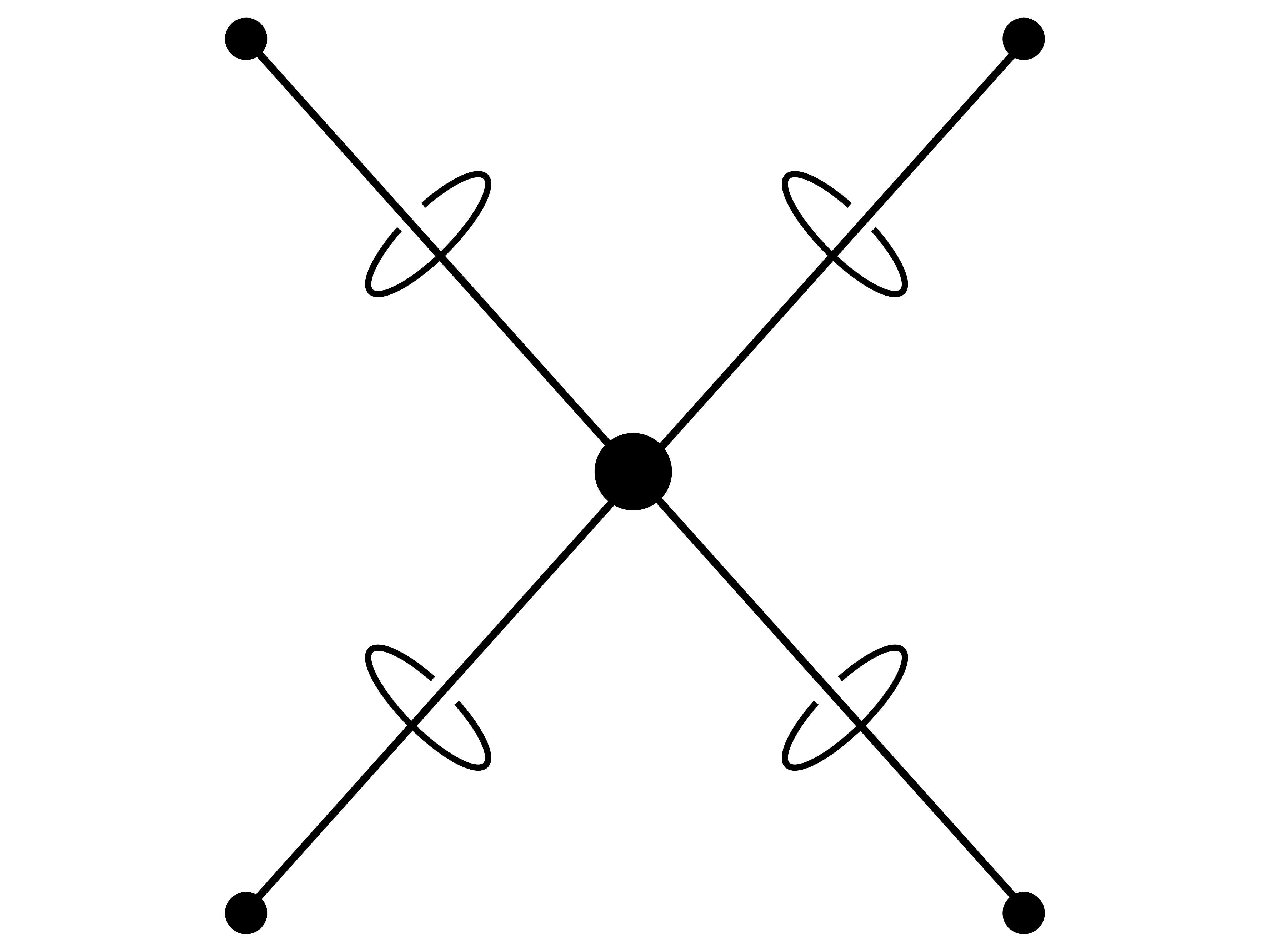}
\caption{The 1-skeleton reduction of the tubular structure encoding the edge modes on the punctured surface.}
\label{fig:network}
\end{subfigure}
\end{figure}

One of the motivations for our work is to use the tools of edge modes and surface charges to revisit the discretization of general relativity, as done for instance in loop quantum gravity (LQG).
Starting with a partition of space in 3D bounded regions of trivial topology, i.e. topologically equivalent to 3-balls with 2-sphere boundaries, we focus on the description of edge modes around each boundary sphere. Considering a single 3D region, we represent its interfaces with the neighboring 3D cells as small disks puncturing its 2D boundary surface, as illustrated on Figure \ref{fig:fullsphere}.  We consider these punctures as the sources of curvature and torsion for the geometry, assuming that the geometry is flat everywhere else. These defects propagate from the 2D boundary within the 3D bulk along tubes, represented in Figure \ref{fig:slimtube}. Reducing this tubular structure to its 1-skeleton, obtained by contracting the disks to points and tubes to lines, yields a graph structure where the whole 3D region is represented by a single vertex to which are attached edges representing the punctures, as drawn on Figure \ref{fig:network}. Such graphs dressed with geometrical data---fluxes and holonomies---define states of geometry in LQG.

In the present work, we do not start with the reduced graph structure as in LQG, but we aim to study  the structure of kinematical edge modes living on the punctured 2D boundary surface following  \cite{Freidel:2016bxd}.
From the LQG viewpoint, this entails seriously taking into account the interpretation of spin networks as discrete geometries with 3D regions dual to network nodes and 2D interfaces dual to network links, without taking a point-like limit in which the 3D regions are contracted into nodes (see also \cite{Markopoulou:1997hu, Sahlmann:2011uh,Sahlmann:2011rv,Haggard:2015kew,Freidel:2015gpa,Freidel:2016bxd,Freidel:2018pvm}). 
This leads to tubular networks, obtained by gluing the tubular structures of figure \ref{fig:slimtube} together. This follows the same intuition as earlier proposal that proposed to  upgrade spin networks to tube networks in LQG. One motivation was to represent spin networks for a $q$-deformed $\SU(2)$ gauge group in terms of conformal blocks \cite{Markopoulou:1997hu} or in terms of the moduli space of flat $\SL(2,\C)$ connections on the tube surfaces \cite{Haggard:2015kew,Han:2016dnt}. Another one was to take into account both $\SU(2)$ holonomies about and around graph edges as double spin networks \cite{Charles:2016xzi} or to use Drinfeld tubes \cite{Dittrich:2017nmq,Delcamp:2018efi}. But instead of postulating a priori algebraic structures living on the tubular network, we derive them directly from an analysis of the symmetries of general relativity in the presence of boundaries and we dress the surface networks with actual gravitational edge modes. In particular, we will show that this reveals charges probing the triad and connection tangentially to the surface and it hence allow us to construct observables that cannot be defined in the standard LQG discretization (nevertheless, see \cite{Sahlmann:2011rv} for a definition of surface integration of the flux). The new observables, missing from LQG, correspond to string vibration modes of the punctures, as derived in the loop gravity string framework introduced in \cite{Freidel:2016bxd}, but also to an important new momentum observable defining the boundary charge induced by the bulk invariance under 3D  diffeomorphism. This last feature solves the long-standing issue in LQG of defining quantum states of geometry carrying a representation of the 3D diffeomorphism constraint of general relativity.

The plot of the paper is as follows.
In the next Section \ref{sec:edge-grav}, we will describe the gravitational edge modes living on the 2D boundary surfaces in the first order formulation of canonical general relativity in terms of flux (co-triad) and Ashtekar connection. We will show that the bulk kinematical  constraints translates into boundary charges and conservation laws. In particular, 3D diffeomorphisms and $\SU(2)$ gauge invariance, put together with the Bianchi identity, lead respectively to momentum, electric and magnetic charges on the boundary surface.
Then, in Section \ref{sec:string}, we introduce sources of curvatures on the punctured surface and describe the phase space of edge modes and observables. In particular, the phase space for the surface naturally factorizes into the phase spaces attached to each puncture, supplemented with global conservation laws for the electric and momentum charges.
Section \ref{sec:QP} tackles the quantization of this phase space, focusing on a single puncture. We describe the quantum puncture in terms of a ladder of Kac-Moody charges describing the geometry excitations around the puncture.
This realizes the flux---the basic LQG observable representing elementary surface areas as spins---as a composite object, whose fine structure we analyze in Section \ref{sec:TotAM}.
This way, LQG is shown to be  a coarse-grained version of our framework, focusing on a small subalgebra of observables in the much richer algebra of edge mode charges  we describe. Finally,  this leads in Section \ref{sec:poincare} to a proposal of lifting the $\SU(2)$ spin networks of LQG to Poincar\'e charge networks, labelled both by momenta and fluxes quantum numbers, and thus carrying a representation of both the $\SU(2)$ gauge transformations and 3D diffeomorphisms at the discrete level.

\section{Gravitational Edge Mode Kinematics}
\la{sec:edge-grav}

This section presents the edge modes on the 2D boundary surface punctured with sources of curvature and torsion.
We show how the bulk gauge invariances become boundary symmetries with non-trivial charges. This leads to electric, magnetic and momentum charges associated respectively to the Gauss law, Bianchi identity and 3D diffeomorphism constraints in the bulk. These boundary charges satisfy conservation laws and reflect the  degrees of freedom of geometry associated to the 3D region.

 \subsection{Symplectic structure, boundary charges and constraints}

Considering a space-like Cauchy hypersurface $M$ in a  3+1 foliation of spacetime, we focus on a bounded 3D region $\cB$ and its 2D boundary $S$.
%
%
Starting from the first order formulation of 3+1 gravity in terms of coframe fields $(e^I)_{I=0,1,2,3}$ and spin connection, 
one can perform the canonical analysis in the time gauge where the coframe field $e^0=n$  is identified with  the timelike normal  to $M$.
It is well-known \cite{Barbero:1994ap,Holst:1995pc,Immirzi:1996di,BarroseSa:2000vx}  that in the presence of an Barbero--Immirzi parameter labelling a topological boundary term to the action,  the bulk phase space can be described  in terms of  a canonical pair of SU$(2)$ connection-flux variables.

This structure is best encoded into the total presymplectic  2-form which contains both bulk and boundary components\footnotemark, as it was shown in \cite{Freidel:2015gpa,Freidel:2016bxd}, namely
\ba \label{symp}
\Omega
&=& \int_{\cB} ( \delta A^{i}\wedge \delta
\Sigma^{\va}_{i} ) + \frac{1}{2  \kappa \gamma } \int_{S} (\delta e_i\wedge\delta e^i)\,,
\ea
where  $\kappa=8 \pi G$ encodes Newton's constant for gravity and $\gamma $ denotes the Barbero--Immirzi parameter.
\footnotetext{
It was also established, in the context of isolated horizons, in  \cite{Engle:2010kt} that the  bulk phase space needs to be extended at the boundary and involves non-commutative  boundary frame fields. We further refer the interested reader to \cite{Corichi:2016zac} for a survey of boundary extensions compatible with isolated horizon boundary conditions. See also \cite{Bodendorfer:2013sja} for a similar extension in higher dimension.
}
The index $i\in\{1,2,3\}$  labels coordinates on the 3D tangent space of the hypersruface $M$, or it equivalently labels a basis of the Lie algebra $\su(2)\sim\R^3$.
The Ashtekar--Barbero $\SU(2)$ connection $A^i$ is explicitly  given by  $A^i = \Gamma^i +\gamma K^i$ in terms of  the 3d spin connection $\Gamma^i$ associated with the bulk frame $e^i$ and   the extrinsic curvature 1-form $K^i$ on $M$. The conjugate momentum to the Ashtekar--Barbero  connection is an $\su(2)$-valued  2-form, called the flux $\Sigma_i$.
It is determined by the  coframe field through the  simplicity constraint,  
\be\la{flux}
\Sigma_i
=
\frac{1}{2\kappa\gamma}\,\epsilon_{ijk}\,e^j\wedge e^k
:=
\frac{1}{2 {\kappa\gamma }}(e\times e)_i
\,,
\ee
where we use a cross product notation\footnote{$(A\times B)_i:=\epsilon_{ijk} A^j B^k$.} on the tangent space.
As we explain below, this constraint, when pulled-back on the boundary surface $S$, becomes an integral part of the Gauss law.

From the symplectic form above, we can read the canonical pairs and commutation relations. In the bulk $\cB$, the Ashtekar--Barbero  connection $A^i$ is conjugate to the flux $\Sigma_i$. And the symplectic form further induces a Poisson structure for the boundary fields on the surface $S$:
\be\label{cr}
\{e^i_A(x), e^j_B(y)\}= { \kappa \gamma}\, \delta^{ij} \epsilon_{AB}\delta^2(x,y)\,,
\ee
where $A,B$ label a set of coordinates on the boundary $S$ and $\epsilon_{AB}$ is the totally skew tensor on two indices.

This phase space  needs to be supplemented by a set of gauge constraints  generating the $\SU(2)$  gauge invariance and the  
spacetime diffeomorphism invariance. We distinguish the physical phase space implementing the full set of gauge constraints from the kinematical phase space implementing the invariance under  $\SU(2)$ gauge transformations and space-like diffeomorphisms. In the present work, we focus on the kinematics, postponing the study of the dynamics  -Hamiltonian constraint generating time diffeomorphisms- to future work.

The bulk components of these constraints  are the Gauss-law and 3D diffeomorphism constraint. 
These two constraints can be written in terms of the bulk phase space variables $(\Sigma_i, A^i)$ as follows\footnotemark{}
\be
\rd_A \Sigma_i=0
,\qquad
F^i\wedge \imath_\xi \Sigma_i=0
\,,
\ee
where the covariant derivative of a $\su(2)$-valued form $\alpha$ is $\rd_A \alpha^i :=\rd \alpha^i+\epsilon_{ijk} A^j \alpha^k$ and 
its curvature is given by $F^i(A)= \rd A^i + \tfrac12 \eps_{ijk}A^j\w A^k$.
The diffeomorphism constraint involves an arbitrary tangent vector $\xi \in T\Sigma$ and  $\imath_\xi$ denotes the interior product.
\footnotetext{
Here, the constraints are naturally written as 3-forms on the canonical hypersurface $M$. We can also write them in terms of the densitized triad instead of the flux 2-form, $E^a_{i}:= \eps^{abc}\Sigma^i_{bc}$, as often done in the loop quantum gravity formalism. The two constraints, $\rd_A \Sigma_i=0$ and $F^i\wedge \imath_\xi \Sigma_i=0$, then read  $D_{A}E_{i}=0$
and $\xi^aE^{b}_{i}F^i_{ab}=0$.
}

We also include in the mix  the Bianchi identity, which is a purely kinematical constraint on the curvature $F^i$. Indeed, putting $\rd_A \Sigma_i=0$ and $\rd_A F^i=0$ at the same level hints towards a duality between electric and magnetic modes in gravity, which seems crucial in the analysis of the symmetries of the theory (see \cite{Freidel:2018fsk} for an example of the importance of duality in the analysis of QED). 
More precisely, the Gauss law and Bianchi identity  are both understood as  {\it charge conservation} equations. Gauss law is about conservation of  the ``$\SU(2)$ electric charge'' or flux $\int_D \Sigma$, while  Bianchi identity is about conservation of  the ``$\SU(2)$ magnetic charge'' $\int_D F$,
for $D$ a small disk.

An important point that we want to stress is that the 3D diffeomorphism constraint can also be written as a charge conservation (which comes in the ADM formalism from the fact that it is defined as the covariant derivative of the extrinsic curvature).
The charge in question is the ``$\SU(2)$  momentum" $\int_D P$ where
\be
P_i:=\frac{\rd_A e_i}{\sqrt{\kappa\gamma }}. 
\ee
This is straightforward to prove, One just need to use the Bianchi identity $\rd_A^2 e=( F\times e)_i$ and the simplicity constraints which implies that $\kappa\gamma \imath_\xi \Sigma_i = (\imath_\xi e\times e)_i$.  This means that 
the diffeomorphism constraint can also be written as the condition 
$(\rd_A P_i) (\imath_\xi e^i) =0$.
This means that the set of kinematical constraints that we need to implement can be rewritten in a fully symmetric manner as 
\be
\rd_A \Sigma_i=0,\qquad \rd_A F^i=0,\qquad \rd_A P_i=0. 
\ee
We can therefore view the Gauss law, the Bianchi identity and the diffeomorphism constraint as flux conservation, monopole conservation and momentum conservation respectively. In order to reveal these new charges, we need to include boundaries in our description and look at the edge modes on the boundary surface.

The understanding of all the gravitational constraints in a symmetrical manner as charge conservation is one of the main underlying theme of our work. As a result, we will discover that  quantum states of geometry are labelled by three quantum numbers: spin $j_p$, momenta $P_p$ and monopole charge $k_p$. Only the spin quantum number is revealed in loop gravity, while the other charges are usually ignored\footnotemark. Our framework allows us to go beyond these restrictions.
\footnotetext{
It should nevertheless be noted that magnetic edge modes have been introduced in loop quantum gravity specifically to deal with isolated horizon boundary conditions and account for black hole entropy  \cite{Ashtekar:1999wa,Engle:2010kt}.
}
Moreover, putting aside the Bianchi identity and magnetic charges, an essential feature of loop quantum gravity is an asymmetric treatment of the Gauss law and of the 3D diffeomorphism constraint. Indeed, on the one hand, the Gauss law is discretized level on graphs into the closure constraints, whose Poisson flow generates $\SU(2)$ gauge transformations. 
This allows the  Gauss law to be first discretized and then quantized.
One construct then an Hilbert space in which it acts covariantly and $\SU(2)$ gauge invariance is then implemented at the quantum level by restricting to spin network states.
 On the other hand, there is no 3D diffeomorphism constraint at the discrete level, thus no 3D diffeomorphism constraint operator at the quantum level (see e.g. \cite{Renteln:1989me,Dittrich:2004bn,Thiemann:2007zz} for details and discussion). What is done instead is that the diffeomorphism constraints is solved classically in the continuum, by averaging over diffeomorphism class of graph. This is a long-standing issue in loop quantum gravity. The 3D diffeomorphism invariance is imposed directly at the quantum level by taking equivalence of graphs embedded in the 3D hypersurface $M$. This method means that, {\it in the standard loop quantum gravity framework},
\begin{enumerate}
\item The 3D diffeomorphism invariance is not treated on the same footing  as the $\SU(2)$ gauge invariance;
\item It is not clear how matter fields, which enters the 
Einstein equations as a source for the diffeomorphism constraints\footnote{In the presence of matter the diffeomorphism constraints become $D_\xi = T_{n\xi}$,
where $T_{n\xi}=T_{ab} n^a \xi^b$ is the matter momenta density, with $n$ the timelike normal to the slice and $T_{ab}$ the matter energy-momentum tensor.} should be discretized and quantized;
\item It is not clear how boundaries, which break 3D diffeomorphism invariance, should be treated and how the resulting charges should be defined.
\end{enumerate}
Our approach, treating both the Gauss law and the 3D diffeomorphism constraint on equal footing, and focusing directly on the boundary charges instead of only considering  the bulk constraints,  should allows to remedy and address those issues.

 \subsection{Edge modes on boundary surface}

Let us study in more details the structure of the boundary charges and their discretization.
So we are considering a bounded region $\cB$ of the canonical hypersurface $M$ with boundary $S$. In the presence of boundary,  the bulk constraints in $\cB$ have to be supplemented  by boundary conditions on $S$. These boundary conditions determine the symplectic structure for  the boundary fields and reveal the edge modes living on $S$. 

\begin{figure}[h!]
\center
\includegraphics[height=4cm]{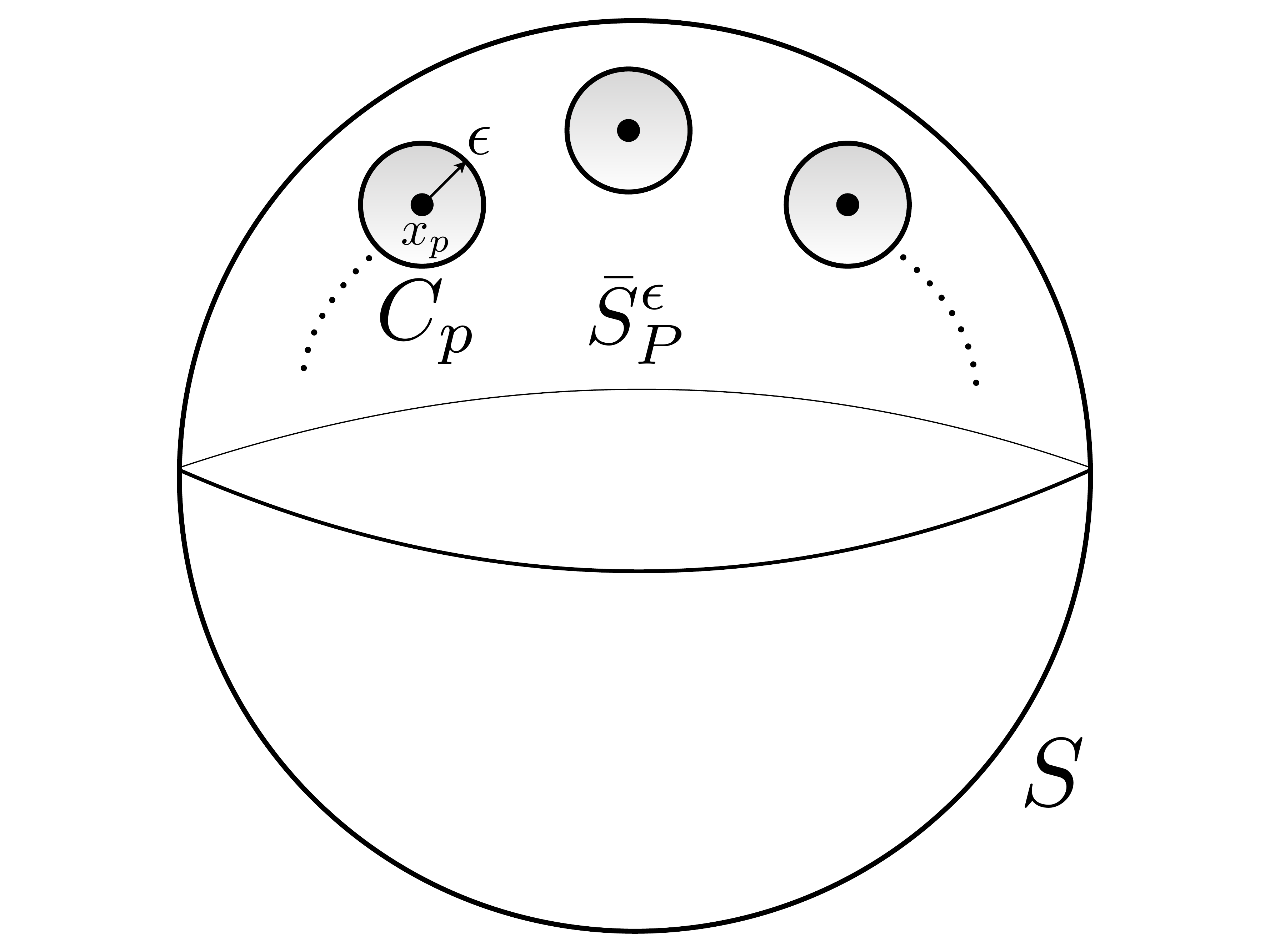}
\caption{Punctured Sphere.}
\label{fig:puncturedsphere}
\end{figure}

In order to analyze these in details, let us assume that the boundary $S_P:= S^2\setminus P $ has  the topology of a punctured 2-sphere,  with    $P\equiv \{x_p\in S\,| p=1,\cdots , N\}$, denoting the set of punctures.
It is convenient to think of each puncture $x_{p}$ as a small disk $D_p^\epsilon$ of radius $\epsilon$, as illustrated on FIG. \ref{fig:puncturedsphere}. We denote $C_p$ the boundary of $D_p^\epsilon$.   $C_p$ surrounds the puncture $x_{p}$ and contracts to it in the limit $\epsilon \to 0$.
Then the punctured boundary sphere  $S_P$ is the limit  $S_P=\lim_{\epsilon\to 0} \bar{S}_P^\epsilon$, where $\bar{S}_P^\epsilon= S^2\setminus \cup_p D_p^\epsilon$ is the complement of a union of the small disks  $D_p^\epsilon$.

As boundary conditions, we impose  simplicity, curvature and staticity constraints, respectively:
\be\label{bdyc}
 \Sigma_i\myeq{S^2}
 \frac1{2{\kappa\gamma }}(e\times e)_i,\qquad F^i(A)\myeq{S_P}0,\qquad P^i \myeq{S_P}0.
\ee
Here $\myeq{S^2}$ denotes an equality for the 2-forms pulled-back on ${S^2}$. The other equality $\myeq{S_P}$ means, on the other hand, that the pulled-back equality is valid away from the punctures.  These boundary conditions  allow for singular sources of curvature and momenta  at the punctures, while the flux remains smooth on the whole boundary surface $S$.
If we denote by $k_p^i$ and $P_p^i$ the source of curvature and momenta at the puncture $p$ and  by $\delta_p(x):= \delta^{(2)}(x,x_p)\,\rd^2 \sigma$ the delta-distribution 2-form on the surface $S^2$, this means that: 
\be\label{curv} 
F^i (A)(x) \myeq{S^2} 2\pi \sum_p k_p^i\,  \delta_p(x)\,,\qquad
P^i(x) \myeq{S^2} {2\pi} \sum_p P_p^i\,  \delta_p(x)\,.
\ee 
Using  $\SU(2)$ gauge invariance we can always diagonalize the curvature  and assume that $k_p^i = k_p \delta^i_3$.
In this work, we will restrict our analysis to the case where each $k_p\in \mathbb{Z}$ is an integer.
This form of curvature is such that the holonomy of the connection around the punctures is trivial: $\exp \oint_{C_p} A =1$.
So although the curvature is Planckian when $k_p\neq 0$, the holonomies are  invisible, in the sense that the holonomies remain trivial due to the compactness of the $\SU(2)$, although there is indeed a non-trivial curvature.
This fact is  the main reason behind the quantization of the area spectra at the quantum level \cite{Freidel:2016bxd}.

Furthermore, keeping trivial holonomies around the punctures  allows us to treat the curvature charges $k_p$ as classical labels.
In general, the charges $k_{p}$ are operators in the quantum theory, on the same footing as the momenta and fluxes on the boundary surface and they satisfy a non-commutative algebra. Nevertheless, for the purpose of this paper, we restrict to the integer case\footnotemark{} $k_p\in \mathbb{Z}$, for which we can  forget that $k_p$ is an operator and treat the curvature as background, and focus on the quantization of fluxes and momenta.
\footnotetext{
Some of our analysis should survives when the spectra of $k_p$ is in  $\mathbb{Z}/N$ for some integer $N$ along the lines sketched in \cite{Freidel:2016bxd}. This affects the periodicity properties of the frame field around the puncture and is reflected, for instance, in its mode expansion \eqref{PhiXPQ}.
}

We now explain how the boundary conditions (\ref{bdyc}) naturally appear  from demanding the constraints to be differentiable in the presence of a boundary.
Let us first illustrate this for the simplicity constraint which  relates the $\SU(2)$ flux to the coframe fields.
The main point is that in the presence of a boundary the naive Gauss Law $-\int_M \alpha^i \rd_A \Sigma_i$ is no longer differentiable (with respect to variations of the coframe field). It can be made differentiable by adding a boundary term $\int_S \alpha^i \Sigma_i$ which defines the bulk Gauss generator \cite{Cattaneo:2016zsq}:
\be
G_\alpha^{B}
:=
-\int_{B} \rd_A\alpha^i \wedge \Sigma_i
=
\int_B \alpha^i \rd_A \Sigma_i
-
\int_S \alpha^i \Sigma_i
\,.
\ee
 The challenge is then that $G_\alpha^{B}=0$   if and only if  $\alpha^i \Sigma_i$ vanish on the boundary $S$,  which means that the gauge invariance is broken by the presence of the boundary\footnote{The distinction between gauge and symmetry stems from the fact that a gauge transformation possess a vanishing  generator. In general, a symmetry transformation possesses a non zero charge.}.
 There are several options to deal with this issue.
The first one, which is the most traditional, would be to simply  accept this fact and  promote the non-vanishing charges $\int_S \alpha^i \Sigma_i$  to the status of symmetry generators.
Another option, the one taken in the LQG setting, is to impose that   $G_\alpha^{B}$ vanishes on $S_P$,   i-e outside a set of measure zero on the boundary  represented by the punctures. This imposes that the geometrical flux is singular and given by   $\Sigma^i_{\mathrm{LQG}}= \sum_p X^i_p \delta_p(x)$. The only symmetry charges left in this case are  the loop gravity fluxes $X_p$ associated with each puncture.
The third option, which leads to the loop gravity string picture \cite{Freidel:2016bxd}, is to follow the idea developed in \cite{Donnelly:2016auv} and to introduce edge modes that restore the gauge symmetry. We follow this path here: The coframe field on the boundary remains dynamical and defines {\it electric} edge modes $e^i$ on the boundary which carry their own symplectic potential given in \eqref{symp}; the gauge invariance is restored by  defining a gauge constraint generator involving a bulk component depending on the bulk fields $(\Sigma,A)$ and a boundary  component depending on the edge modes $e^i$.

Given a $\su(2)$-valued field $\alpha^i$, the $\SU(2)$ gauge constraint in the presence of boundary now reads:
\beq
\la{Ga}
G_{\alpha}&:=& -\int_{B} \rd_A\alpha^i \wedge \Sigma_i +\frac{1}{2\kappa \gamma} \int_{S}  \alpha^i (e\times e)_i
\\
&=&
\int_B \alpha^i \rd_A \Sigma_i
+
\int_S \alpha^i \,\left[\frac{1}{2\kappa \gamma}  (e\times e)_i-\Sigma_i\right]
\,.
\nn
\eeq
It is clear, after integration by part, that the condition $G_\alpha=0$ imposes both the Gauss law in the bulk
and the simplicity constraint on the surface. The gauge invariance is therefore restored by the presence of the edge modes.
In addition, we have  new symmetry charges given by $\Sigma_\alpha:= (2\kappa\gamma)^{-1}\int_S \alpha^i(e\times e)_i$,  which act only on the edge mode variables and  rotate them. 
The key difference between this picture and the naive symmetry breaking picture is that it becomes clear why the symmetry charges $\Sigma_\alpha$ satisfy a non-commutative current algebra:
\be
\{\Sigma_\alpha,\Sigma_\beta\}=  \Sigma_{\alpha\times \beta}.
\ee
In the usual loop quantum gravity framework, this non-trivial commutation relation remains mysterious if one thinks of the symmetry charge as defined directly from the bulk flux $\Sigma_i$ which commutes with itself according to the bulk symplectic potential (see \cite{Ashtekar:1998ak} for the original presentation of this problem and \cite{Cattaneo:2016zsq} for an interesting recent discussion of the puzzle). 

An important subtlety is that the bulk term in (\ref{Ga}) vanishes identically if $\alpha$ is assumed to be covariantly constant.
This means that  the generator $G_\alpha$ without addition of edge modes is a gauge constraint for $\alpha$ such that $\rd_A\alpha=0$.
We demand that this property is still satisfied in the presence of edge modes, which yields the extra condition $\int_S \alpha^i \Sigma_i=0$ for $\rd_A\alpha=0$, which we refer to as {\it the closure condition} .

\medskip

The same logic can be followed for the Bianchi identity, which defines the curvature constraint. The naive curvature constraint
$\int_B \beta_i \rd_A F^i(A)$ is not differentiable (with respect to field variations) in the presence of a boundary. It can be improved by adding a boundary term $\int_S \beta_i F^i$ which defines the bulk curvature constraint $F_\beta^{B} := \int_{B} \rd_A\beta_i \wedge F^i$.
A new feature is that this generator is still not integrable in the presence of punctures, but one needs to add an additional term $ \oint \beta_i  A^i $ around each puncture. This means that the full differentiable\footnotemark{} curvature constraint in the presence of boundary and punctures is finally given by 
\be
F_{\beta}
\,=\,
 - \underbrace{\int_{B} \rd_A\beta^i \wedge F(A)_i }_{gauge}
 \,+\,
 \sum_{p\in P} \underbrace{\oint_{C_p} \beta_i A^i}_{symmetry}.
\ee
\footnotetext{
The differentiability of the curvature constraint follows from the evaluation
\be
\delta F_\beta 
=
- \int_{S} \beta_i \rd_A \delta A^i + \sum_{p\in P} \oint \beta_i \delta A^i = \int_{S} \rd_A \beta_i \wedge \delta A^i
\,.
\nn
\ee}
This means that there are two layers in the gauge symmetry breaking pattern. The first layer is due to codimension $1$  edge modes and the second layer to codimension $2$  corner modes around punctures.
From now on we will impose that $F_\beta=0$ for all $\beta$ such that 
$\beta(x_p)=0$ for every puncture $x_p \in P$. 
This means that the physical charges associated with the curvature constraint are given by the holonomies
$K_\beta=\oint_{C_p}\beta_p\cdot A$ around each puncture\footnotemark. The remark made for the flux applies also here. Since the curvature constraint is identically satisfied when $\beta$ is covariantly constant we demand that $\sum_p \oint_{C_p} \beta_i A^i=0$, when $\rd_A \beta=0$, which is a magnetic analog of the closure condition. 
\footnotetext{Another option which will be explored elsewhere is to introduce magnetic edge modes.}

\medskip

Similarly, boundary and corner terms are needed in order to insure the differentiablity of the diffeomorphism constraint. Given a vector $\xi \in TM$ which restricts to a vector  in $TS_P$ on the boundary and a vector in $TC_p$  around the punctures, we find that  the differentiable\footnotemark{} generator of diffeomorphism  is 
\be
D_\xi
\,=\,
-\underbrace{\int_{B} F^i(A)\wedge \imath_\xi\Sigma_i
-\frac1{\kappa \gamma}
 \int_{S_P}   (\imath_\xi{e}^i )  \rd_A e_i}_{gauge\,=\,D_\xi^{\cG}}
 \,+ \,\,
 \frac1{2{\kappa \gamma}} \underbrace{\sum_{p\in P}  \oint_{C_p} { e}^i \imath_\xi e_i}_{symmetry}
 \,.
\ee
%
\footnotetext{
We compute the differential of the diffeomorphism generator:
\be
\delta D_\xi = \int_B \left( 
 \imath_\xi F^i \wedge \delta \Sigma_i-\delta A \wedge \rd_A \imath_\xi \Sigma \right)
+\frac1{\kappa \gamma} \int_S  \delta e_i \wedge \left(\rd_A \imath_\xi+\imath_\xi \rd_A\right) e^i 
\,.
 \nn
\ee
}
%
The first part of the constraint $D_\xi^{\cG}$ is the gauge constraint which vanishes on-shell.
In other words, we demand that $D_\xi^{\cG}=0$ or equivalently that  $D_\xi=0$ for all $\xi$ such that $\xi|_{C_p}=0$.
This imposes the diffeomorphism constraint away from the punctures. 

As we have already explained, it is possible to introduce a momentum constraint which is equivalent to the diffeomorphism constraint. More precisely, one defines the  translational generator
\be
P_\varphi 
\,:=\, 
- \underbrace{ \int_{\Sigma}   (\rd_A\varphi)_i\wedge P^i}_{gauge\,=\,P_\varphi^{\cG}} 
\,+ \,
\frac1{\sqrt{\kappa \gamma}} \underbrace{\sum_{p\in P}  \oint_{C_p} \varphi^i e_i}_{symmetry}
\,.
\ee
The bulk component $P_\varphi^{\cG}$ defines the translational gauge constraint, generating the  frame transformation 
$\delta_{\varphi} e^i=\rd_A \varphi^i$.
Demanding $P_\varphi^{\cG}=0$ is equivalent to the diffeomorphism constraint $D_\xi^{\cG}=0$ for invertible frame.
This follows from the fact that the diffeomorphism constraint is  a field dependent translation with parameter $\varphi^i ={(\kappa\gamma)}^{-\f12}\,\imath_\xi e^i$:
\be
 D_\xi^{\cG} = \frac{P_{\imath_\xi e}^{\cG}}{\sqrt{\kappa\gamma}}.
\ee 
Once the diffeomorphism or translational constraints are imposed, this leaves us with non-zero charges associated with each puncture that represent the $\SU(2)$  energy and momenta. These are given by 
\be
\label{Diff}
D_{\xi}
\,:=\,
\frac1{2\kappa \gamma} \oint_{C_p} e^i \imath_\xi e_i
\,,\qquad
P_\varphi 
\,{=}\,
\frac1{\sqrt{\kappa \gamma}}\oint_{C_p} e^i \varphi_i
\,.
\ee
As  explained in \cite{Freidel:2016bxd}, $P_\varphi$ define $U(1)^3$ Kac--Moody currents while $D_\xi$ define the Virasoro generators of  deformations of the puncture contour $C_{p}$.

\subsection{Charges and their conservation}
\la{sec:conservation}

\begin{figure}[h!]

\centering

\begin{tikzpicture}[scale=1.6]

\coordinate(p) at (0,0) node[below]{$p$};
\draw (p) node {$\bullet$};
\centerarcnodes[thick](0,0)(6:354:.5)(a1,a2);
\draw (a1)++(0.4,.18) node{$\theta=2\pi$};
\draw (a2)++(0.32,-.15) node{$\theta=0$};

\coordinate(q) at (2.5,0) ;
\draw (q) node {$\bullet$}node[below]{$r=r_{\star}$};
\centerarcnodes[thick](2.5,0)(174:130:.5)(b1,c1);
\centerarcnodes[dotted](2.5,0)(-174:120:.5)(b2,c2);
\draw[thick] (a1)--(b1);
\draw[thick] (a2)--(b2);
\coordinate(d1) at ($(2.5,0)+({1.5*cos(130)},{1.5*sin(130)})$);
\coordinate(d2) at ($(d1)+(c2)-(c1)$);
\draw[dashed] (c1)--(d1);
\draw[dashed] (c2)--(d2);

\draw[dotted] (p)--(q);
\coordinate(pr) at ($(0,0)+({.5*cos(100)},{.5*sin(100)})$);
\draw[dotted] (p)--(pr);
\draw (pr) node[above]{$r=\epsilon_p$};

\end{tikzpicture}
\caption{Boundary around a given puncture $p$. The dashed lines represent the (beginning of the) same boundary for the next puncture, while the dotted circle denotes all the other punctures boundary.}
\label{fig:puncture}

\end{figure}
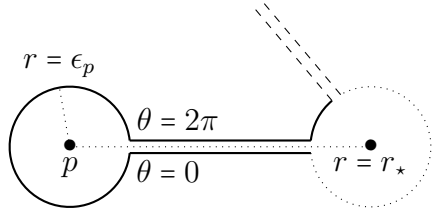

We have seen in the previous section that, in presence of  a boundary surface $S$, there are three type of boundary charge densities 
$(\Sigma_i,F^i,P_i)$ which represent spin, magnetic charge and momenta. We demand that  these satisfy on $S_P$ (outside the punctures) the following boundary conditions:
\be
\Sigma_{i}\myeq{S_P} \frac1{2{\kappa \gamma}} (e\times e)_{i}
\,,\qquad
F(A)^i \myeq{S_P} 0
\,,\qquad
P^i \myeq{S_P} 0
\,.
\ee
To solve  these constraints on $S_P$ it is convenient to chose a regularization of the punctures from $S_P$ to $S_P^\epsilon=S^2\setminus \cup_p D_p^\epsilon$.
We choose a point $*$ on $S^2$ and  we define a cut of the surface given by a system of paths that depart from  the chosen root $*$, run around the infinitesimal circles $C_p$'s and  then come back to the root $*$, as shown in Figure  \ref{fig:puncture}.

The cut defines a flower graph $\Gamma_P$, whose petals are the infinitesimal circles $C_p$'s and whose stems run from $*$ to the $p$'s. Outside the cut, i-e in the boundary region $S_{\Gamma_P}=S^2\setminus \Gamma_P$, the connection is flat and we have 
$A = g^{-1}\rd g$ for a function $g: S_{\Gamma_P} \to \mathrm{SU}(2)$.

We now choose a complex structure on the 2-sphere and denote  $(z,\bar{z})$  the complex directions tangential to the surface $S$. 
Let us  introduce  as well a complex basis $\tau_a$ with $(a=3,+,-)$  of the $\su(2)$ Lie algebra\footnotemark, and its dual basis $\tau_{\bar{a}}$ with  $\bar{a}=(3,-,+)$, normalized so that $[\tau_3,\tau_\pm]=\pm i \tau_{\pm}, [\tau_+,\tau_-]=i\tau_3$.
\footnotetext{
$\SU(2)$ group elements $g$ are then obtained by exponentiating Lie algebra vectors as $g=\exp[u_{\bar{a}}\tau_{a}]$ with $u_{3}\in\R$ and $\overline{u_{-}}=u_{+}$.
}
 Around each puncture $p$ with complex coordinates  $(z_p,\bar{z}_p)$, we introduce local coordinates   $(r,\theta)$  such that the point  $z-z_p=re^{i\theta}$. Here,  $r$ denotes the radial coordinate, with the circle $C_p=\partial D_p^\epsilon$ representing the boundary  of the disc $D_p^\epsilon$  located at $r=\epsilon_p$, and $\theta \in[0,2\pi[$ is the angle parametrizing $C_p$.
 The first condition in (\ref{curv}) on the curvature defines the source of curvature at the punctures, $F^i=2\pi\sum_{p}k_{p}^i\delta_{p}$ on the surface $S$. It means that the group element  $g$ satisfies  around $p$ the condition $ g_p( \theta) = \exp (\theta k_p^a\tau_a )$. The holonomy of the connection around the puncture is simply given by the group element $g_p( 2\pi) =\exp (2\pi k_p^a\tau_a )$ which equals the identity $\mathbb{I}$ if we restrict our analysis to the case with integer norm $|\vec{k}_p| \in \mathbb{Z}$.

 The second condition in (\ref{curv}) locates the source of momenta at the punctures, $P^i=2\pi\sum_{p}P^i_{p}\delta_{p}$ on the surface $S$, and define the boundary staticity constraint on the punctured boundary.
This constraint was obtained in   \cite{Freidel:2016bxd} as a boundary equation of motion
 \be
 P^a=\frac{\rd_A e^a}{\sqrt{\kappa \gamma}}\myeq{S_P}0
 \,.
 \ee
 Away from the punctures, where the connection is flat, we then have
 \be
 \la{ephi}
 e^a=\sqrt{\frac{\kappa \gamma}{2\pi}} \,(g^{-1} \rd \Phi\, g)^a\,,
 \ee
  where we have introduced three scalar fields $\Phi^a$ and a convenient normalization has been adopted.
The momenta $P^a_p$ associated with each puncture can be recovered from the holonomy of the frame field along the boundary circle $C_p$ :  
\be
\frac1{2\pi} \oint_{C_p} (ge_\theta g^{-1})^a \rd\theta = \sqrt{\kappa \gamma}\,P_p^a.
\ee 
%
%
Now that we have solved the curvature and momentum constraint on the boundary and specified the boundary conditions in terms of holonomies, we can also show that the total flux can be expressed as a sum of 
fluxes associated with each puncture.
Let us consider a covariantly constant $\su(2)$ element  $\hat{\alpha}$.
This means that $\hat{\alpha}= g^{-1}\alpha g$, where $\alpha$ is constant on $S_{\Gamma_P}$.
As we have seen above, the total flux  vanishes in this case due to the closure condition, $\Sigma_S(\alpha)=\int\alpha^a\Sigma_a =0 $.
On the other hand, we have that it can be written as a sum of fluxes at each puncture\footnotemark{}:
\be
\Sigma_S(\alpha) =\sum_p \alpha_a\Sigma_p^a,\qquad \Sigma_p^a :=   \oint_{C_p}  (\Phi \times \rd \Phi)^a.
\ee
\footnotetext{
In order to prove this we need to use that the holonomies $g_p(2\pi)g_p^{-1}(0)$ around each punctures are trivial. Otherwise we would get extra terms due to the 
integrals $\int_*^p   (\mathrm{Ad}_{g_p}({2\pi})-\mathrm{Ad}_{g_p}(0))(\Phi \times \rd \Phi) $.
} 
The scalar fields $\Phi$ contain one translational zero mode  whose information is not contained in the frame field $\rd\Phi$.
Under the shift $\Phi\to \Phi+X $ with constant $X$, we get that 
\be\la{Shift}
\Sigma_p \to \Sigma_p + X\times P_p. 
\ee
The flux closure condition 
\be\la{fluxcons}
\sum_p \Sigma_p^a=0
\ee
 therefore implies the 
momenta conservation condition 
\be\la{Pcons}
\sum_p P_p^a=0.
\ee

\section{The loop gravity string framework}
\la{sec:string}

The `loop gravity string' framework was introduced in \cite{Freidel:2016bxd} to study the richness of the gravitational edge modes appearing on 2D boundaries, as unfolded in the previous section. We here review this framework in order to set the stage for quantization in the rest of the manuscript.

 \subsection{Phase space structure}\la{sec:sppp}

\begin{figure}[h!]
\center
\begin{subfigure}{40mm}
\includegraphics[width=35mm]{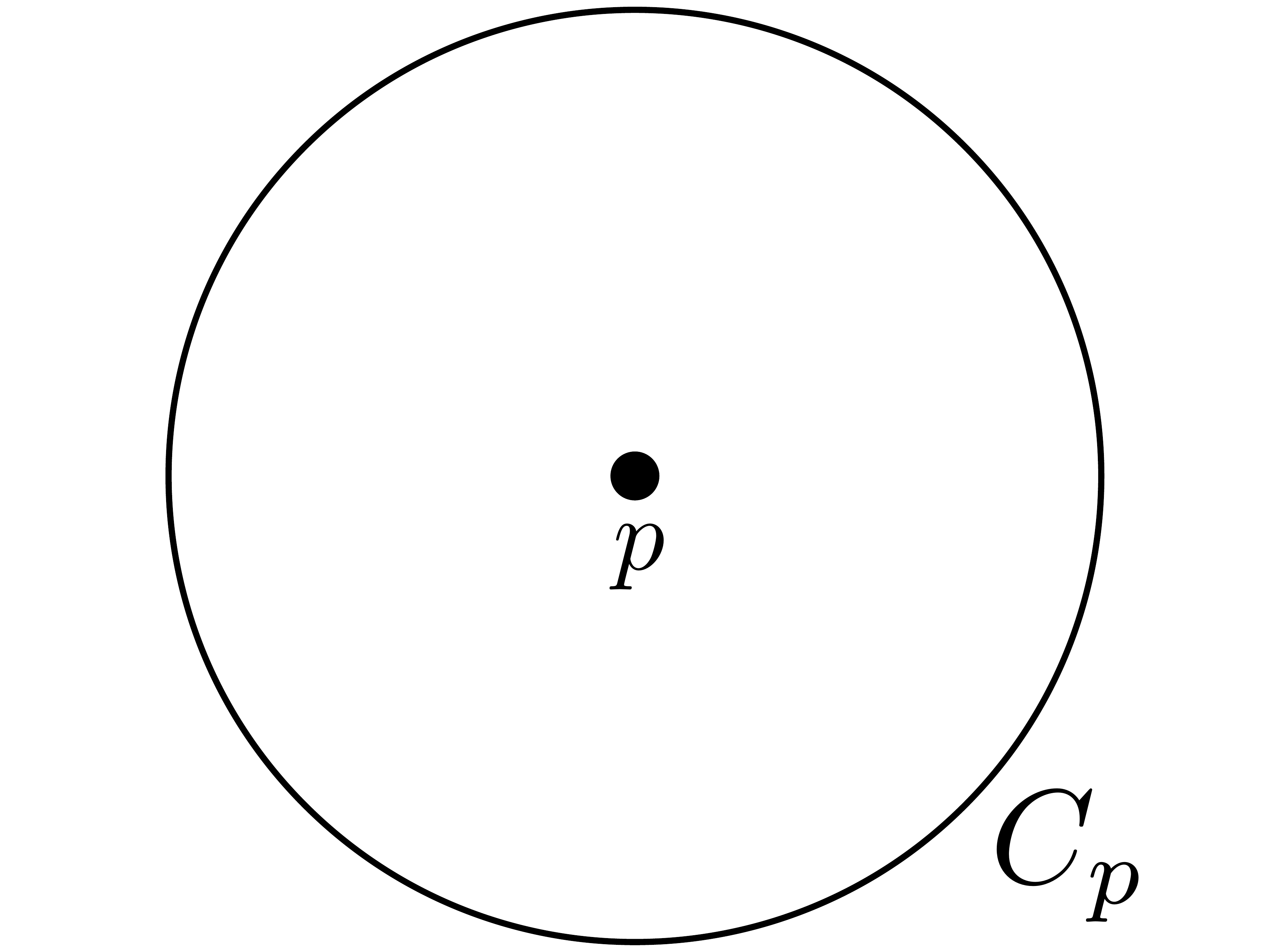}
\end{subfigure}
\hspace*{4mm}
\begin{subfigure}{40mm}
\includegraphics[width=35mm]{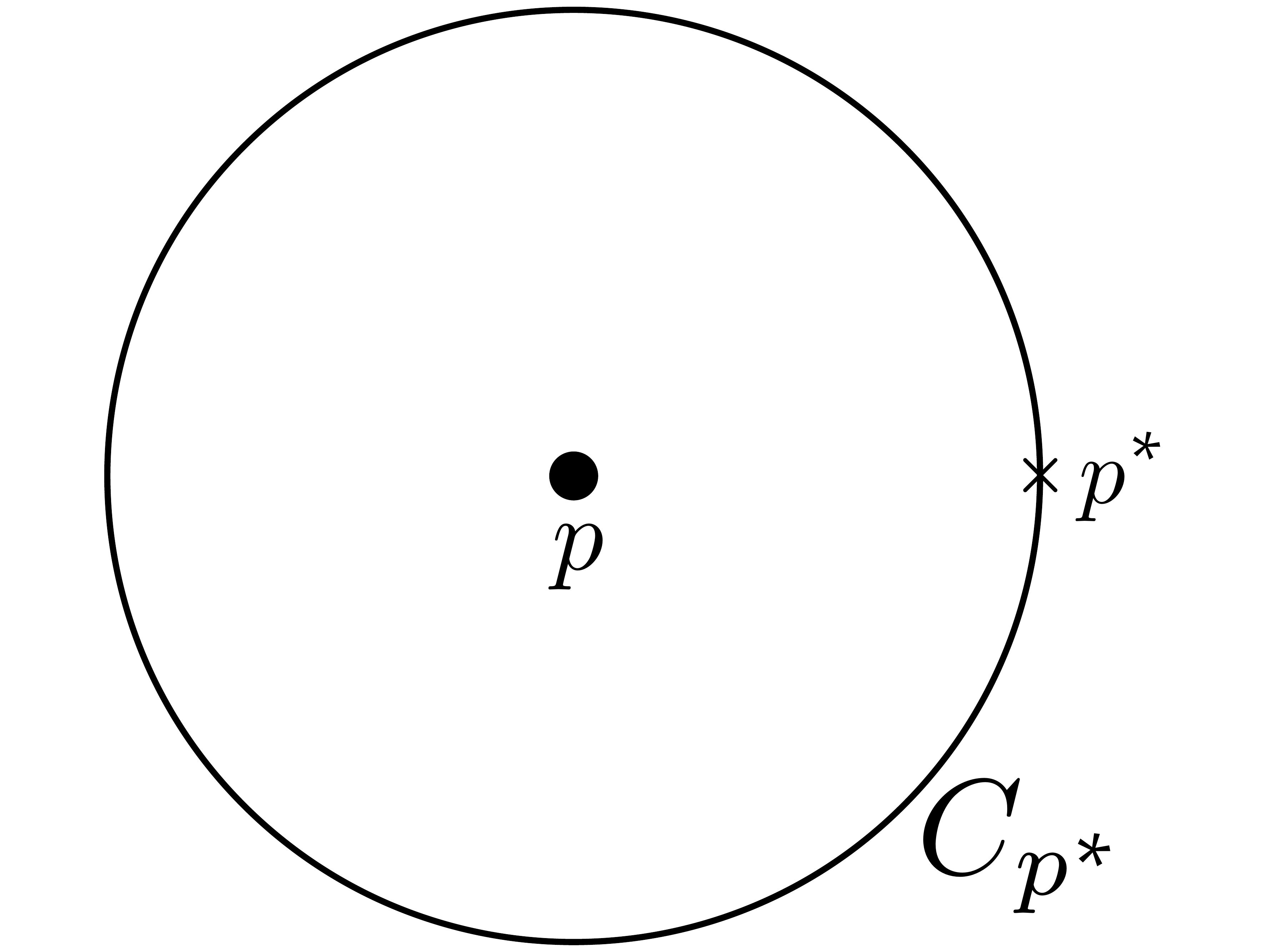}
\end{subfigure}
\hspace*{5mm}
\begin{subfigure}{50mm}
\includegraphics[height=35mm]{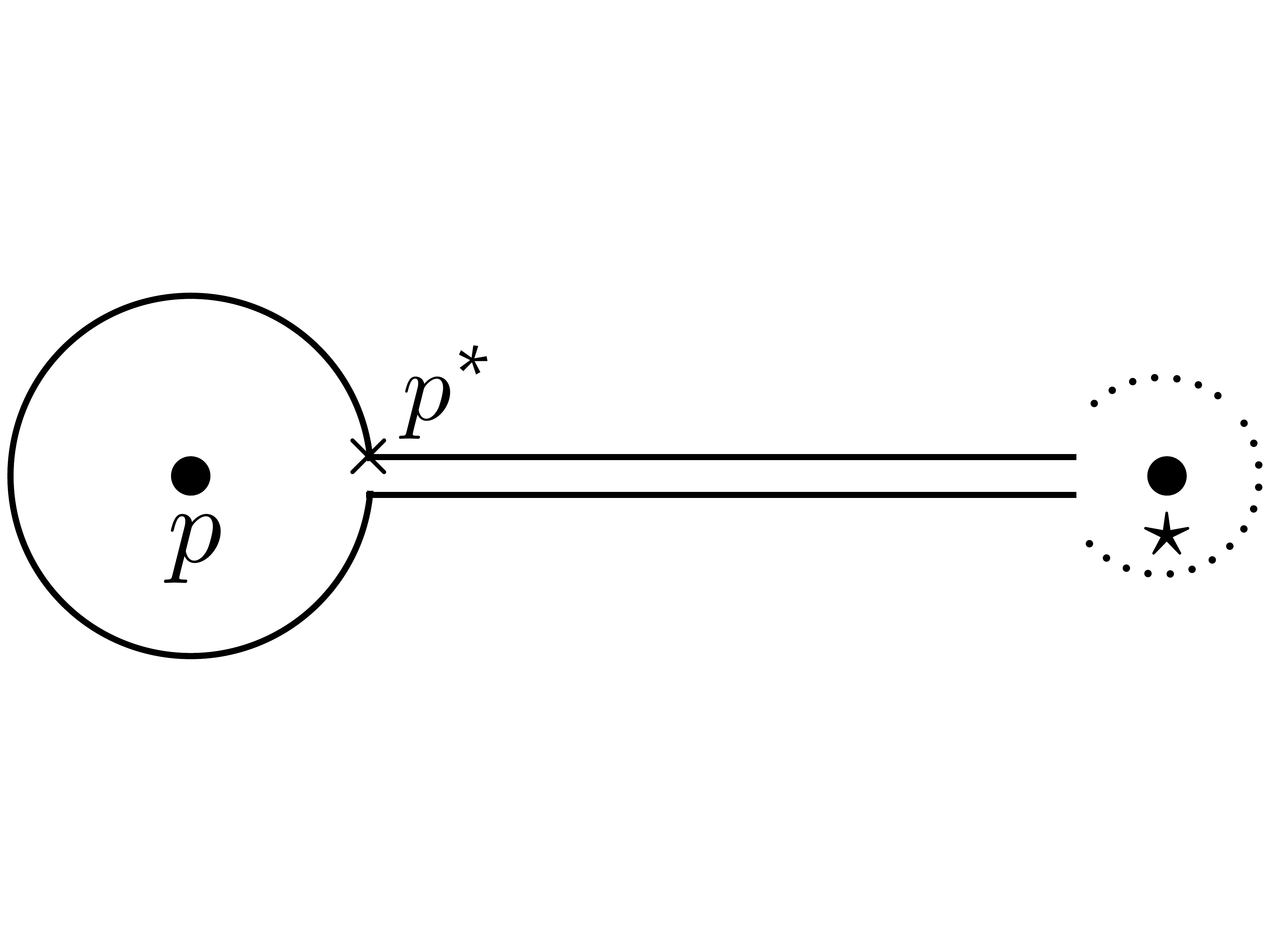}
\end{subfigure}
\caption{
The structure around the puncture $p$ :
we choose an anchor point $p^*$ on the boundary circle $C_{p}$ in order to define the integration contour $C_{p^*}$  around the puncture $x_{p}$, then the  integral contour  over the whole set of punctures is obtained by linking the contours $C_{p^*}$ together to the root point $*$.
}
\label{fig:pstar}
\end{figure}

 Once the bulk constraints are satisfied inside the 3D region $\cB$, the degrees of freedom are localized on its boundary $S$.
 Moreover, the boundary  symplectic structure $\Omega_{S}=\tfrac1{2\kappa\alpha} \int_{S} (\delta e^i\wedge \delta e_i)$ can be written as a sum of contributions from the punctures thanks to the staticity constraint and  momentum conservation:
 \be\label{Op}
 \Omega_{S}=-\sum_p  \Omega_p
 \,,\qquad 
 \Omega_p
 =
 \frac1{4\pi} \int_{C_{p^*}} \delta \Phi^a  \curlywedge \delta \rd \Phi_a
 \,-\f12 \delta P_p^a \curlywedge \delta \Phi_a(p^*)\,,
 \ee
 where the curly wedge symbol denotes a wedge in field space. The point $p^*$ is an arbitrary point on the circle $C_p$ around the puncture $x_{p}$, from which we start and end the integral, as shown in Figure \ref{fig:pstar}.  The additional term associated with $\Phi(p^*)$ is necessary in order to impose that $\Omega_p$ is independent of the position of the starting point $p^*$ on $C_p$. In other words, the correction due to $\Phi(p^*)$ insures the invariance of $\Omega_p$ under reparametrization of the circle\footnotemark{} (this mechanism was first observed in \cite{Freidel:2017wst} in the context of string theory).
 \footnotetext{
One can check that the corrected symplectic potential  $\Theta_\alpha:=\frac1{2\pi} \int_{\alpha}^{2\pi+\alpha} \Phi_a \delta \Phi^a- P_a \delta \Phi^a(2\pi+\alpha)$ is indeed independent of $\alpha$, i.e. $\pa_\alpha \Theta_\alpha=0$.
 }

To prove \eqref{Op}, we start with the formula for the frame field  $ e=\sqrt{\tfrac{\kappa\gamma}{2\pi}} g^{-1}\rd\Phi g$ on $S_{\Gamma_P}$.
Moreover, since $\Gamma_P$ is a set of measure zero, we can replace the integral over $S$ as an integral over $S_{\Gamma_P}$ and write: 
 \ba
\Omega_S&=&\Omega_{S_{\Gamma_P}}=
\frac1{2\kappa \gamma}\int_{S_{\Gamma_P}} \delta e^i\curlywedge \delta e_i =\frac1{4\pi}\int_{\pa S_{\Gamma_P}} \delta \Phi^a \curlywedge \rd \delta  \Phi_a \n\\
& =& -\frac{1}{4\pi} \sum_p \left( \int_{C_{p^*}}  \delta \Phi^a \curlywedge \rd \delta  \Phi_a - \int_*^p \left[ \delta \Phi^a \curlywedge \rd \delta  \Phi_a\right]_0^{2\pi}\right) \,,
 \ea
 where the last integral comes from the cuts going from the reference point $*$ to the puncture $p$, $0$ and $2\pi$ refer to the two sides of the cut (going from the root $*$ on the surface to the anchor point $p^*$ around the puncture and back from $p^*$ to $*$).
 This integral can be easily evaluated since $\rd\Phi(r,2\pi)=\rd\Phi(r,0)$ and $\Phi^a(r,2\pi)-\Phi^a(r,0)= 2\pi P^a_p$, with $r$ a coordinate along the cut  $(*,p)$. One gets that 
 \be
 \frac{1}{4\pi} \sum_p  \oint_*^p \left[ \delta \Phi^a \curlywedge \rd \delta  \Phi_a\right]_0^{2\pi} =  
 \frac{1}{2} \sum_p \delta P_p^a \curlywedge \left(\oint_*^p  \rd \delta  \Phi_a\right)=
 \frac{1}{2} \sum_p \delta P_p^a \curlywedge \left(\delta \Phi_a(p^*) - \delta \Phi_a(*)\right)
 \,.
 \ee
The term associated with $\delta\Phi(*)$ vanishes thanks to momentum conservation $\sum_p P^a_p=0$. Overall this shows \eqref{Op}.

To probe and further uncover the structure of the symplectic form, we can expand the field $\Phi^a$ in modes around each puncture $p$.
To this purpose, it is convenient to use a gauge where $k_p^a \tau_a=k_p \tau_3$.  In this gauge, we have that $g_p^{-1}(\theta) \tau_a g_p(\theta) = e^{i k_p^a \theta} \tau_a$, where $(k_p^3,k_p^+,k_p^-)=(0,+k_p,-k_p)$. The scalar fields $\Phi^a(z,\bar{z})$ then admit
 a mode expansion around the  puncture $p$ given by  \cite{Freidel:2016bxd}:
\be
\la{PhiXPQ}
\left.\Phi^a(z_p+\epsilon_p e^{i\theta},\bar z_p+\epsilon_p e^{-i\theta})\right.
=  X_p^a+ \theta P_p^a+iQ_p^a(\theta)\,,
\ee 
which involves the modes $Q_n$ that appear in the expansion of  the frame field:
 \be\label{defQ}
 Q_{p,n}^a:= \oint_{C_p} e^{in\theta} e^a_\theta \rd \theta\,,
 \qquad
 \bar{Q}_{p,n}^a =  Q^{\bar{a}}_{p,-n}\,,
 \ee
\be
\la{Qsum}
P_p^a:=Q^a_{p,-k_p^a}, \qquad 
Q_p^a(\theta):= \sum_{n+k^a\neq 0}\frac{Q^{a}_{p,n} e^{-i\theta (n+ k_p^a)}}{(n+ k_p^a)}\,.
\ee
%
%
%
The integration constant $X_p^a$ will turn out to play the role of a zero mode  `position'  observable  conjugated to the momentum $P_p^a$.
%
The $Q_{p,n}^a$'s represent the modes of the frame field $e^a$ pulled back on the boundary of the disk $D_p$ at coordinate radius $\epsilon_p$,
while  $Q^a_{p,-k^a}$ is  the puncture momenta $P_p^a$.


We inject this mode expansion \eqref{PhiXPQ} in the expression of  the symplectic potential \eqref{Op}.
A straightforward calculation yields
\ba
\Omega_p&=&
 \frac{1}{4\pi}  \oint_p  \delta \Phi^a \curlywedge \rd \delta  \Phi_a-\frac12 \delta P_p^a \curlywedge \delta \Phi_a(p^*)  \n\\
&=&\sum_a
\left[ \frac{1}{4\pi}\int_{0}^{2\pi}  ( \delta X_p^a + \theta \delta P_p^a+i\delta Q_p^a(\theta ))  \curlywedge (\delta P_p^{\bar a}+i\partial_\theta \delta Q_p^{\bar a}(\theta ))\rd\theta\right] -
 \frac{1}{2} \sum_p \delta P_p^a \curlywedge \delta \Phi_a(p^*)
\n\\
&=&\sum_a \left[\frac12\, \delta  X_p^a \curlywedge \delta P_p^{\bar a} -\frac{1}{4\pi}\sum_a \int_{0}^{2\pi} \delta Q_p^a(\theta)\curlywedge\partial_\theta \delta Q_p^{\bar a}(\theta)  \right]
\n\\
& &
\,+\, \frac12 \, \delta P^a   \curlywedge 
\delta \left(\frac{i}{2\pi} \int_{0}^{2\pi} \theta \partial_\theta  Q_a(\theta )-\Phi_a(p^*)\right)\n\\
&=&\sum_a \left[\delta  X_p^a \curlywedge \delta P_p^{\bar a} -\frac{1}{4\pi}\sum_a \int_{0}^{2\pi} \delta Q_p^a(\theta)\curlywedge\partial_\theta \delta Q_p^{\bar a}(\theta)  \right]\,,\la{OO}
\ea
where we used that $\oint Q =\oint \pa_\theta Q=0$. The tricky identity that allows us to prove the last equality  is  the integral
\be 
\frac{i}{2\pi}\int_{0}^{2\pi} \theta \partial_\theta  Q_a(\theta )\rd \theta
= \frac{i}{2\pi}  \left[\theta  Q_p(\theta )\right]_0^{2\pi} - \frac{i}{2\pi} \int_{0}^{2\pi}   Q_a(\theta )\rd \theta
= iQ_p(2\pi  )=(\Phi(p^*)-X_p).
\ee
One can finally expand the last integral in \eqref{OO} in terms of the modes  \eqref{Qsum} and obtain the symplectic structure associated to each puncture $p$:
\be 
\Omega_p = 
\sum_a \left[\delta  X_p^a \curlywedge \delta P_p^{\bar a} +
\frac{i}2 
\sum_{n+k^a\neq 0}\frac{\delta Q^{a}_{p,n} \curlywedge \delta Q^{\bar{a}}_{p,-n}}{(n+ k_p^a)}
 \right]
 \,.
\ee
%
This gives the Poisson brackets for the phase space associated to a single puncture:
\ba
\{X_p^a,P_{p'}^{ b}\}&=& \delta_{pp'} {\delta^{a\bar{b}}}\,,\la{xP}
\\
\{Q_{p,n}^a,{Q}_{p',m}^b\} &=& -i \delta_{pp'} {\delta^{a\bar{b}}}  (n+k_p^a) \delta_{n+m}\,.\la{QQ}
\ea
Since we have assumed that the curvature charges $k_p^a$ are integers, we can entirely re-absorb them as shifts of the mode expansion and define shifted charges
\be
\alpha_{p,n}^a:= Q^a_{p,n-k_p^a}
\,,\qquad
\bar{\alpha}_n^a= \alpha_{-n}^{\bar{a}}
\,,\qquad
\{\alpha_n^a\,,\,\bar{\alpha}_m^b\}
\,=\,
-i\,n \,\delta^{ab} \delta_{n,m}
\,.
\ee
We recognize this tower of oscillators as forming a U$(1)^3$ Kac-Moody algebra.

\medskip

The overall symplectic structure can be provided with a simple interpretation.
Instead of thinking of the punctures as mere punctual sources on the boundary surface, we have solved carefully and explicitly the boundary conditions on the punctured surface excising a small disk $D_{p}^\epsilon$ around every puncture $p$. As a consequence, the structure defining each puncture is not 0-dimensional anymore but is the 1-dimensional boundary $C_{p}$ of the disk $D_{p}^\epsilon$, which can be thought as a string. This leads to a zero-mode string position and momentum, $X_{p}$ and $P_{p}$, which will be become conjugate operators in the quantum theory, and to an infinite tower of higher/lower modes $Q_n^a$ describing the vibration modes of the string and generating a Kac-Moody algebra.
From the viewpoint of loop quantum gravity, this means that the thickening of the spin network edges  as tubes gives rise to vibration modes of the 1d-boundary around punctures.

\subsection{Flux and spin observables}
\la{sec:fluxclassical}

We can use the new fundamental field $\Phi^a$ to  express the integrated  flux \eqref{flux} over the closed surface $S$. As in the case of the symplectic form \eqref{Op} derived in the previous section, we rely on the decomposition  $S = \bar{S}_P^\epsilon \cup_p D_p^\epsilon$  and the fact that the connection is flat on $S_{\Gamma_P}$. Working out a similar calculation leads to a decomposition of the total flux  as the sum of contributions associated to each puncture,
\be\la{Sphi}
\Sigma^a(S)=\int_{S_{\Gamma_P}} \Sigma^a
=- \sum_p  \Sigma^a_p =0\,,
\ee
with the flux for each single puncture  given by
\be
\Sigma_p^a
=
 \frac1{4\pi} \int_{C_{p^*}} (\Phi\times \rd \Phi)^a - \frac12 (P_p\times  \Phi(p^*))^a
:=
\frac1{4\pi}\oint_{C_{p}}(\Phi\times \rd \Phi)^a
 \,.
 \ee
 We have introduced the convenient notation $\oint_{C_p}$ to underline that this combination of integration over the cut contour $C_{p^*}$ plus the correction term in the momentum $P_{p}$ does not depend on the chosen anchor point $p^*$ on the $C_{p}$ and defines a truly reparametrisation invariant integration.
Indeed, it is important to keep in mind that in the expansion \eqref{PhiXPQ} around each puncture of the fields $\Phi^a=  X_p^a+ \theta P_p^a+iQ_p^a(\theta)$, the Kac-Moody charges $Q_p^a$ are periodic in $\theta$ and the only term creating a discontinuity on the contour around the puncture is the linear term $\theta P_p^a$ depending on the momentum. This is exactly the term that we subtract above in the definition of the flux $\Sigma_{p}$ associated the puncture $p$.

The flux at $p$ can further be evaluated in terms of the mode expansion.
One finds that it is given by the sum of an angular-momenta contribution plus a spinning contribution\footnote{ A similar decomposition of the flux appeared also in \cite{Freidel:2013bfa}, which led to the interpretation of spinning geometries for the dual cellular spaces. }: 
\be\la{Sigmap}
 \Sigma^a_p=A^a_p+S^a_p\,,
\qquad
A^a_p:= (X_p\times  P_p)^a,\qquad S^a_p:=\frac{i}{2} \sum_{n\neq 0}\frac{(\alpha_{p,n}\times \alpha_{p,-n})^a}{n}
\,,
\ee
in terms of the shifted oscillators $\alpha_{p,n}^a:= Q^a_{p,n-k_p^a}$.

We would like to point out that, in general, \eqref{Sphi} contains an extra term depending on the position of the reference point  $\star$ on the  2-sphere, through $ \Phi(*) \times (\sum_p P_p)$. This term nevertheless disappears thanks to the  total momentum conservation relation, as already noted at the end of Section \ref{sec:conservation}. We thus omit to explicitly write this term in the single puncture flux as it simply represents a shift in the string position with no physical relevance.

From the Poisson brackets \eqref{QQ}, it is straightforward to show that the $A^a_{p}$'s and $S^a_p$'s commute with each other and that they each form a $\su(2)$ algebra. We thus recover the expected $\su(2)$ Poisson algebra for the flux  $\Sigma_{p}$.
Comparing to the usual framework for loop quantum gravity, we see that the $\SU(2)$ flux is now a composite object obtained as the sum of the string (0-mode) angular momentum (in internal space) plus a spin  encoded in the higher vibration modes.
We analyze this in detail at the quantum level in the next sections.

\section{The Quantum Puncture}
\la{sec:QP}

In this section, we quantize the space phase of a single puncture. The Kac-Moody charges become the elementary operators and we reconstruct the geometric flux operator, giving the physical area, as a composite operator resulting from the recoupling of the spins for every  mode of the string around the puncture.

\subsection{The ladder of Kac-Moody operators}
\la{sec:kacmoody}

Focusing on a single puncture, according to the field decomposition \eqref{PhiXPQ} giving the triad in a neighborhood of the puncture on the boundary surface, the basic observables are the 0-mode variables $(X,P)$ and the charges $Q_{n}$, which satisfy the Poisson brackets \eqref{xP} and \eqref{QQ}:
\ba
\{X^a,P^{ b}\}= {\delta^{a\bar{b}}}
\,,\qquad
\{Q_n^a,{Q}_m^b\} = -i {\delta^{a\bar{b}}}  (n+k^a) \delta_{n+m}.
\ea
We also have that 
\be 
P^a=Q_{-k^a}^a
\,,\qquad
\{X^a,Q_n^b\}=\{P^a,Q_n^b\}=0
\,,
\ee
with the reality conditions $\bar{Q}^a_{n}=Q^{\bar{a}}_{-n}$ (recall that we work in the complex diagonal basis  $a=(3,+,-)$ , with $\bar{a}=(3,-,+)$ denoting the conjugate basis).
We will work in  the gauge-fixed choice $k^a=(0,+k,-k)$. 
Quantizing these observables and their Poisson algebra, the charges $Q_n^a$ form a twisted Kac-Moody algebra \cite{DiFrancesco:1997nk}, which can be conveniently repackaged as  a collection of harmonic oscillators. Moreover, the integral twist $k\in\Z$ can be entirely re-absorbed as a mere shift in the labelling of the harmonic oscillators and does not play any role in representing the algebra of the puncture observables at the quantum level. 

Raising the observables $(X,P,Q_{n})$ to quantum operators, we could use the standard convention to write $(\hat{X},\hat{P},\hat{Q}_{n})$ to distinguish them from the classical observables. We will nevertheless omit the extra $\hat{ }$ to alleviate the notations.
First of all, we start by quantizing the 0-mode observables:
\be
[X^{3},P^{3}]=i
\,,\qquad
[X^{\pm},P^{\mp}]=i
\,.
\ee
The momentum coordinates $P^a$ are Casimir operators for the whole Kac-Moody algebra:
\be
P^3=Q^3_{0}
\,,\quad
P^\pm=Q^\pm_{\mp k}
\,,\quad
[P^a,Q^b_{n}]=0
\,,\,\,
\forall a,b,n
\,.
\ee
Then it is convenient to repackage the higher mode operators by re-absorbing the $\pm k$ shifts, for $n\in\N$, $n\ge 1$, thus defining a tower of decoupled harmonic oscillators\footnotemark{}:
\be
\alpha_n^a \equiv Q^a_{n-k^a}
\,,\qquad
\alpha_n^a{}^\dagger= \alpha_{-n}^{\bar{a}}=Q^{\bar{a}}_{-n-k^{\bar{a}}}
\,,
\ee
with canonical commutation relations
\be
\la{alphas}
\left[\alpha_n^a\,,\,\alpha_m^b{}^\dagger\right] = n \delta^{ab} \delta_{n,m}\,,
\ee
and all other commutators vanishing, in particular $[\alpha^a_{n},\alpha^b_{n}{}^\dagger]=0$ for $a\ne b$ and $[\alpha^a_{n},\alpha^b_{m}{}^\dagger]=0$ for $n\ne m$.
\footnotetext{
For our gauge-fixed choice of $k^a=(0,+k,-k)$ with $k\in\Z$, the explicit shifts in the mode labelling of the Kac-Moody read $\km^3_{n}=Q^3_{n}$ and $\km^+_{n}=Q^+_{n-k}$, and $\km^-_{n}=Q^-_{n+k}$, with the commutation relations $[\km^3_{n},\km^3_{n}{^\dagger}]=[\km^+_{n},\km^+_{n}{^\dagger}]=[\km^-_{n},\km^-_{n}{^\dagger}]=n$ for $n\ge 1$.
}
%

This confirms that  an integral twist $k$ does not play an important role in the algebraic structure and simply creates an offset in identifying the frequency of the vibration modes.
It is remarkable that taking the curvature $k$ to be an integer (in Planck units) does not affect the structure of the charges and quantum observables for the puncture. This means that as far as the quantum theory is concerned, a shift of the curvature is invisible and the connection is essentially compact. This can be understood as reflecting the periodicity of the $\SU(2)$ holonomy around the puncture ${\cal P}\exp\oint A$ as we vary the Ashtekar--Barbero connection $A$.
The compactification of the connection also implies that its conjugated variables which are  geometrical operators possess discrete spectra.
Therefore, we witness here a new mechanism behind the quantisation of geometry: the invisibility of the integral shift of the connection, even if we do not work with holonomy operators.
What is remarkable is that a shift of $k$ correspond to a Planckian curvature
excitation, a very non-classical configuration.
It would thus be interesting to investigate in the future the case of a non-integral twist $k\in ]0,1[$, defining a non-trivial curvature at the Planck scale.

For each mode $n\ge 1$, the operators $\alpha_{n}$ and $\alpha_{n}^\dagger$ represent a canonical pair of conjugate 3-vectors $(\vx_{n},\vp_{n})\in\R^3\times \R^3$ written in the ``light-cone'' basis  $a=(3,+,-)$. Indeed,  starting from the basic commutation relations $[x_n^j,p_n^k]=in\,\delta^{jk}$, for Hermitian operators $x_n=x_n^\dagger$ and $p_n=p_n^\dagger$, we consider the linear combinations:
\be
\la{xpalpha1}
x_n^{\pm}=\f1{\sqrt{2}}\big{[}x_n^{1}\pm i x_n^{2}\big{]}
\,,\quad
p_n^{\pm}=\f1{\sqrt{2}}\big{[}p_n^{1}\pm i p_n^{2}\big{]}
\,,
\ee
From which we can construct the oscillators
\be
\la{xpalpha2}
\alpha_n^{\pm}=\f1{\sqrt{2}}\big[ x_n^{\pm}+ip_n^{\pm}\big]
\,,\quad
\alpha^{3}_n=\f{1}{\sqrt{2}}\big{[}x^{3}_n+ip^{3}_n\big{]}
\,,
\ee
which leads back to the wanted Poisson  brackets $[\alpha_n^a,\alpha_n^b{}^\dagger]=n \delta^{ab}$.
The mode frequency $n$ is then taken into account directly in the commutation relation between $x$'s and $p$'s.

\medskip

We naturally introduce the Hilbert space of states for the quantum puncture, with basis states labelled by the momentum $\vp\in\R^3$ for the 0-mode and diagonalizing the oscillator energies $E^a_{n}=\alpha^a_{n}{^\dagger}\alpha^a_{n}$ in terms of the numbers of quanta  ${N}^a_{n}\in\N$ for the higher modes $n\ge 1$:
\be
P^a\,|\vp,\{N^a_{n}\}\ket=p^a\,|\vp,\{N^a_{n}\}\ket
\,,\quad
E^a_{n_{0}}\,|\vp,\{N^a_{n}\}\ket=n_{0}N^a_{n_{0}}\,|\vp,\{N^a_{n}\}\ket
\,,
\ee
\be
\left|\begin{array}{l}
\alpha_{n_{0}}^a\,|\vp,\{N^a_{n}\}\ket
=
\sqrt{n_{0}N^a_{n_{0}}}\,|\vp,N^a_{n_{0}}-1,\{N^a_{n}\}_{n\ne n_{0}}\ket
\vspace*{2mm}\\
(\alpha_{n_{0}}^a{})^\dagger \,|\vp,\{N^a_{n}\}\ket
=
\sqrt{n_{0}(N^a_{n_{0}}+1)}\,|\vp,N^a_{n_{0}}+1,\{N^a_{n}\}_{n\ne n_{0}}\ket
\end{array}
\right.
\,,
\ee
where
\be
|\vp,\{N^a_{n}\}\ket=
\prod_{a,n\ge1}
\f{(\km^a_{n}\dag)^{N^a_{n}}}{\sqrt{N^a_{n}!}}
\,|\vp\ket\,,
\ee
and the state $|\vp\ket$ is a shorthand for $|\vp,\{0\}\ket$.

\subsection{Virasoro generators}

As introduced in \cite{Freidel:2016bxd},  using the Sugawara construction, we can  define the generators of the Virasoro algebra  by a normal ordering:
\ba
\la{L}
&&  L_{n} = \frac12 \sum_a\sum_{m\in\,\Z}\! : \alpha^a_{m}  \alpha^{\bar a}_{n-m}:\,,
\ea
 where $:\ \ :$ stands for the normal ordering defined by
\be
:{\alpha}^a_{n}{\alpha}^b_{m}\!:\,=\begin{cases}
         \alpha^b_{m} \alpha^a_{n}\quad\text{if}\;n>0\,,\\
          \alpha^a_{n} \alpha^b_{m}\quad\text{if}\;n\leq0\,.\end{cases}  
\ee
These operators  satisfy the reality conditions $L^\dagger_{n}=L_{-n}$
and their commutators gives the Virasoro Lie algebra with a central charge $c=3$:
\ba\la{Vir}
[ L_{n},  L_{m}]=(n-m)  L_{n+m} + \frac{c}{12}n(n^2-1)\delta_{n+m,0}
\,.
\ea
These generate reparametrization of the string around the puncture.
They appear from the charge
\be
L_n = \tfrac1{2\kappa \gamma} \oint e^{in\theta} e_\theta^a e_{\theta a}
\ee which encodes  all local angular deformations of the geometry.

The zero mode generator $L_{0}$ is as expected the total energy:
\be
L_{0}=\f12 \vP^2+\sum_{n\ge 1}E_{n}
\qquad\textrm{with}\quad
\left|\begin{array}{l}
\vP^2=\sum_{a} P^aP^{\bar{a}}
\vspace{1mm}\\
E_{n}=\sum_{a}E_{n}^a=\sum_{a}\km^a_{n}{}^\dagger\km^a_{n}{}
\end{array}
\right.
\,,
\ee
while the positive Virasoro modes $n\ge1$ can be recast as
\be
L_{n\ge 1}
=
P^a\km^a_{n}
+\sum_{m\ge 1}\alpha^a_{m}{}^\dagger\alpha^a_{m+n}
+\f12\sum_{m=1}^{n-1}\alpha^a_{n-m}\alpha^a_{m}
\,.
\ee
We compute the action of the Virasoro generators on the Kac-Moody operators,
\be\la{LQ-1}
[ L_{n},  \alpha^a_{m}]=-m \alpha^a_{n+m}\,,\qquad {[}L_{n},X^a{]}=-i\alpha^a_{n}\,,
\ee
which we can translate into an action of the 0-mode canonical coordinates $(X^a,P^a)$ and the harmonic oscillators. Besides the momentum $P^a$ which is invariant under the transformation generated by the Virasoro generators, $[L_{n},P^a]=0$ for all $n\in\Z$, we get the following commutators:
\be
{[}L_{0},X^a{]}=-iP^a
\next
{[}L_{0},\km^{a}_{m}{]}=-m\km^{a}_{m}
\next
{[}L_{0},\km^{a}_{m}\dag{]}=m\km^{a}_{m}\dag
\,.
\ee

The states $|\vp,\{N^a_{n}\}\ket$ define a highest weight representation of the Virasoro algebra for each fixed value of the momentum $\vp\in\R^3$. The highest weight vector $|\vp\ket\equiv |\vp,\{N^a_{n}=0\}\ket$ has vanishing number of quanta for all the oscillators $n\ge 1$,
\be\la{primary}
P^a\,|\vp\ket=p^a\,|\vp\ket
\,,\qquad
L_{0}\,|\vp\ket
\,=\,
\f12\vp^2|\vp\ket
\,,\qquad
L_{n\ge1}\,|\vp\ket
\,=\,
0
\,.
\ee
The other states are obtained by populating the number of quanta for the oscillators $N^a_{n}\in\N$:
\be
\la{L0}
L_{0}\,|\vp,\{N^a_{n}\}\ket
\,=\,
\left(
\f12\vp^2
+\sum_{n}nN_{n}
\right)
\,|\vp,\{N^a_{n}\}\ket
\,,\qquad\textrm{with}\quad
N_{n}=\sum_{a}N^a_{n}
\,.
\ee
We distinguish the contribution $\f12{p^2}$ of the 0-mode from the higher mode contribution $\sum_{n}nN_{n}$ with the number of quanta $N_{n}$ for each frequency $n\in\N,n\ge 1$.

\subsection{The quantum flux operator}
\label{sec:spinflux}

Let us now represent the geometrical flux  $\Sigma^a$ as quantum operators acting on the Hilbert space of the quantum puncture.
As described earlier in Section \ref{sec:fluxclassical},  the LQG flux can be decomposed in terms of the string orbital angular momentum $A^a$ of the 0-mode of the string around the puncture and the collective spin angular momentum $S^a$ of the higher string modes $n\ge 1$:
\be
\Sigma^a=
A^a
+
S^a
\,,\qquad
\vA=\vX\w\vP
\,,\qquad
S^a=\sum_{n\ge1} S^a_{n}
\,,\qquad
[P^a,S_{n}^b]
=
[X^a,S_{n}^b]
=0
\,,
\ee
where the operator ordering for the 0-mode angular momentum does not matter, $A^i=\eps^{ijk}X^iP^j=-\eps^{ijk}P^jX^k$, and where the spin operators for each mode $n\ge1$ are defined in terms of the Kac-Moody operators as:
\ba
\la{Sn}
S^3_{n}
&=&
\f1n
\left[
\alpha^-_{n}\dag\alpha^-_{n}
-
\alpha^+_{n}\dag\alpha^+_{n}
\right]
\,,\\
S^+_{n}
&=&
\f1n
\left[
\alpha^3_{n}\dag\alpha^+_{n}
-
\alpha^-_{n}\dag\alpha^3_{n}
\right]
\,,\n\\
S^-_{n}
&=&
\f1n
\left[
\alpha^+_{n}\dag\alpha^3_{n}
-
\alpha^3_{n}\dag\alpha^-_{n}
\right]
\,.\n
\ea
The angular momenta for different modes commute with each other 
and each form an independent $\so(3)$ Lie algebra:
\be
\,[A^i,S^a_{n}]=0
\,,\qquad
[A^i,A^j]=\eps^{ij}{}_k A^k,\qquad 
[S^a_{n},S^b_{m}]=\delta_{n,m}\, \epsilon^{ab}{}_c S^c_n.
\ee
The epsilon tensor is totally skew and in the complex basis reads 
$\epsilon^{3+-}=1=\epsilon^{3+}{}_+$.
The explicit expression for the commutator in this basis is 
\be
[S^3_{n},S^\pm_{n}]=\pm S^\pm_{n}
\,,\quad
[S^+_{n},S^-_{n}]=S^3_{n}
\,.
\ee
The quadratic Casimir of the $\so(3)$ algebra for each spin angular momentum is given by 
\be\label{Cas}
\vS^2_{n}=(S^3_{n})^2+(S^+_{n}S^-_{n}+S^-_{n}S^+_{n})=S^3_{n}(S_n^3+1)+ 2S^-_{n}S^+_{n}
\,.
\ee
Just as the orbital angular momentum operators $A^i$ generates $\SO(3)$ rotations on the pair of 3-vectors $(\vX,\vP)$, the spin angular momentum $\vS_{n}$, for each mode $n\ge 1$, gives the $\so(3)$ generators of the 3d rotations acting on the pair of 3-vectors\footnotemark{} $(x_n^a,p_n^a)$ defined by the harmonic oscillators.
\footnotetext{
As explained in Section \ref{sec:kacmoody}, for each mode $n$, the harmonic oscillators $(\km^a_{n},\km^a_{n}\dag)$ define a complex 3-vector, which can be written as a pair of canonically conjugate real 3-vectors $(x_{n}^i,p_{n}^i)$ upon changing from the $i=(1,2,3)$ basis to the $a=(3,+,-)$ basis as given in  \eqref{xpalpha1} and \eqref{xpalpha2}.
We define the angular momentum operators $S_n^{a}=\eps^{a}{}_{bc}x_n^{b}p_n^{c}$, with commutators $[S_n^{a},S_n^{b}]=i\eps^{ab}{}_c S_n^{c}$  valid in any basis.
}
This translates into the following commutators between the spin operators $S^a$ and the harmonic oscillator creation operators\footnote{
We  also give the Hermitian conjugate of those commutators for the sake of completeness:
\be
\begin{array}{lll}
{[}S^3,\km^3_{n}\dag]=0
\,,&
{[}S^3,\km^+_{n}\dag]=-\km^+_{n}\dag
\,,&
{[}S^3,\km^-_{n}\dag{]}=\km^-_{n}\dag
\,,\vspace{1mm}\\
{[}S^+,\km^3_{n}\dag{]}=-\km^-_{n}\dag
\,,&
{[}S^+,\km^+_{n}\dag{]}=\km^3_{n}\dag
\,,&
{[}S^+,\km^-_{n}\dag{]}=0
\,,\vspace{1mm}\\
{[}S^-,\km^3_{n}\dag{]}=\km^+_{n}\dag
\,,&
{[}S^-,\km^+_{n}\dag{]}=0
\,,&
{[}S^-,\km^-_{n}\dag{]}=-\km^3_{n}\dag
\,.
\end{array}
\nn
\ee
}
:
\be
\begin{array}{lll}
{[}S^3,\km^3_{n}]=0
\,,&
{[}S^3,\km^+_{n}]=\km^+_{n}
\,,&
{[}S^3,\km^-_{n}{]}=-\km^-_{n}
\,,\vspace{1mm}\\
{[}S^+,\km^3_{n}{]}=-\km^+_{n}
\,,&
{[}S^+,\km^+_{n}{]}=0
\,,&
{[}S^+,\km^-_{n}{]}=\km^3_{n}
\,,
\vspace{1mm}\\
{[}S^-,\km^3_{n}{]}=\km^-_{n}
\,,&
{[}S^-,\km^+_{n}{]}=-\km^3_{n}
\,,&
{[}S^-,\km^-_{n}{]}=0
\,.
\end{array}
\ee

Now that we have realized the flux operator as an infinite sum of independent spin operators, $\Sigma^a=A^a+\sum_{n\ge1}S_{n}^a$, this means that the LQG spin will result from the recoupling of the spins coming from all the string modes $n\ge 0$. We study in detail this fine structure in the next section, which is dedicated to exhibiting the change of basis from the  states $|\vp,\{N^a_{n}\}\ket$ to states diagonalizing the spins $\vA^2$ and $\vS_{n}^2$ and the total recoupled spin $\overrightarrow{\Sigma}^2$.

\section{The Fine Structure of the Flux}
\la{sec:TotAM}

This section explores the fine structure of the geometrical flux $\overrightarrow{\Sigma}$ as the sum of the orbital angular momentum $\vX\w\vP$ of the string 0-mode and the spins $\vS_{n}$ of every higher mode $n\ge1$, thus implying that the total loop quantum gravity (LQG) $\su(2)$ structure defined by the $\Sigma^a$ is obtained from the recoupling of all those spins. 
The goal of this section is to perform explicitly the change of basis
from the string states $|\vp, \{N_n^a\}\ket$ diagonalizing the energy operators $E_n$ and the spin basis diagonalizing the spin operators $S_n^2$. 


\subsection{Diagonalizing the 0-mode angular momentum}

We start with the 0-mode angular momentum $\vA=\vX\w\vP$ and we describe the change of basis from the momentum basis $|\vp\ket$ to the rotation basis $|\Delta,l,m\ket$ diagonalizing 
$(\vP^2, \vA^2, A_3)$ with respective  eigenvalues $(\Delta, l(l+1), m)$.
In practice, this is achieved by diagonalizing the 3d Laplacian in spherical coordinates.

\begin{figure}[h!]
\centering

\includegraphics[width=70mm]{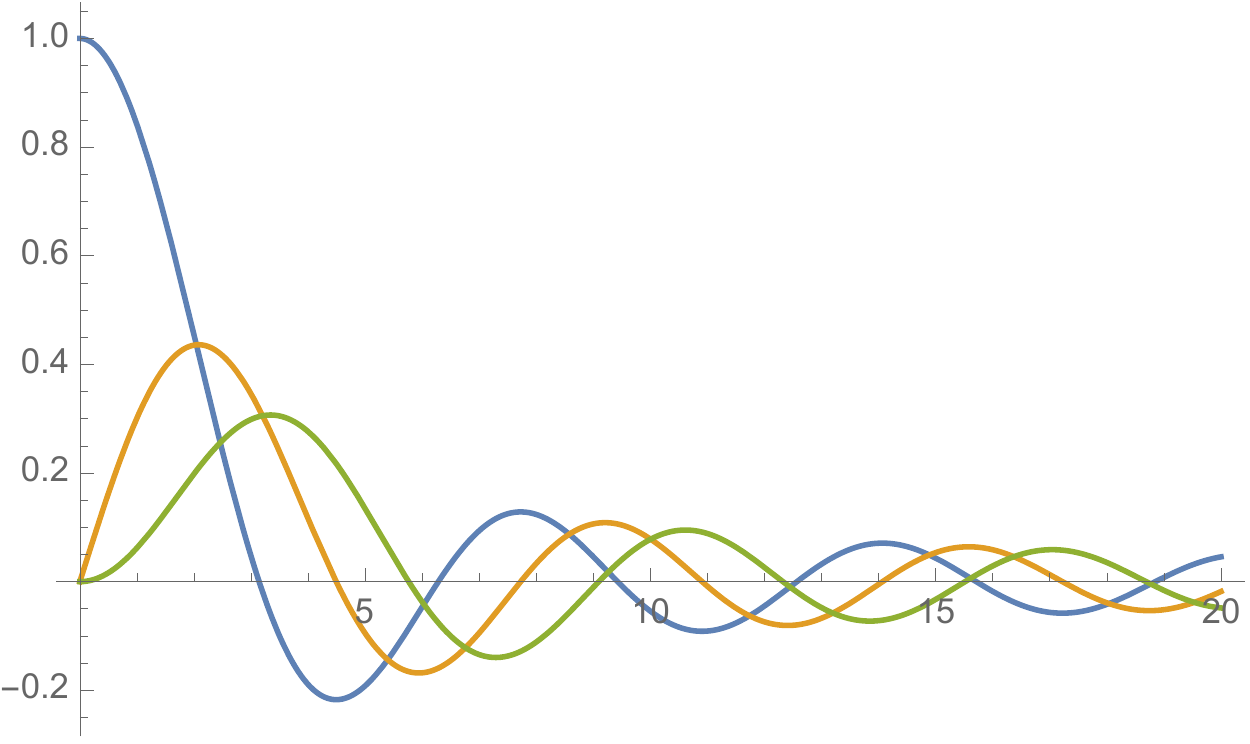}

\caption{Plots of the spherical Bessel functions $\rho_{l}(r)$ for $l=0,1,2$ (with the amplitude decreasing as $l$ increases) giving the radial solution to Helmholtz equation for $\Delta=1$.
When $l=0$, the spherical Bessel function $\rho_{0}(r)$ gives back the sine cardinale function $\sin r /r$.
 \label{fig:Besselplot}}

\end{figure}

More precisely, let us work in the $X$-polarization so that the momentum is quantized as $P^i=-i\pp_{i}$, where $\pp_{i}$ stands short for $\pp_{X^{i}}$, and the angular momentum operators read $A^i=-i\eps^{ijk}X_j\pp_{k}$. The operator $\vP^2$ is the 3d Laplacian $-\sum_{i=1}^3\pp_{i}\pp_{i}$ while the $\so(3)$ Casimir $\vA^2$ is the Laplacian on the 2-sphere. It is convenient to switch to spherical coordinates $(r,\theta,\phi)$ with the radius $r=\sqrt{\vX^2}$, the zenith angle (or co-latitude) $\theta$ and the azimuthal angle $\phi$:
\be
\pp^2
\equiv
\pp_{i}\pp_{i}
=
\f1{r^2}\pp_{r}\Big{[}r^2\pp_{r}\Big{]}+\f1{r^2\sin\theta}\pp_{\theta}\Big{[}\sin\theta\pp_{\theta}\Big{]}+\f1{r^2\sin^2\theta}\pp_{\phi}
\,,
\ee
It is well-known \cite{Varshalovich:1988ye} that in these coordinates, the angular momentum operator is given by 
\be
\vA^2
= r\pa_r (r\pa_r+1) -r^2\pp^2
= r^2\Big{[}\pp_{r}^2+\f2r\pp_{r}-\pp^2\Big{]}
=
-\bigg{[}
\f1{\sin\theta}\pp_{\theta}\big{[}\sin\theta\pp_{\theta}\big{]}
+
\f1{\sin^2\theta}\pp_{\phi}
\bigg{]}
\,.
\ee
In order to diagonalize both the Laplacian $\pp^2$ and the angular momentum $\vA^2$, we separate the radial and angular dependences and  factor the 0-mode wave-function as $F(r)Y(\theta,\phi)$.
We first diagonalize the angular momentum using the spherical harmonics on the sphere,
\be
\vA^2\,Y^l_{m}(\theta,\phi)=l(l+1)\,Y^l_{m}(\theta,\phi)
\,,\qquad
A_{3}\,Y^l_{m}(\theta,\phi)=i\pp_{\phi}\,Y^l_{m}(\theta,\phi)=m\,Y^l_{m}(\theta,\phi)\,,
\ee
where $l\in\N$ is an integer and $m$ runs by integer steps between $-l$ and $+l$.
Then the radial dependent component  satisfies the Helmholtz equation
\be
\pp_{r}^2F+\f2r\pp_{r}F-\f{l(l+1)}{r^2}F=-\Delta F
\,,
\ee
which is solved by the spherical Bessel functions\footnotemark{}
\footnotetext{
If we do not require that the function be regular at $r=0$, then we have extra modes given by the second type of spherical functions, $\tilde{\rho}_{l}(r\sqrt{\Delta})\,Y^l_{m}(\theta,\phi)$ with $\tilde{\rho}_{l}(y)=Y_{l+\f12}(y)\sqrt{\f{\pi}{2y}}$.
}
$F(r)=\rho_{l}(r\sqrt{\Delta})$, with
\be
\rho_{l}(y)
=
\sqrt{\f{\pi}{2y}}\,J_{l+\f12}(y)
=
(-y)^l\left(y\f{\pp}{\pp y}\right)^l\,\f{\sin y}y
=
2^l\,\sum_{n\in\N}
y^{l+2n}
\f{(-1)^n\,(n+l)!}{n!(2n+2l+1)!}
\,,
\ee
which we plot in  Figure \ref{fig:Besselplot}.
These spherical Bessel functions are usually noted $j_{l}$, but we call them instead $\rho_{l}$  in order to avoid confusion with other notations (e.g. the spins). 

Combining the radial and angular dependences gives the Helmholtz basis states\footnotemark{}
$\bra \vX |\Delta,l,m\ket=\rho_{l}(\sqrt{\Delta}\,r)\,Y^l_{m}(\theta,\phi)$,
which diagonalize both the momentum norm $\vP^2$ and the orbital angular momentum:
\be
\vP^2 \,|\Delta,l,m\ket=\Delta \,|\Delta,l,m\ket
\,,\quad
\vA^2\,|\Delta,l,m\ket=l(l+1)\,|\Delta,l,m\ket
\,,\quad
A_{3}\,|\Delta,l,m\ket=m\,|\Delta,l,m\ket
\,.
\ee
\footnotetext{
Using the spherical coordinate factorization of the Fourier modes in terms of the Bessel functions $\rho_{l}$ and Legendre polynomials ${\cal P}_{l}$,
\be
e^{i\vX\cdot\vP}
=
\sum_{l\in\N}i^l (2l+1) \rho_{l}(XP){\cal P}_{l}(\hat{X}\cdot\hat{P})
=
4\pi\sum_{l\in\N}i^l  \rho_{l}(XP)Y^l_{m}(\hat{P})\bar{Y}^l_{m}(\hat{X})
\,,
\nn
\ee
where we distinguish the norm and unit direction of the coordinate and momentum vectors, $\vP=P\hat{P}$ and $\vX=X\hat{X}$,
we can also give the decomposition of the states $|\Delta,l,m\ket$ in the momentum basis:
\be
\bra \vP |\Delta,l,m\ket
=
\f{2\pi^2 \,i^l}\Delta \delta(p-\sqrt{\Delta})
\,Y^l_{m}(\hat{P})
\,.
\nn
\ee
}
The scalar product between the Helmholtz basis states are given by the choice of normalization for the spherical harmonics and the spherical Bessel functions:
\beq
\bra \Delta,l,m|\tilde \Delta,\tl,\tm\ket
&=&
\int \rd^2\Omega\,\,\overline{Y^l_{m}}(\theta,\phi)\,Y^{\tl}_{\tm}(\theta,\phi)
\,
\int_{0}^{+\infty}r^2\rd r\,\,\overline{\chi_{l}}\left(r\sqrt{\Delta}\right)\chi_{l}\left(r\sqrt{\tilde \Delta}\right)
\nn\\
&=&
\delta_{l\tl}\delta_{m\tm}\,
\f{\pi}{2\Delta}\delta\left({\sqrt{\Delta}-\sqrt{\tilde \Delta}}\right)
\,.
\eeq
Compared to the loop quantum gravity puncture (i.e. the end of a spin network edge) carrying a state $|j,m\ra$ in a $\SU(2)$ irreducible representation, the 0-mode state of the puncture now carries an extra quantum number $\Delta$, which gives the norm squared of the 0-mode momentum $\vP$.
On top of this, the puncture will carry an infinite tower of extra quantum numbers characterizing the states of the higher mode of the string, as we describe below.

\subsection{Spin diagonalization for the higher modes and $\sll_{2}(\R)$-degeneracy}

The higher modes of frequency $n\ge 1$,  consist in  three harmonic oscillators $(\alpha^a_{n},\alpha^a_{n}{^\dagger})$ with $a\in\{3,+,-\}$ and the Hilbert space is spanned by the basis states $|\{N^a_{n}\}\ra$ labeled by the number of quanta $N^a_{n}\in\N$ for each oscillator. 
We would like to switch basis and identify states that diagonalizes both the angular momentum $\vS_{n}$ defined earlier in \eqref{Sn} and the total energy of the oscillators $E_{n}=\sum_{a}E^a_{n}$. This energy gives the contribution of the $n$-mode to the Virasoro generator $L_{0}$ as written in \eqref{L0}.

The setting is very similar to the case of the 0-mode described in the previous section. As explained in Sections \ref{sec:kacmoody} and \ref{sec:spinflux}, the triplet of harmonic oscillators $(\alpha^a_{n},\alpha^a_{n}{^\dagger})$  represent a canonical pair of 3-vectors, which we can denote $(\vx_{n},\vp_{n})$. Then the spin angular momentum $\vS_{n}$ is simply $\vx_{n}\times\vp_{n}$. Just as with the 0-mode, we would like to diagonalize the squared angular momentum $\vS_{n}^2$ and its component $S_{n}^3$, respectively with eigenvalues $j_{n}(j_{n}+1)$ in terms of the spin $j_{n}\in\N$ and the magnetic moment $m_n$ ranging between $-j_{n}$ and $+j_{n}$. And as in the case of the 0-mode, these two quantum numbers are not enough to label basis states and the spin $j_{n}$ has a non-trivial degeneracy. However, unlike for the 0-mode, we are not interested in labelling this degeneracy by the eigenvalues of the momentum $\vp_{n}^2$ but by eigenvalues of the energy $E_{n}=\vp_{n}^2+ n \vx_{n}^2$. This leads to a different construction of basis states.

More precisely, we wish to diagonalize the total energy carried by the $n$-mode, $E_{n}=\sum_{a}\alpha^a_{n}{^\dagger}\alpha^a_{n}=nN_{n}$, and the angular momentum operators $\vS_{n}^2$ and $S^3_{n}$. First, the energy $E_{n}$ obviously commutes with the $\so(3)$ Lie algebra generated by the angular momentum  components. In fact, we can identify two other operators quadratic in the basic oscillator operators, that we denote $f_n$ and its conjugate $f_n^\dagger$ which commute with the angular momentum.
It is convenient to rearrange the basis of $\so(3)$ operators as follows and define
\ba
h_{n}
&:=&
\f1{2n} \sum_{a}(\alpha^a_{n}\dag\alpha^a_{n}+\alpha^a_{n}\alpha^a_{n}\dag)
=
N_{n}+\f32
\,,\cr
f_{n}
&:=&
\f1n\sum_a\alpha^a_{n}\alpha^{\bar{a}}_{n}
=
\f1n\big{[}\alpha_{3}^2+2\alpha_{-}\alpha_{+}\big{]}
\,.
\ea
Interestingly one finds that the three elements  $(h_n,f_n,f_n^\dagger)$ form a $\sll_{2}(\R)$ algebra:
\be
[h_{n},f_{n}]=-2f_{n}
\next
[h_{n},f_{n}^\dagger]=+2f_{n}^\dagger
\next
[f_{n},f_{n}^\dagger]=4 h_{n}
\,,
\ee
and that they all commute with $S^a_n$.
This last fact is obvious since $(h_n,f_n,f_n^\dagger)$ are linear combination of the rotational invariants $\vx_{n}^2$, $\vp_{n}^2$ and $\vx_{n}\cdot\vp_{n}$ which  are  the squeezing operators for the triplet of harmonic oscillators. 
This means that those three operators, $h_{n}$, $f_{n}$ $f_{n}^\dagger$, act on states without affecting the action of the angular momentum operators  $\vS_{n}^2$ and $S^3_{n}$. 
In the following one denotes $j_n(j_n+1)$ the eigenvalue of $\vS_{n}^2$ and $m_n$ the eigenvalue of $S^3_{n}$.
The $\sll_{2}(\R)$ operators acts diagonally on states $|j_{n},m_{n}\rangle$
and therefore can be used to characterise the 
 degeneracy of such spin states.

The key to understanding the action of the $\sll_{2}(\R)$ operators on spin states is determining in which $\sll_{2}(\R)$-representation the spin states live. This is given by a Casimir balance equation equating the $\so(3)$ quadratic Casimir to the $\sll_{2}(\R)$ quadratic Casimir.
The $\so(3)$ Casimir is decribed in (\ref{Cas}) and given by 
$\vS_{n}^2
=
S^3_{n}(S_n^3+1)+ 2S^-_{n}S^+_{n}$  the $\sll_{2}(\R)$ Casimir is given by
\be
C:= h_{n}^2-\f12(f_{n}^\dagger f_{n}+f_{n}f_{n}^\dagger)+\f{3}4=N_n(N_n+1) - f_n^\dagger f_n.
\ee
The balance equation that follows from their expression in terms of the oscillators is simply
\be
\vS^2=C.
\ee
Such a $\su(2)\times\sll_{2}(\R)$ algebraic structure with the same balance equation already appeared  in \cite{Freidel:2015gpa,Freidel:2018pvm}, where it related the $\su(2)$ geometrical flux to the 2d surface metric.
Here this relation allows us to go from the string quanta basis $|\{N_{n}^a\} \ket$  at level $n$ to a basis of states $|j_{n},m_{n},d_{n}\ket$ diagonalizing $\so(3)$ Casimir $\vS_{n}^2=j_{n}(j_{n}+1)$, the $\so(3)$ generator $S^3_{n}=m_{n}$ and the total  number of quanta $N_{n}$. The extra quantum number 
\be 
d_{n} := \frac{(N_n - j_n)}2
\ee labels the degeneracy of the spin states. 
We now show that $d_n\in \mathbb{N}$ is a positive integer and we construct a map
\be
 |N^a_{n} \ket \to |j_{n},m_{n},d_{n}\ket,
\ee
whic is a unitary isomorphism.  

First, in a basis that diagonalises $\vS^2$ and $N_n$,
the Casimir balance equation reads
\be
 f_n^\dagger f_n = N_n(N_n+1)-j_n(j_n+1).
\ee
Since the operator $f_n^\dagger f_n$ is positive, this  implies that the spin is bounded from above by the number of quanta, $0\le j_{n}\le N_{n}$.
It is therefore natural to consider $d_n:= (N_n-j_n)/2$.
We can easily see that $N_n$ and $j_n$ have the same parity, therefore $d_n$  is a positive integer.
Then, on the highest states $|\psi\rangle$ annihilated by the lowering $\sll_{2}(\R)$-operator, $f_n |\psi\rangle=0$, the balance equation further implies that the spin $j_{n}$ is exactly equal to the number of quanta $N_{n}$:
\be
\left|
\begin{array}{l}
f_n \,|\psi\rangle=0 \vspace*{1mm}\\
\vS_{n}^2 \,|\psi\rangle=j_{n}(j_{n}+1)\,|\psi\rangle \vspace*{1mm}\\
h_{n} \,|\psi\rangle=(N_{n}+\f32)\,|\psi\rangle 
\end{array}
\right.
\qquad\Longrightarrow\quad
 j_n(j_n+1) 
=
N_n(N_n+1)
\,.
\ee
Furthermore, acting on such a state with the $\sll_{2}(\R)$ raising operator $f_{n}^\dagger$ does not  affect the spin $j_{n}$ but it increases the number of quanta, thereby allowing to create a shift between $j_{n}$ and $N_{n}$:
\be
\left|
\begin{array}{l}
f_n \,|\psi\rangle=0 \vspace*{1mm}\\
\vS_{n}^2 \,|\psi\rangle=j_{n}(j_{n}+1)\,|\psi\rangle \vspace*{1mm}\\
h_{n} \,|\psi\rangle=(N_{n}+\f32)\,|\psi\rangle 
\end{array}
\right.
\qquad\Longrightarrow\quad
\left|
\begin{array}{l}
\vS_{n}^2 \,(f_{n}^\dagger)^{d_n}\,|\psi\rangle
=
j_{n}(j_{n}+1)\,(f_{n}^\dagger)^{d_n}\,|\psi\rangle|\psi\rangle
\vspace*{1mm}\\
h_{n} \,(f_{n}^\dagger)^{d_n}\,|\psi\rangle
=
(N_{n}+2d_n+\f32)\,(f_{n}^\dagger)^{d_n}\,|\psi\rangle
\end{array}
\right.\,.
\ee
This method allows us to recover all the states of the Hilbert space with arbitrary numbers of quanta. As shown in more details in appendix \ref{app:su2xsl2}, the power $d_n$ of the raising operator $f_{n}^\dagger$ labels the spin degeneracy  for the $n$-th string mode. Then we can define states $|j_{n},m_{n},d_{n}\rangle$ forming an orthonormal basis of the Hilbert space for the triplet of harmonic oscillators and related by a unitary map to the original basis $|\{N_{n}^a\}\rangle$.

To construct the new basis, we start with the states $(\km^-_{n})^{j_n}\,|0\ket$ acting $j_n$ times with the creation operator $\km^+_{n}$ on the vacuum state $|0\ket$ anihilated by $\alpha_n$  with no quanta of energy $N^3_{n}=N^\pm_{n}=0$,
\be
|j_{n},j_{n},0\ket
=\f{(\km^-_{n}\dag)^{j_n}}{\sqrt{n^{j_n} j_n!}}\,|0\ket
\,,\qquad
\bra j_{n},j_{n},0|j_{n},j_{n},0\ket=1
\,,
\ee
which corresponds to populating quanta of the $\km^-_{n}$ oscillator, with $(N^3,N^+,N^-)=(0,0,j_{n})$.
This turns out to be both a highest weight vector for the $\so(3)$ Lie algebra and a lowest vector for the $\sll_{2}(\R)$ algebra:
\be
\left|\begin{array}{l}
S^+_{n}\,|j_{n},j_{n},0\ket=0
\vspace{2mm}\\
f_{n}\,|j_{n},j_{n},0\ket=0
\end{array}\right.
\,,\qquad\qquad
\left|\begin{array}{l}
\vS^2_{n}\,|j_{n},j_{n},0\ket
=
j_n(j_n+1)\,|j_{n},j_{n},0\ket
\vspace{1mm}\\
S^3_{n}\,|j_n,j_n,0\ket=j_n\,|j_{n},j_{n},0\ket
\vspace{1mm}\\
h_{n}\,|j_{n},j_{n},0\ket=(j_n+\f 32)\,|j_{n},j_{n},0\ket
\end{array}\right.
\,.
\ee
Then we obtain the other states by acting with the $\so(3)$ lowering operator $S^-_{n}$ and with the $\sll_{2}(\R)$ raising operator $f_{n}^\dagger$.
The $\so(3)$ representation is determined by the spin $j_{n}$, and  the $\sll_{2}(\R)$ representation is also determined by that same spin $j_{n}$ due to the Casimir balance equation. 
So, on the one hand, acting with the $\so(3)$ operators $S^a_{n}$ on this highest weight vector explores the spin $j_n$ representation of $\so(3)$ (with dimension $2j_n+1$).
On the other hand, acting with the $\sll_{2}(\R)$ operators $h_{n}$ and $f_{n},f_{n}^\dagger$ on $|j_{n},j_{n},0\ket$ explores the infinite-dimensional irreducible representation of $\sll_{2}(\R)$ with spin $j_n$ from the positive discrete series of unitary representations.
The important technicality is that the two sets of operators commute with each other. So  acting with $f_{n}^\dagger$ does not change the $\so(3)$ spin, i.e. the eigenvalue of the $\so(3)$ Casimir $\vS^2_{n}=j_n(j_n+1)$. And reciprocally acting with $S^-_{n}$ does not change the $\sll_{2}(\R)$ spin, i.e. the eigenvalue of the $\sll_{2}(\R)$ Casimir. This way, repetitively acting with $f_{n}^\dagger$ on the initial state $|j_{n},j_{n},0\ket$ leads to an infinite tower of $\so(3)$ irreducible representations with the same spin $j_n$.
Details can be found in appendix \ref{app:su2xsl2}.

In practice, this gives states $|j_{n},m_{n},d_{n}\ket$ labeled by  three integers $j_n,d_n\in\N^2$ and  $m_n\in\Z$ satisfying $-j_n \le m_n \le +j_n$:
\be
|j_{n},m_{n},d_{n}\ket
\propto
(f^\dagger_{n})^{d_n}\,(S^-_{n})^{j_n-m_n}\,|j_n,j_n,0\ket
\qquad
\,,
\left|\begin{array}{l}
N_{n}\,|j_{n},m_{n},d_{n}\ket
=
(j_n+2d_n)\,|j_{n},m_{n},d_{n}\ket
\vspace{1mm}\\
\vS^2_{n}\,|j_{n},m_{n},d_{n}\ket
=
j_n(j_n+1)\,|j_{n},m_{n},d_{n}\ket
\vspace{1mm}\\
S^3_{n}\,|j_{n},m_{n},d_{n}\ket=m_n\,|j_{n},m_{n},d_{n}\ket
\end{array}\right.\,.
\ee
The eigenvalue of the energy operator $E_{n}$ gives the contribution of the $n$-th vibration mode to the total energy $L^0$.
The exact proportionality factors in the definition above  ensure that the states are normalized and can be worked out from the corresponding $\so(3)$ and $\sll_{2}(\R)$ actions (see Appendix \ref{appendix:a} for details):
\beq
|j_{n},m_{n},d_{n}\ket
&=&
\sqrt{\f{(j_n+d_n)!(2j_n+1)!}{j_n!d_n!(2j_n+2d_n+1)!}}
(f^\dagger_{n})^{d_n}
\,
\sqrt{\f{2^{j_n-m_n}(j_n+m_n)!}{(2j_n)!(j_n-m_n)!}}
(S^-_{n})^{j_n-m_n}
\,|j_{n},j_{n},0\ket
\nn\\
&=&
\sqrt{\f{2^{j_n-m_n}(2j_n+1)(j_n+d_n)!(j_n+m_n)!}{j_n!d_n!(2j_n+2d_n+1)!(j_n-m_n)!}}
(f^\dagger_{n})^{d_n}
\,
(S^-_{n})^{j_n-m_n}
\,|j_{n},j_{n},0\ket
\,,
\eeq
so that 
\be
\bra j'_{n},m'_{n},d'_{n}|j_{n},m_{n},d_{n}\ket
=
\delta_{j_n,j_n'}\delta_{m_n,m_n'}\delta_{d_n,d'_n}
\,.
\ee
%
For $d_n=0$, we also provide an explicit formula for the state corresponding to the $\so(3)$ state $|j_n,m_n\ket$ and with no $\sll_{2}(\R)$ excitation as a quantum superposition of states $|\{N^a_{n}\}\ket$  with $(N^{3}_n+N^{-}_n+N^{+}_n)=j_n$ and $(N^{-}_n-N^{+}_n)=m_n$, namely\footnote{
The normalization can be checked from the summation formula:
\be
\sum_{a=0}^{\f{j_n-m_n}2}
\f{1}{2^{2a}\,a!(j_n-m_n-2a)!(m_n+a)!}
=
\f{2^{j_n+m_n}\Gamma(j_n+\f12)}{\sqrt{\pi}(j_n+m_n)!(j_n-m_n)!}\,,
\ee
with
$\Gamma(j_n+\f12)
=
\sqrt{\pi}\left(j_n-\f12\right)\left(j_n-\f32\right)..\f12
$.
}:
\ba
|j_{n},m_{n},0\ket
&=&
\sqrt{\f{2^{j_n-m_n}j_n!(j_n+m_n)!(j_n-m_n)!}{(2j_n)!}}\n\\
&&\times
\sum_{a=0}^{\f{j_n-m_n}2}
\f{(-1)^a}{2^a}
\,
\f{(\alpha^+\dag)^{a}(\alpha^-\dag)^{m_n-a}(\alpha^3\dag)^{j_n-m_n-2a}}{a!(j_n-m_n-2a)!(m_n+a)!}
\,\,|0\ket
\,.
\ea

In the general case, when $d_n\ne 0$, the state $|j_{n},m_{n},d_{n}\ket$ is a superposition of states $|\{N^a_{n}\}\ket$ with $N^3_n=j_n-m_n-2(a-b)$, $N^+_n=d_n+(a-b)$ and $N^-_n=d_n+m_n+(a-b)$ with $a$ running from 0 to $(j_n-m_n)/2$ and $b$ running from 0 to $d_n$. The total number of quanta is now the shifted spin $N_n=j_n+2d_n$ and the magnetic index remains $m_n=(N^{-}_n-N^{+}_n)$, while the spin $j_n$ itself is not a simple combination of the quanta of energy of the three oscillators (a similar but considerably simpler construction  in 3D gravity appeared in \cite{Wieland:2018ymr}  to quantize the boundary geometry). As shown in appendix \ref{app:su2xsl2}, it is possible to reciprocally decompose the states of the original basis $|\{N^a_{n}\}\ket$ in terms of the new basis states $|j_{n},m_{n},d_{n}\ket$ labeled by the spin $j_{n}$, the $\so(3)$  level $m_{n}$ and the $\sll_{2}(\R)$ level $d_{n}$.
This means that the map 
\be
 |\{N^a_{n}\}\ket \to |j_{n},m_{n},d_{n}\ket,
\ee
is a unitary isomorphism. 

Now that we have the spin basis for the higher modes $n\ge 1$, we can tensor together them with the 0-mode states to get a basis of the whole Hilbert space for the quantum puncture.

\subsection{Flux as effective spin from $\SU(2)$ recoupling}\la{sec:spinbasis}

We have started with states $|\vp,\{N^a_{n}\}\ket$ labeled by the momentum eigenvalues and number of quanta for every string modes, and we have shown how to perform a change of basis to go to a spin basis $|\Delta,l,m,\{j_{n},m_{n},d_{n}\}_{n}\ket$ diagonalizing the mass operator $\vP^2$ and the $\so(3)$ Casimir operators of each string mode. Explicitly, we have:
\be
\left|
\begin{array}{lcl}
\vP^2\,|\Delta,l,m,\{j_{n},m_{n},d_{n}\}_{n}\ket
=
&\Delta&
\,|\Delta,l,m,\{j_{n},m_{n},d_{n}\}_{n}\ket\,,
\vspace{1mm}\\
N_{n}\,|\Delta,l,m,\{j_{n},m_{n},d_{n}\}_{n}\ket
=
&(j_{n}+2d_{n})&
\,|\Delta,l,m,\{j_{n},m_{n},d_{n}\}_{n}\ket\,,
\vspace{1mm}\\
L_{0}\,|\Delta,l,m,\{j_{n},m_{n},d_{n}\}_{n}\ket
=
&\left[\Delta+\sum_{n\ge 1}n(j_{n}+2d_{n})\right]&
\,|\Delta,l,m,\{j_{n},m_{n},d_{n}\}_{n}\ket\,,
\vspace{1mm}\\
\vA^2\,|\Delta,l,m,\{j_{n},m_{n},d_{n}\}_{n}\ket
=
&l(l+1)&
\,|\Delta,l,m,\{j_{n},m_{n},d_{n}\}_{n}\ket\,,
\vspace{1mm}\\
\vS^2_{n}\,|\Delta,l,m,\{j_{n},m_{n},d_{n}\}_{n}\ket
=
&j_{n}(j_{n}+1)&
\,|\Delta,l,m,\{j_{n},m_{n},d_{n}\}_{n}\ket\,.
\end{array}
\right.
\ee
The states $|\Delta,l,m,\{0,0,0\}_{n}\ket$ are the Virasoro lowest weight vectors and are all annihilated by the generators $L_{n\ge 1}$.

Since this basis diagonalizes the angular momenta of every modes, the 0-mode angular momentum $\vA^2$ as well as the higher modes $\vS^2_{n}$ for $n\ge 1$, one can wonder why we use a different basis for the 0-mode than for the higher modes, labelled on the one hand by the quantum numbers $\Delta, l, m$  and on the other hand by the quantum numbers $j_{n},m_{n},d_{n}$. The difference lays in the contribution of each mode to the Hamiltonian $L_{0}$. The energy term for a higher mode $n\ge 1$ is the harmonic oscillator Hamiltonian $\vp_{n}^2+n\vx_{n}^2$ with energy quanta proportional to $n$, while the 0-mode with $n=0$ simply contributes  $\vP^2$. This leads to the different construction of the basis states, as explained in the previous sections, labelled by $\Delta$ for the 0-mode and by the $\sll_{2}(\R)$ degeneracy label $d_{n}$ for higher modes.

As a summary we can conclude that the state  we have constructed so far are  tensor product  states
\be
 |\Delta,l,m,\{j_{n},m_{n},d_{n}\}_{n}\ket =
 |\Delta,l,m\ket \otimes |\{j_{n},m_{n},d_{n}\}_{n}\ket
\ee
where $ |\Delta,l,m\ket$ diagonalizes $P^2,A^2,A_3$ and the the oscillator Hilbert space is 
\be
 |\{j_{n},m_{n},d_{n}\}_{n}\ket \in \bigotimes_{n=1}^\infty (V_{j_{n}} \otimes \tilde{V}_{j_{n}}),
\ee
where $V_{j_n}$ is an $\so(3)$ representation of spin $j_n$ and  $\tilde{V}_{j_n}$ is an $\sll(2,\R)$ discrete representation of matching spin.
One can think of $\tilde{V}_{j_n}$  as a degeneracy label  for the $\so(3)$ representation.

The geometrical flux, which is the main object in the loop quantum gravity framework, now appears as a composite object, being the total angular momentum of all the string vibration modes, $\vSigma=\vA+ \vS$ where $\vS:= \sum_{n\ge 1}\vS_{n}$. Since  we want to diagonalize the overall $\so(3)$ Casimir $\vSigma^2$, we need to label the states by eigenstates of the total angular momenta $\Sigma^2$ and $\Sigma_3$ instead of labelling the states by the eigenstates of $A^2$ and $A_3$.
Before describing the change of basis from the angular momentum eigenstates  to the flux eigenstates, we would like to stress a point which will become essential in the final section \ref{sec:M-S} of this work when analyzing the reduction of the gravity string to loop quantum gravity. Indeed, although the flux components $\Sigma^a$ clearly form a closed $\so(3)$-algebra, it is natural from the perspective of the formalism developed here to consider it as part of the larger Poincar\'e algebra $\mathfrak{iso}(3)$ formed with the momentum $\vP$:
\be
[\Sigma^a,\Sigma^b]= i\epsilon^{abc} \Sigma^c,\qquad
[\Sigma^a,P^b]=  i\epsilon^{abc} P^c,\qquad
[P^a,P^b]=0.
\ee
The $\mathfrak{iso}(3)$ Lie algebra has two Casimirs, $\vP^2$ and $\vSigma\cdot\vP$.
Therefore, here we would like to introduce a change of basis from the decoupled oscillator states $ |\Delta,l,m,\{j_{n},m_{n},d_{n}\}_{n}\ket$ to basis states adapted to the flux, and thus adapted to loop quantum gravity, which diagonalize the Poincar\'e Casimirs, $\vP^2$ and $\vSigma\cdot\vP$, as well as the $\so(3)$ Casimir $\vSigma^2$.
We proceed in two steps:
\begin{itemize}
\item First, we recouple all the higher mode spins together in a single overall spin $\vS:= \sum_{n\ge 1}\vS_{n}$. We introduce a basis diagonalizing $\vS^2$ and the magnetic moment $\vS\cdot\vP$ along the momentum direction. Since $\vSigma\cdot\vP=\vS\cdot\vP$, this diagonalizes the second Poincar\'e Casimir.

\item Second, we recouple $\vA$ and $\vS$ into the flux $\vSigma$ and introduce the final basis diagonalizing $\vP^2$, $\vSigma\cdot\vP$ and $\vSigma^2$ as wanted.

\end{itemize}

Let us work out this procedure in details.
As announced above, we first  recouple the spins of the higher vibration modes, $\vS=\sum_{n}\vS_{n}$. We introduce eigenstates of $\vS^2$ and the spin component $s=\vS\cdot\vP$  along the $\vP$-direction:
\be
 |\{j_{n},m_{n},d_{n}\}_{n}\ket  \rightarrow |S,s\ket \otimes |I\ket
 \quad\textrm{such that}\quad
 \left|
 \begin{array}{clccl}
\vS^2&|S,s\ket \otimes |I\ket
&=&
S(S+1)&|S,s\ket \otimes |I\ket
\,,
\\
\vS\cdot\vP&|S,s\ket \otimes |I\ket
&=& 
s&|S,s\ket \otimes |I\ket
\,.
 \end{array}
 \right.
 \nn
\ee
While  $|S,s\ket$ describes how the state transforms under $\SO(3)$ rotations, the second factor  $|I\ket$ encodes all the $\so(3)$-invariant degrees of freedom.
More precisely, the state $|I\ket$ is an intertwiner between the overall $S$ and arbitrary spins $\{j_{n}\}_{n\in\N}$ with arbitrary degeneracies $\{d_{n}\}_{n\in\N}$:
\be
|I\ket
\in 
\mathrm{Inv}_{\SO(3)}
\left( \bigotimes_{n\in\N}\left( \bigoplus_{j_{n}\in\N}(V_{j_{n}} \otimes \tilde{V}_{j_{n}})\right), V_{S} \right),
\ee 
where $\mathrm{Inv}_{\SO(3)}$ denotes the sets of invariant maps.
This Hilbert space of $\so(3)$ intertwiners, cleanly defined as the limit $N\to\infty$ of the truncations to modes $n\le N$, contains all the states which are created and annihilated by all the operators commuting with $\vS$. As shown in \cite{Girelli:2005ii,Freidel:2009ck,Freidel:2010tt} this space carries, quite remarkably,  a representation space of the unitary group $U(N)$ with $N\to \infty$. It would be interesting to develop this remark further in the near future.

The second step of the procedure is to recouple the spin $\vS$ with the total flux $\vSigma$. The angular momentum will then be reconstructed a posteriori as $\vA= \vSigma-\vS$.
We want  the states diagonalizing $\vP^2$, $\Sigma\cdot \vP$, $\vSigma^2$ and $\Sigma_3$. The first operator $\vP^2$ is the Kac-Moody Casimir and already diagonalized by the quantum number $\Delta$. The second operator $\vSigma\cdot \vP$ is a Poincar\'e Casimir and is already diagonalized by the quantum $s=\vSigma\cdot \vP=\vS\cdot \vP$ since $\vA\cdot \vP=0$. Then we construct a basis of states diagonalizing $\vSigma^2$ and $\Sigma_3$ by finally performing the spin recoupling $\vSigma=\vA+\vS$:
This fusion leads to a change of basis from the original decoupled oscillator states $ |\Delta,l,m,\{j_{n},m_{n},d_{n}\}_{n}\ket$ to recoupled spin states $|\Delta, s, \cJ, \cM, I\ket$. These final states satisfy:
\be
\left|\begin{array}{clccl}
\vP^2
&
|\Delta, s, \cJ, \cM, I\ra
&=&
\Delta&
|\Delta, s, \cJ, \cM, I\ra\,,
\vspace*{1mm}\\
\vSigma\cdot\vP
&
|\Delta, s, \cJ, \cM, I\ra
&=&
s&
|\Delta, s, \cJ, \cM, I\ra\,,
\vspace*{1mm}\\
\vSigma^2
&
|\Delta, s, \cJ, \cM, I\ra
&=&
\cJ(\cJ+1)&
|\Delta, s, \cJ, \cM, I\ra\,,
\vspace*{1mm}\\
\Sigma^3
&
|\Delta, s, \cJ, \cM, I\ra
&=&
\cM&
|\Delta, s, \cJ, \cM, I\ra\,,
\end{array}\right.
\ee
with $\vS\,|I\ra=0$.
The states $|\Delta, s, \cJ, \cM\ket$ form a representation of the Poincar\'e algebra: the operators $\Sigma_{a}$ and $P_{a}$ act only on the state $|\Delta, s, \cJ, \cM\ket$ and do not act on the intertwiner state $|I\ra$.
$\Delta $ and $s$ are the Poincar\'e Casimirs and the total spin $\cJ$ is the LQG spin.

At the end of the day, the quantum puncture derived here carries much more data than the puncture in the standard loop quantum gravity framework. The geometrical  flux $\vSigma$ and its spin $\cJ$ are composite objects resulting from the recoupling of all the modes of the string winding around the puncture. On top of that, the puncture also carries a momentum $\vP$, whose norm $\Delta=\vP^2$ gives the energy carried by the 0-mode,  and an internal intertwiner $I$ encoding the $\SO(3)$-invariant charges carrying by the puncture (among which, the degeneracy numbers $d_{n}$ resulting from the $\sll_{2}(\R)$ structure of every higher mode $n\ge 1$). 

This concludes our analysis of a quantum puncture and the fine structure of the flux in the loop gravity string framework. The next task is to put punctures together on the boundary surface around every network vertex and then link those surfaces together into a generalized spin network. We discuss those structures in the next section. We will especially underline the role of the momentum $\vP$, which remains together with the flux $\vSigma$, in the particle limit of the boundary surface and punctures, and leads to a crucial extension of  the LQG spin networks taking into account the symmetry under 3d diffeomorphisms.

\section{From the  gravity string to Poincar\'e networks}
\la{sec:poincare}

A careful canonical analysis of general relativity has revealed dynamical edge modes living on boundary surfaces within the canonical space-like hypersurface. Studying those edge modes on a punctured surface with point-like sources of curvature and torsion leads to stringy charges forming a Kac-Moody algebra around each puncture. This string edge modes described by a tower of harmonic oscillators $(\km^a_{n},\km^a_{n}\dag)$ at each puncture of the boundary surface have unveiled  the fine structure of geometrical fluxes of the loop quantum gravity description of quantum geometry excitations. The LQG framework thus appears as a coarse-grained formalism overlooking these finer degrees of freedom.

These Kac-Moody charges, missing from the traditional LQG theory, offer a refreshing CFT perspective to gravitational edge modes and the LQG discretization of geometry. For instance, the 0-mode of the Kac-Moody algebra carries a momentum $\vP$ which defines  the conformal weight $\Delta=\vP^2$ of the primary field associated to the vacuum state, as shown in \eqref{primary}.
Moreover, our reformulation of the kinematical constraints of gravity as conservation laws presented in Section \ref{sec:string}
further provides a physical interpretation of this new quantum number $\Delta$ as a boundary charge associated to 3D diffeomorphisms. 
 
This richer structure of the quantum puncture on the boundary leads to a new realization of quantum geometry states as networks of surface charges, obtained by gluing the boundary edge modes of bounded space regions together to form the quantum space. These edge mode networks now live on a thickened tubular version of the LQG spin networks,  carrying Kac-Moody charges instead of the mere spin encoding the overall $\SU(2)$ flux. These structures provide a fuller description of kinematical degrees of freedom and carry a representation of boundary $\SU(2)$ symmetry transformations as well as boundary diffeormorphisms.
 
 \subsection{Edge mode networks as new quantum geometry states}
  
We call the {\it gravity string}  the structure formed by the collection of quantum punctures on the boundary surface $S$.
With the quantum puncture Hilbert space described in the previous sections, it provides a new picture of the quantum space. More precisely, the gravity string can be understood as the quantum geometry of the local subsystem---the bounded region $\cB$---with excitations of quantum space localized at  punctures on the boundary surface $S=\pp \cB$ and encoded in the Kac-Moody charges associated to each  puncture.
The edge modes on the boundary surface are represented as the string modes around each puncture, they contain all the information about  quantum geometry in $\cB$ and its boundary $S$.  Then the degrees of freedom around each puncture allow to glue bounded regions  together so to reconstruct the total Hilbert space associated to the whole canonical hypersurface $M$.

\begin{figure}[h!]
\centering
\includegraphics[width=0.38\textwidth]{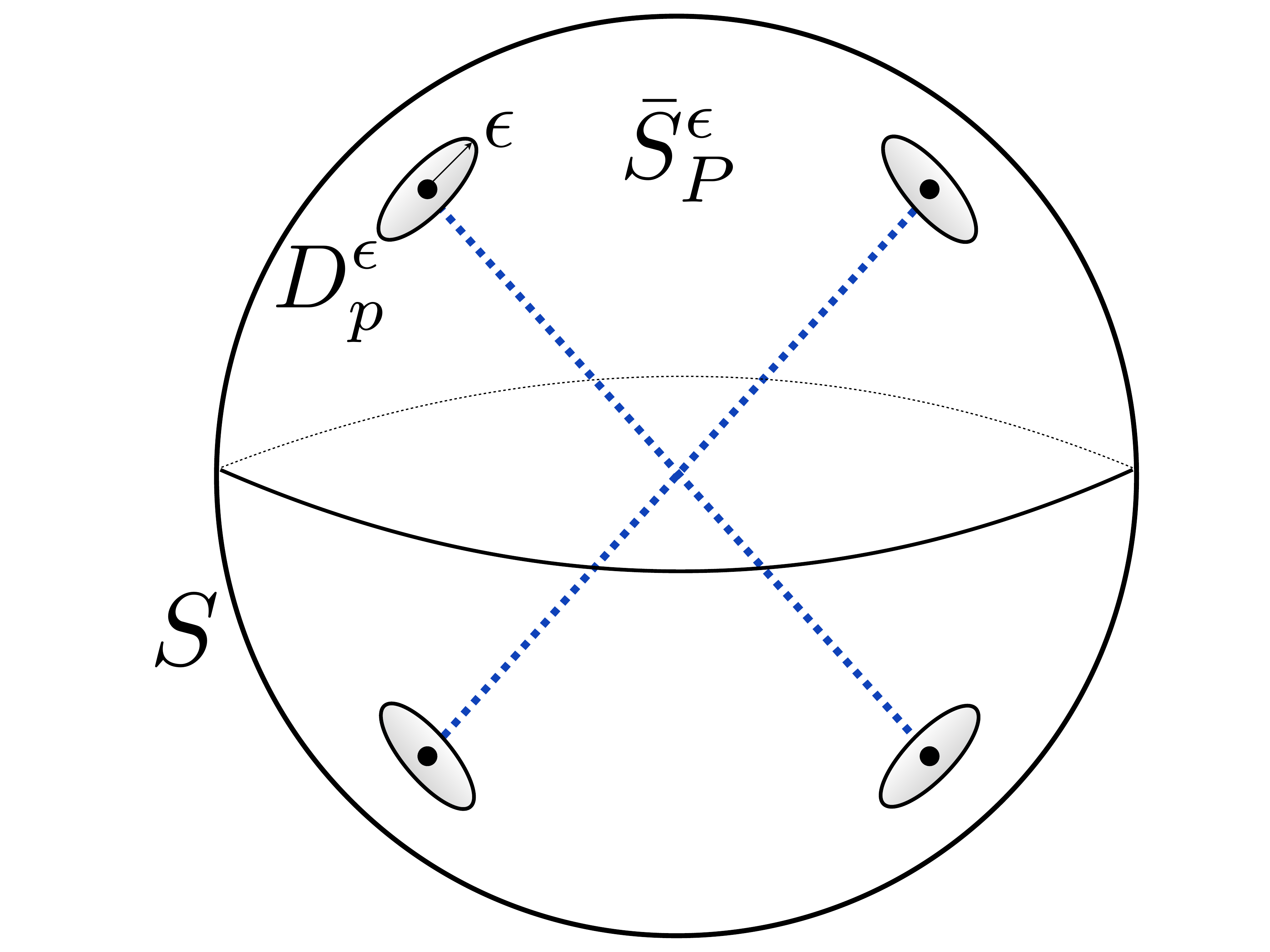} 
\includegraphics[width=0.18\textwidth]{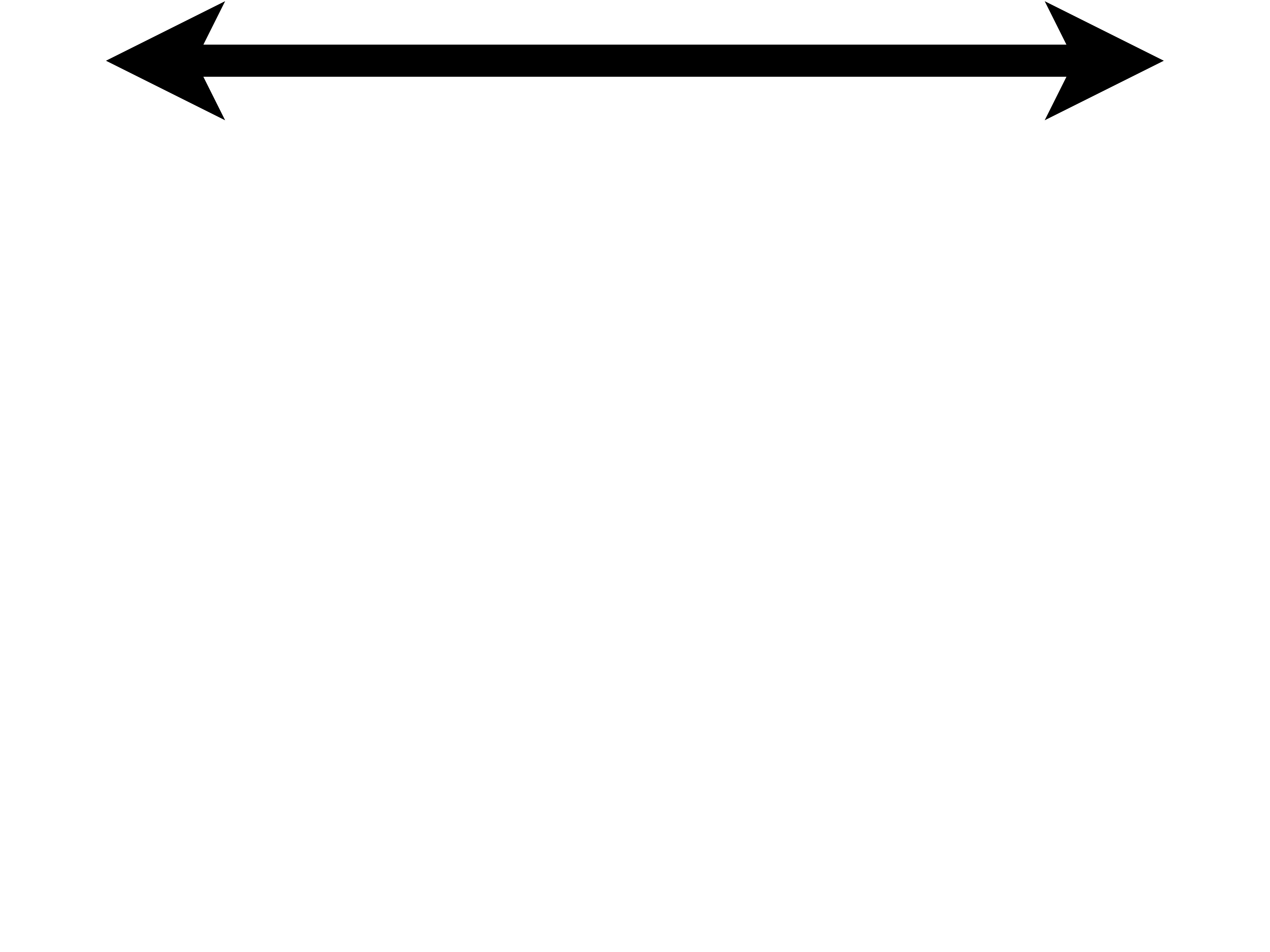} 
\includegraphics[width=0.38\textwidth]{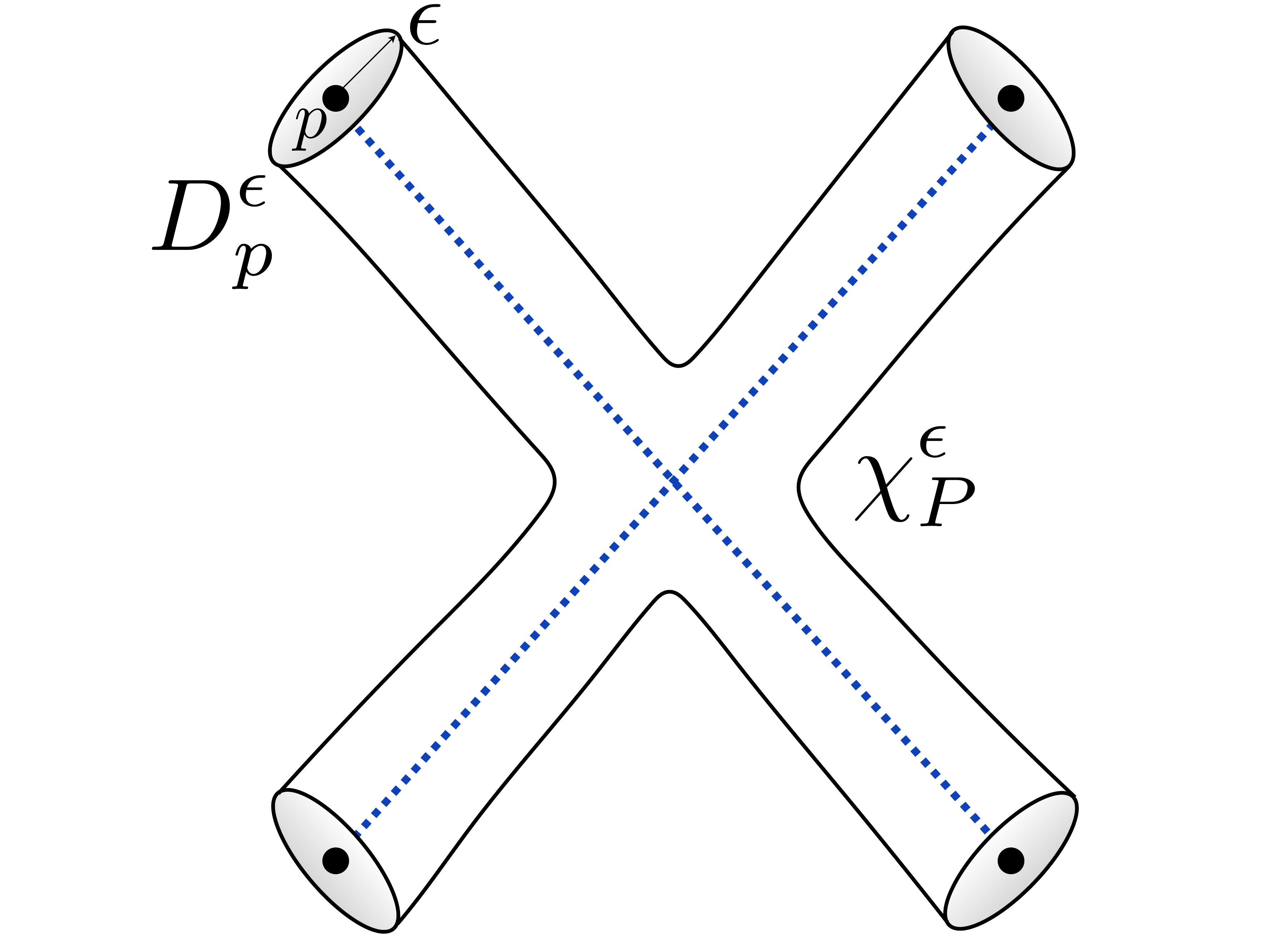}
\caption{The  gravity string: flattening of the punctured surface into tubes linking the punctures together and localizing the sources of curvature in the bulk of the bounded region.}
\label{fig:net}
\end{figure}
This gluing process can be more easily described by `flattening' the gravity string.
Indeed, since we have assumed that the sources of curvature are localized at the punctures, we have $F(A)=0$ on the boundary surface away from the punctures, $ \bar{S}_P^\epsilon$, which means that the connection is pure gauge on that domain, namely $A=g^{-1}\rd g$. We   now extend this distributional curvature to the bulk region enclosed by the 2-sphere $S$. More precisely, we  consider a 2D tubular structure connecting all the circles $C_p$'s around the punctures on $S$ and formed by tubes departing from the boundary circles and all joining in the bulk. Curvature propagates from one puncture to  another inside the tubes, but it vanishes outside, i.e. we have $F(A)=0$ in the bulk away from the tubes.

We illustrate this on Figure \ref{fig:net} for the 4-valent gravity string. 
The boundary surface  is  the 2-sphere $S$ from which we cut out 4 disks $D^\epsilon_p$  of radius $\epsilon$ around the 4 punctures $p$'s. We then carve out a tubular structure $\chi^\epsilon_P$ from the ball  enclosed by $S$. In the ball outside of   $\chi^\epsilon_P$ the connection is also pure gauge and we can thus  perform a flattening process of the string as shown in Figure \ref{fig:net}. We refer to this procedure as ``flattening'' because we remove the parts where the connection is flat, $F(A)=0$, and focus on the zone with non-trivial curvature.

\begin{figure}[h!]
\centering
\includegraphics[width=0.7\textwidth]{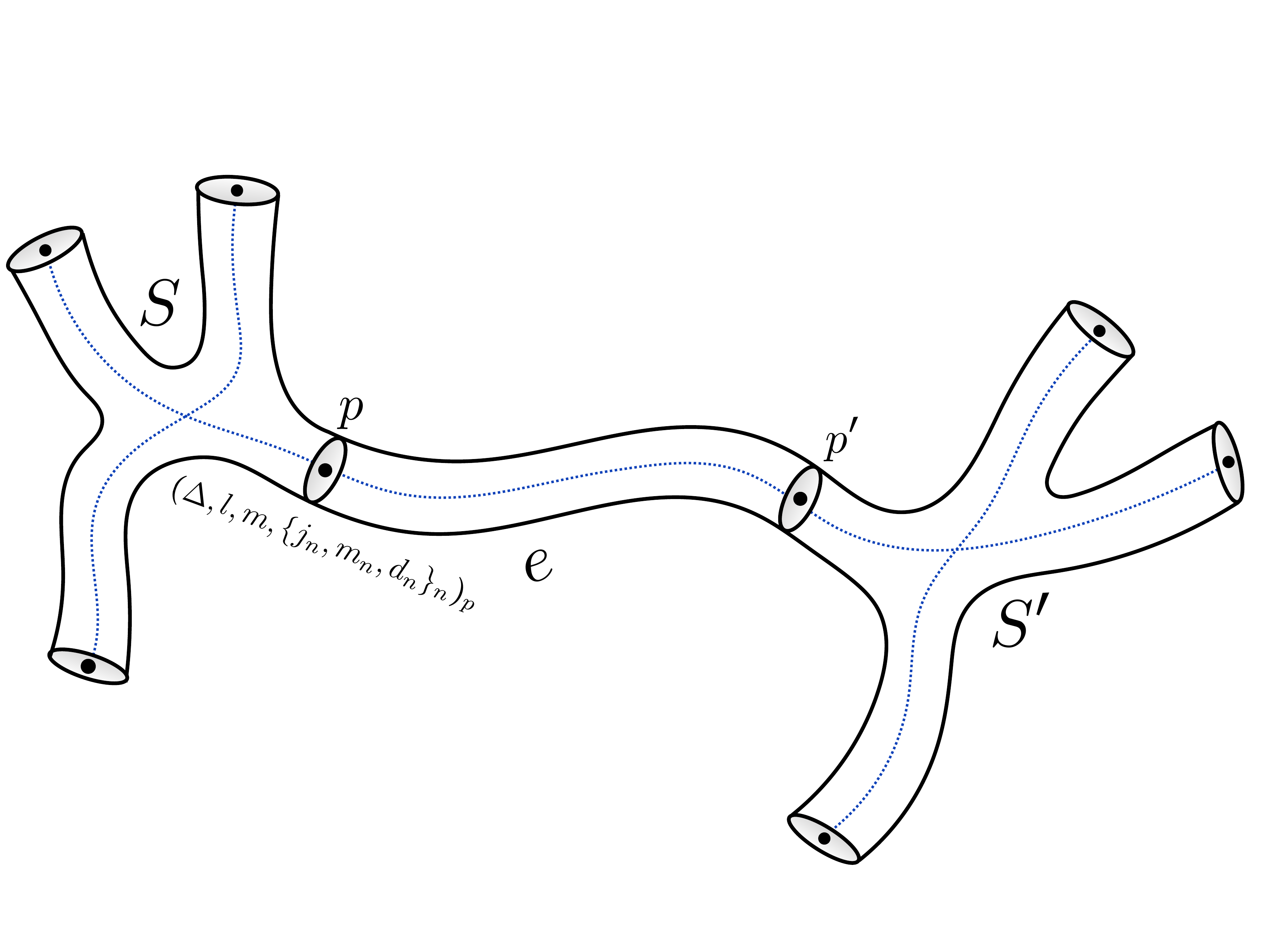} 
\caption{
Gluing gravity strings into a network of quantum edge modes:
the edge modes on each punctured surface - or corner-  are represented by quantum states labelled by quantum numbers $(\Delta,l,m,\{j_{n},m_{n},d_{n}\})_p$ for each puncture; these punctured surfaces are then glued through tubes linking their punctures.
}
\label{fig:glue}
\end{figure}

Each tube now carries the quantum Kac--Moody charges, as described in the previous sections.  This leads to new quantum states of geometry in terms of this tubular structure with the tubes around each gravity string dressed with quantum puncture states, in the Kac-Moody basis with quantum numbers $(\Delta,l,m,\{j_{n},m_{n},d_{n}\})_p$ or in the recoupled spin basis with quantum numbers $(\Delta,s,\cJ,\cM,I)_{p}$.
Those chunks of quantum space are then patched together by gluing the cylinders around the curvature defects through appropriately matching and propagating  the quantum numbers of the puncture where the gluing happens (accounting for the change of frame and coordinates from one punctured sphere to the next). This gluing process is graphically represented in Figure \ref{fig:glue}.

\subsection{Poincar\'e networks as a discretization limit}
\la{sec:M-S}

Our analysis has revealed a wealth of new quantum numbers associated to the edge mode charges that can be constructed out of the fundamental Kac-Moody oscillators $\alpha^a_n$'s. We have seen above how these can be organized into a set of generalized quantum geometry states carrying a representation of  the full kinematical constraint algebra, both $\SU(2)$ Gauss law and the 3D diffeomorphism constraint. 
The natural question is then: in which limit can the standard LQG spin network states be recovered from the gravity string framework?

It is clear that the LQG picture correspond to a `particle limit' of the general construction, where we focus on the 1-skeleton of the punctured boundary surface with the disk around the punctures contracted to the punctures themselves, the tubes  contracted to lines and thus the string vibration modes frozen or overlooked.
We propose\footnotemark{}  to discard the higher Kac-Moody modes of the quantum punctures and focus their 0-mode sectors, effectively truncating the full Kac--Moody algebra to the sub-algebra generated by the 0-mode momentum $P$ and the $\SU(2)$ flux $\Sigma$.
We switch from the decoupled oscillator basis  $|\Delta,l,m,\{j_{n},m_{n},d_{n}\}\ra$ diagonalizing the Kac-Moody charge operators to the recoupled spin basis $|\Delta,s ,\cJ, \cM, I \ket$ diagonalizing $\vP^2$, $\vP\cdot\vSigma$ and $\vSigma^2$ as described earlier in section \ref{sec:spinbasis}. Then we consider the state truncation
\be\label{forget}
|\Delta,s ,\cJ, \cM, I \ket
\quad\longrightarrow\quad
|\Delta,s ,\cJ, \cM \ket,
\ee
which simply ignore the information contained in the higher intertwinner modes since it is inaccessible to the zero mode sub-algebra.
\footnotetext{
Another option would be to kill all the higher mode oscillators by sending them to zero and identifying the LQG flux to the string orbital angular momentum of the 0-mode, namely $\vec \Sigma = \vec A$. However, this imposes the implicit restriction that $\Sigma\cdot P=0$. Although this does not affect the range of the flux $\Sigma$, it implies that one of the two Poincar\'e Casimirs vanishes. It seems more natural to allow for a non-trivial spin for the higher modes,  $\vec \Sigma = \vec A+\vec S$ with a priori $\vS\ne 0$, focusing on the observables $\vP$ and $\Sigma$ while considering the higher modes as decoupled from their dynamics, hence ``frozen''.
}

We can now  envision two levels of truncation. Starting from the whole ladder of Kac-Moody charges $\alpha^a_{n}$ for a puncture, we consider the (much) smaller subalgebra of observables defined by the $\SU(2)$ flux $\Sigma^a$ and  the 0-mode momentum $P^a$, which form a closed Poincar\'e algebra. In this truncation, quantum punctures live in representations of the Poincar\'e algebra ${\mathfrak{isu}}(2)$. Considering the punctures around every boundary surface and gluing them together, we obtain networks of {\it Poincar\'e charges} or simply Poincar\'e networks. 
The forgetting operation described in (\ref{forget}) corresponds to a coarse graining operation where  we trace over the intertwiner $I$ (containing the $\SO(3)$-invariant degrees of freedom internal to the puncture)  when computing expectation values.
This is a coarse-grained version of the gravity string framework, where the fine structure of the Kac-Moody higher modes is effectively hidden within the flux $\Sigma$.

The second level of truncation which is implied in loop gravity, is even more drastic and consists in further discarding the 0-mode momenta $P^a$. Focusing on the closed $\su(2)$ algebra formed by the flux $\Sigma^a$ yields back the standard LQG framework with punctures dressed with $\su(2)$ representations and spin network quantum states. We find this level of truncation or coarse-graining too drastic. Indeed, the momentum $P^a$ encodes the boundary charge induced by the bulk gauge invariance under 3D diffeomorphisms, while the flux $\Sigma^a$ encodes the boundary charge associated to the $\SU(2)$ gauge invariance. From this perspective, it does not make sense to discard one and keep the other. It is more reasonable to keep both $P^a$ and $\Sigma^a$ and the whole Poincar\'e algebra associated to each puncture, and thereby obtain quantum states carrying a representation of both symmetries under $\SU(2)$ transformations and diffeomorphisms. 
\begin{figure}[h]
\centering
\includegraphics[width=0.38\textwidth]{String-flat} 
\includegraphics[width=0.18\textwidth]{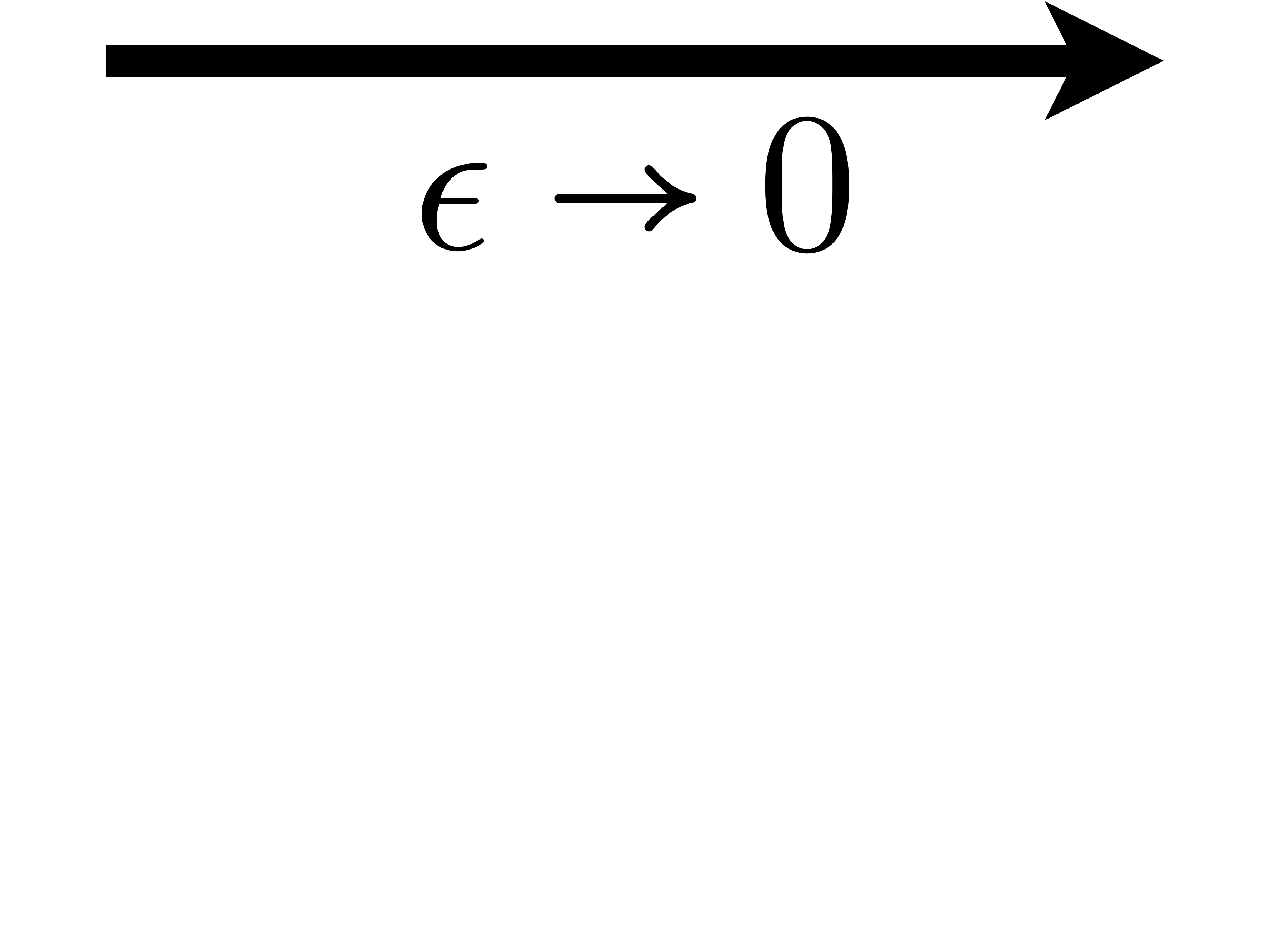} 
\includegraphics[width=0.38\textwidth]{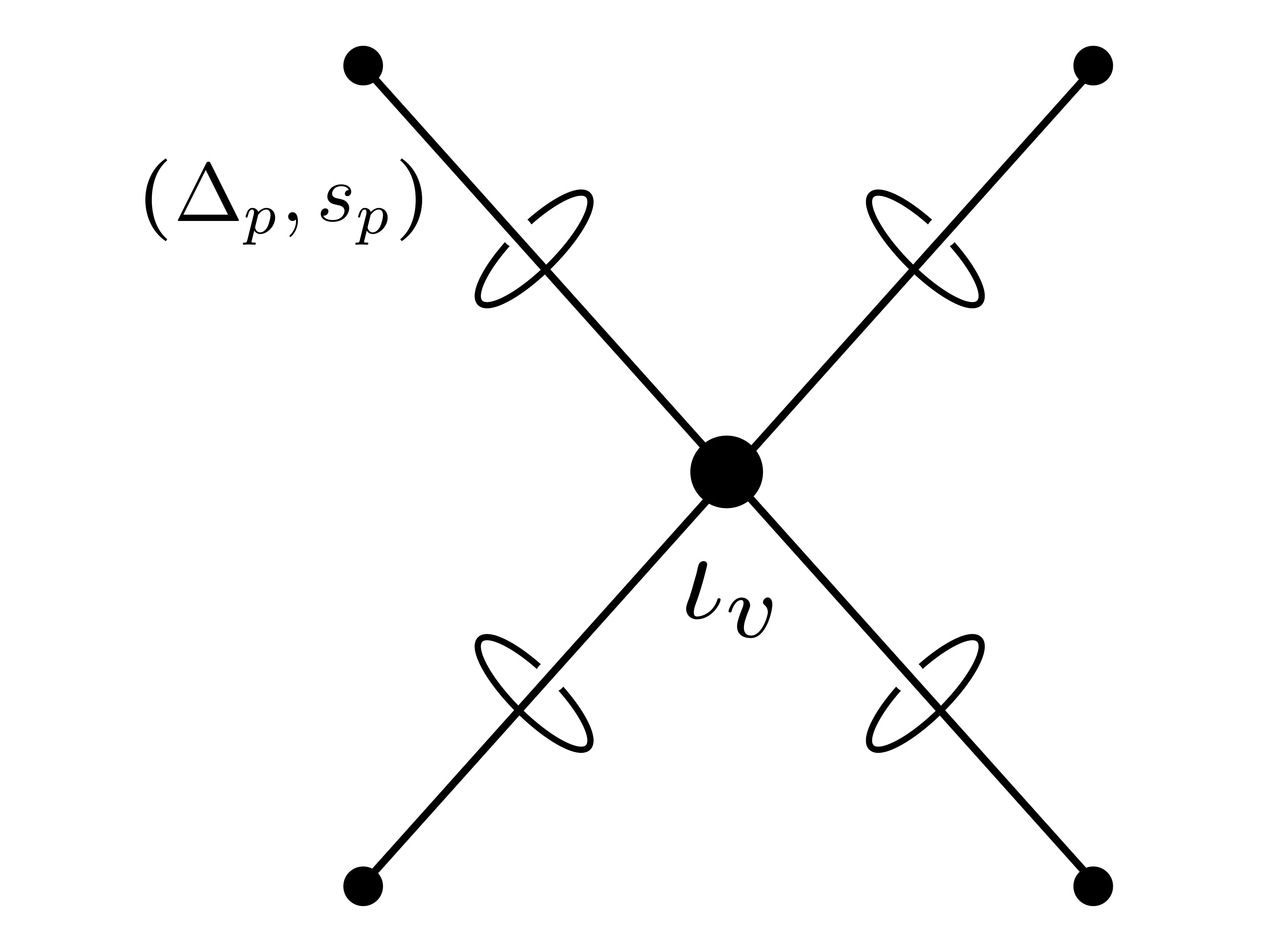}
\caption{
Poincar\'e network truncation of tubular edge mode networks:
the punctured surface of a gravity string is collapsed to its 1-skeleton with the punctures represented by lines attached to a central vertex; each of these lines  carry a flux $\Sigma_{p}^a$ and momentum $P_{p}^a$, respectively the boundary charges carried by the puncture for the $\SU(2)$ bulk gauge transformations and 3D diffeomorphisms; the flux and momentum form a $\mathfrak{isu}(2)$ Poincar\'e algebra so that each quantum puncture $p$ lives in a Poincar\'e representation labelled by the two Casimirs $(\Delta_{p},s_{p}):=(\vP_{p}^2,\vec{\Sigma}_{p} \cdot \vP_{p})$; we do not forget the curvature $k_{p}$ carried by each puncture and graphically represented by the loops winding around the network edges.
}
\label{fig:poinet}
\end{figure}

Let us describe the particle limit of the gravity string in more details. It consists of two steps.
First, we take the limit  $\epsilon\rightarrow 0$ so that the tubes shrink to links meeting at a vertex $v$. This amounts to considering the 1-skeleton of the punctured sphere, keeping the cut from the root point $*$ to every (anchor point $p^*$ around the) punctures $p$ on the surface. This contracts the punctured surface to a single central node $v$ with the punctures represented by links attached to it, as illustrated on Figure \ref{fig:poinet}.
Second, we truncate the algebra of edge mode charges on the punctured surface to the flux-momentum sector formed by the fluxes $\vSigma_{p}$ and momenta $\vP_{p}$ associated to every punctures $p$. These form individual $\mathfrak{isu}(2)$ Poincar\'e algebras for each puncture,
\be
[\Sigma_p^a,\Sigma_p^b]= i\epsilon^{abc} \Sigma_p^c,\qquad
[\Sigma_p^a,P_p^b]=  i\epsilon^{abc} P_p^c,\qquad
[P_p^a,P_p^b]=0.
\ee
This algebra has two Casimir operators, the momentum norm $\vP_{p}^2$ and helicity $s_{p}:=\vec{\Sigma}_{p} \cdot \vP_{p}$. The value of these two Casimirs determines a unique irreducible representation of the Poincar\'e group $\ISU(2)$. We recognize the first Casimir as the conformal weight of the punctures $\vP_{p}^2=\Delta_{p}$, which is also a Casimir of the whole Kac-Moody algebra. On the other hand, the second Casimir is  not a Casimir of the full  Kac-Moody algebra and diagonalizing it involves a change of basis from the states $|\Delta,l,m,d,\{j_{n},m_{n},d_{n}\}_{n}\ket$ to $|\Delta,s,\cJ,\cM,I\ra$ which we introduced in Section \ref{sec:spinbasis}.
The quantum puncture then lives in the Poincar\'e representation with Casimirs $\Delta_{p}$ and $s_{p}$ and  quantum basis states are further labeled by the spin $\vec{\Sigma}_{p}^2=\cJ(\cJ+1)$ and the magnetic moment $\Sigma_{p}^3=\cM$.

This dresses the punctures around the vertex $v$  with states $|\Delta_{p},s_{p},\cJ,\cM\ra$. Moreover, the flux and momenta on the punctured surface satisfy conservation laws, as shown in Section \ref{sec:conservation}, which require to  impose constraints coupling the punctures:
 \be
 \la{consv}
\sum_{e\supset v} \Sigma_e=0\,,\qquad \sum_{e\supset v} P_e=0
\,.
\ee
At the quantum level, this imposes the invariance under the 3D Poincar\'e group $\ISU(2)$ of the tensor product states of the quantum punctures around the vertex. The vertex $v$ is thus dressed with a Poincar\'e intertwiner state $\iota_{v}$, living in the tensor product of the Poincar\'e representations $(\Delta_{p},s_{p})$ dressing the punctures attached to it, as graphically summarized by Figure \ref{fig:poinet}. This defines Poincar\'e charge networks, upgrading loop quantum gravity's spin network to graphs dressed with $\ISU(2)$ irreducible representations and intertwiners. 

\medskip

This method shows how the quantization of a subalgebra of observables, clearly exploring a small sector of the full theory, leads to Poincar\'e networks. 
First, we would like to underline that the truncation to the 0-mode of the strings around the punctures does not lead to LQG and $\SU(2)$ spin networks but instead to networks of Poincar\'e charges. And we stress that these Poincar\'e networks carry a representation of  boundary charges for both the $\SU(2)$ gauge transformations and 3D diffeomorphisms.

Second, we would like to insist on the advantage of having a bigger algebra of observables.
%
Technically, this means that instead of attaching only a quantum  $\SO(3)$ frame to each vertex and allowing transformations along each edge that are  rotations of the flux $\vSigma$ preserving the spin $\vSigma^2$;
we can, in a Poincar\'e network,  attach a quantum Poincar\'e frame to each vertex and a $\mathrm{ISO}(3)$ transformation along each edge. These general Poincar\'e transformation do not necessarily  keep the spin $\vSigma^2$ invariant anymore (similarly to what was noticed in the context of coarse-graining LQG in \cite{Livine:2019cvi}). 
This allow us to describe spin network whose vertices are moving relatively to another.
%

Finally, to summarize, we see that our approach 
provides a clear understanding of which observables are 
missing when working with a truncation of the theory.
The Poincar\'e networks states neglect the oscillation modes of the geometry around punctures and the loop gravity truncation further neglect the momenta charge.
One hope is that the truncation to the zero mode charge Poincar\'e-algebra might be a truncation robust under coarse-graining. Only a finer understanding of the dynamics  will tell.

%
%

\section*{Conclusion \& Outlook}
We have started this work by the critical observation
that the kinematical constraints of gravity in the connection-flux variables formulation can be understood as conservation laws for boundary charges.
We have shown how this understanding opens the way to a new description of quantum geometry where features of the holographic and canonical approaches are combined and generalized. 
This cross-fertilization of ideas and techniques thrives in the arena provided by the edge mode formalism. This formalism captures the nature of local degrees of freedom involved in the division of the system into subsystems. Moreover, it realizes this process as a creation of boundary defect and the gluing of regions 
as a fusion of the boundary charges.

In particular, we have revealed a general pattern in the formulation of kinematical gravity constraints where differentiability in the presence of boundaries, representing local subsystems, requires a nested structure of boundary contributions for surfaces of different codimensions. The surface integral of highest codimension defines symmetry charges encoding the transformation properties of the boundary, while all the lower codimension terms can be recast in the form of a conservation law for these edge modes.

The reason why this beautiful pattern has eluded us for so long could be partly blamed on the chameleonic nature of edge modes. As pointed out in Section \ref{sec:string}, edge modes can enter the stage in different ways. They can appear as a manifest extension of the boundary phase space, necessary to restore gauge symmetry  (as in the case of the boundary frame field in the extended symplectic structure \eqref{cr}  and yielding the boundary simplicity constraint in \eqref{bdyc}).  Or they can be disguised as singular sources of geometrical quantities playing as a proxy of symmetry charges in a discretized setting (as we saw for the case of curvature and momenta \eqref{curv}). Either way, the algebra of new observables they define is generally anomalous, and where there is a central extension, there is life.

We have exploited the appearance of a central charge in the algebra of the fundamental harmonic oscillators \eqref{alphas} to construct representations of the infinite dimensional boundary symmetry algebra. This led us to the introduction of new quantum geometry states labeled by quantum numbers associated with both the geometrical flux and the diffeomorphism charges. We have shown how a truncation of the loop gravity string description of quantum geometry to the zero mode sector allows us to put the diffeomorphism constraint on the same footing as the Gauss law. It becomes implemented as the closure condition for the edge momenta at each node of the network. The possibility to achieve infinitesimal diffeomorphism invariance at the operational level represents a significant improvement of the standard LQG description and, at the same time, reanimate the hope to construct a well-defined dynamics for quantum geometry.

As usual in physics, new quantum numbers call for new spectroscopy, a new description of physical phenomena. The black hole entropy calculation represents a natural environment where the richness of our new machinery can be put at work.
As edge modes carry physical information needed to reconstruct the total Hilbert space of the system from the Hilbert space for
the subsystems, it is natural to expect the new string-like degrees of freedom described here to play an important role to account for the horizon entropy. The formalism of isolated horizons, providing the framework for the standard LQG black hole entropy calculation (see \cite{DiazPolo:2011np} for a review), is characterized by a set of boundary conditions which represent a compatible but more restrictive condition on the curvature condition than the one introduced in \eqref{curv}. Therefore, on the one hand, the LQG approach to the Bekenstein--Hawking entropy derivation
can naturally be  embedded in the edge modes formalism \cite{Ashtekar:1997yu, Engle:2010kt}.
On the other hand, our analysis lends itself to new construction of the Hilbert space for isolated quantum horizons in terms of Poincar\'e networks. There the new quantum numbers can be taken into account employing CFT techniques, like in \cite{Ghosh:2014rra, Carlip:2014bfa}, and describe bulk and boundary degrees of freedom within a unified framework, in analogy to previous proposals   (see e.g. \cite{ Pranzetti:2014tla}).
Enlarging the quantum horizon Hilbert space can have important implications on the role of the Barbero--Immirzi parameter in recovering the exact numerical factor in the Bekenstein--Hawking entropy formula, as advocated from different perspectives \cite{Ghosh:2013iwa, Asin:2014gta, Oriti:2018qty}.
More generally, the inclusion of edge modes in entanglement entropy calculation in gauge theory is a necessary ingredient to recover consistent results \cite{Lewkowycz:2013laa, Donnelly:2014gva, Donnelly:2014fua}.

One of the most exciting prospects of our work is the possibility it opens to finally have a representation of local matter inside a quantum spacetime.
Now that we have a representation of space-like diffeomorphism as an operator, one can finally think about the insertion of momenta as the introduction of a topological defect inside the charge conservation law $\sum_e P_e=0\to \sum_e P_e= P_m $ \cite{Freidel:2006hv}, a discrete analog of the Einstein equation $D_\xi = \kappa T_{n\xi}$ where the right-hand side of the diffeomorphism constraint is the insertion of matter momenta density. The matter momenta will naturally appear at the vertices of our network. This strategy has been successfully applied in 3d gravity at the quantum level, where consistency of the constraint algebra insured that the gravitational defects behave as relativistic quantum particles \cite{Freidel:2004vi}.

To conclude, let us mention that several open questions are left unanswered. First, we need to better understand the role of the magnetic constraint in the boundary charge algebra. As we have treated the magnetic edge mode classically at this stage, by trivializing their holonomy,
it is essential to understand what happens when they are treated on equal footing as quantum operators.
Also, we have conjectured that the charge algebra gives the right set of variables to sustain the coarse-graining operations effectively.
It would be essential to provide more quantitative evidence in favor of this hypothesis.
Last but not least, in this work, we have only discussed the kinematical constraints. Even if we now have a definition for the quasi-local momentum operator, we do not have a description of the quasi-local energy.
New ideas are presumably needed, maybe along the lines of \cite{Wieland:2019hkz}, to generalize the gravitational conservation laws described here to the gravitational energy.



\section*{Acknowledgement}

Research at Perimeter Institute for Theoretical Physics is supported in part by the Government of Canada through NSERC and by the Province of Ontario through MRI.

\appendix

\section{Spin basis for a triplet of harmonic oscillators: the $\so(3)\times\sll_{2}$ structure}
\la{app:su2xsl2}

\subsection{Spin from vector operators}
\la{appendix:b}

Considering a triplet of harmonics oscillators, with annihilation operators $\alpha_{\pm},\alpha_{3}$ and their corresponding creation operators $\alpha_{\pm}\dag,\alpha_{3}\dag$, which represents at the quantum level a canonical pair of 3-vectors\footnote{
Considering two conjugate 3-vectors, $\{x_{i},p_{j}\}=\delta_{ij}$, we define their angular momentum $S_{i}=\eps_{ijk}x_{j}p_{k}$. Their Poisson brackets are $\{S_{i},S_{j}\}=\eps_{ijk}S_{k}$, which we can also write in the $3,\pm$ basis:
\be
S_{\pm}=\f1{\sqrt2}\big{[}S_{1}\pm i S_{2}\big{]}
\,,\qquad
S_{-}=\overline{S_{+}}
\,,\qquad
\left|\begin{array}{lcl}
\{S_{3},S_{\pm}\}
&=&
\mp iS_{\pm}\\
\{S_{+},S_{-}\}
&=&
-iS_{3}
\end{array}\right.
\,.\nn
\ee
These generate the $\SO(3)$ action by 3d rotations on the two vectors $x$ and $p$.
We can also switch to the $3,\pm$ basis for the pair of canonical 3-vectors, and define their annihilation and creation observables:
\ba
&&x_{\pm}=\f1{\sqrt2}\big{[}x_{1}\pm i x_{2}\big{]}
\,,\quad
p_{\pm}=\f1{\sqrt2}\big{[}p_{1}\pm i p_{2}\big{]}\n\\
&&
\alpha_{\pm}=\f1{\sqrt{2}}\big{[}x_{\pm}+ip_{\pm}\big{]}
\,,\quad
{\alpha}^\dagger_{\pm}=\f1{\sqrt{2}}\big{[}x_{\mp}-ip_{\mp}\big{]}
\,,\quad
\alpha_{3}=\f1{\sqrt{2}}\big{[}x_{3}+ip_{3}\big{]}
\,,\quad
{\alpha}^\dagger_{3}=\f1{\sqrt{2}}\big{[}x_{3}-ip_{3}\big{]}
\,,\nn
\ea
with Poisson  brackets:
\ba
\begin{array}{l}
\{x_{-},p_{+}\}=\{x_{+},p_{-}\}=-i \\
\{x_{-},p_{-}\}=\{x_{+},p_{+}\}=0
\end{array}
\,,\quad
\{\alpha_{+},{\alpha}^\dagger_{-}\}
=
\{\alpha_{-},{\alpha}^\dagger_{+}\}
=
0
\,,\quad
\{\alpha_{+},{\alpha}^\dagger_{+}\}
=
\{\alpha_{-},{\alpha}^\dagger_{-}\}
=
\{\alpha_{3},{\alpha}^\dagger_{3}\}
=
-i
\,.\nn
\ea
It is straightforward to check the expression of the $\so(3)$ generators in terms of those creation and annihilation observables:
\be
S_{3}={\alpha}^\dagger_{-}\alpha_{-}-{\alpha}^\dagger_{+}\alpha_{+}
\,,\qquad
S_{+}={\alpha}^\dagger_{3}\alpha_{+}-{\alpha}^\dagger_{-}\alpha_{3}
\,,\qquad
S_{-}={\alpha}^\dagger_{+}\alpha_{3}-{\alpha}^\dagger_{3}\alpha_{-}
\,.\nn
\ee
}, we form spin operators generating a $\so(3)$ Lie algebra:
\be
\left|\begin{array}{lcl}
S_{3}&=&\alpha_{-}\dag\alpha_{-}-\alpha_{+}\dag\alpha_{+}
\\
S_{+}&=&\alpha_{3}\dag\alpha_{+}-\alpha_{-}\dag\alpha_{3}
\\
S_{-}&=&\alpha_{+}\dag\alpha_{3}-\alpha_{3}\dag\alpha_{-}=S_{+}\dag
\end{array}
\right.
\,,\qquad
[S_{3},S_{\pm}]=\pm S_{\pm}
\,,\qquad
[S_{+},S_{-}]=S_{3}
\,,
\ee
with quadratic Casimir operator:
\be
\mathfrak{C}=S_{3}^2+S_{+}S_{-}+S_{-}S_{+}=S_{3}(S_{3}+1)+2S_{-}S_{+}
\,,\qquad
[\mathfrak{C},S_{a}]=0
\,.
\ee
So we can obtain spin states from the triplet of harmonic oscillators. More precisely, this means that we can organize the energy states $|N_{-},N_{+},N_{3}\ra$ of the harmonic oscillators, diagonalizing the energy operators $\hat{N}_{a}=\alpha_{a}\dag\alpha_{a}$, in terms of their spin, i.e. diagonalizing the spin operators $\mathfrak{C}$ and $S_{3}$. Since $S_{3}$ is the difference of number of quanta between the $-$ and $+$ oscillators, its eigenvalue $m$ is necessarily an integer. Therefore we only get integer spins, $\mathfrak{C}=j(j+1)$ with $j\in\N$.

\subsection{The $\sll_{2}(\R)$ fin structure of spin states}

To switch basis from the $|N_{-},N_{+},N_{3}\ra$ states to $|j,m\ra$ states diagonalizing  $\mathfrak{C}$ and $S_{3}$, it turns out that there is a non-trivial degeneracy. This degeneracy is best described by a $\sll_{2}(\R)$  algebra of operators commuting with the spin operators, defined by the squeezing operators for the oscillators\footnote{
The classical counterpart of these $\sll_{2}(\R)$ generators are the observables $x\cdot p$, $x^2$ and $p^2$, which are all $\SO(3)$-invariants:
\be
N=\sum_{a}{\alpha}^\dagger_{a}\alpha_{a}=\f12(x^2+p^2)
\,,\quad
f=\f12\alpha_{3}^2+\alpha_{-}\alpha_{+}=\f12(x^2-p^2)+ix\cdot p
\,,\qquad
\{N,f\}=2if
\,\quad
\{f,f^\dagger \}=-4iN
\,.\nn
\ee
The balance equation for the Casimir corresponds to the norm of the angular momentum in terms of the norms and scalar product,
\be
S^2=|x\w p|^2=x^2p^2-(x\cdot p)^2=N^2-f^\dagger f
\,.\nn
\ee
}
:
\be
\left|\begin{array}{lcl}
h&=&\f32+N
\vspace*{1mm}\\
f&=&\alpha_{3}^2+2\alpha_{-}\alpha_{+}
\vspace*{1mm}\\
f\dag&=&\alpha_{3}^2\dag+2\alpha_{-}\dag\alpha_{+}\dag
\end{array}
\right.
\,,\qquad
\left|\begin{array}{l}
{[}h,f{]}=-2f
\vspace*{1mm}\\
{[}h,f\dag{]}=+2f\dag
\vspace*{1mm}\\
{[}f,f\dag{]}=4h
\end{array}
\right.
\,,\qquad
[h,S_{a}]=[f,S_{a}]=[f\dag,S_{a}]=0
\,,
\ee
where we have introduced the operator giving the total number of quanta $\hat{N}=\sum_{a}\hat{N}_{a}$.
The key point is a balance equation between the $\so(3)$ Casimir and the $\sll_{2}(\R)$ Casimir:
\be
S_{3}^2+S_{+}S_{-}+S_{-}S_{+}
=
\mathfrak{C}
=
N(N+1)-f\dag f
=
h^2-\f12\Big{[}
ff\dag+f\dag f
\Big{]}
+\f34
=
h(h-2)-f\dag f+\f34
\,,
\ee
\be
[\mathfrak{C},S_{a}]=[\mathfrak{C},h]=[\mathfrak{C},f]=[\mathfrak{C},f\dag]=0
\,,
\ee
which we can also write as $j(j+1)=N(N+1)-f\dag f$.
In particular, this means that $N\ge j$ and implies that the eigenvalues of $f\dag f$ are necessarily even integers, $f\dag f=N(N+1)-j(j+1)=(N-j)(N+j+1)$.

Now, since the $\sll_{2}(\R)$ operators commute with the spin operators $S_{a}$, we can use them to generate more states without changing $j$ or $m$. Actually it is possible to check that they fully describe the spin degeneracy  of the triplet of harmonic oscillators. More precisely, we introduce a basis which diagonalizes $\mathfrak{C}$, $S_{3}$ and $\hat{N}$ (or equivalently $h$):
\be
\label{jmd-eigenstates}
\left|\begin{array}{clccl}
\mathfrak{C}&\,|j,m,d\ra
&=&
j(j+1)&\,|j,m,d\ra
\vspace*{1mm}\\
S_{3}&\,|j,m,d\ra
&=&
m&\,|j,m,d\ra
\vspace*{1mm}\\
\hat{N}&\,|j,m,d\ra
&=&
(j+2d)&\,|j,m,d\ra
\end{array}\right.
\,,
\ee
with $j,d\in\N$ and $m\in\Z$ running between $-j$  and $+j$.

The subtle point is that $N-j=2d$ is always even. This reflects the fact that there is no creation operator linear in the $\alpha_{a}\dag$ that commutes with the spin generators. In fact, more generally, there is no polynomial of odd degree\footnotemark{} in the creation operators  $\alpha_{a}\dag$ that commutes with the spin generators $S_{a}$. So a creation operator that  don't change the spin $j$ will always increase the overall number of quanta $N$ by an even number. We can check this directly by a state counting, as shown below.
\footnotetext{
At the classical level, we have two 3-vectors $x$ and $p$. If we have a set of $n$ 3-vectors $v_{1}$, .., $v_{n}$, any polynomial in those vectors that is $\SO(3)$ invariant can be written as a linear combination of products of scalar products $v_{i}\cdot v_{j}$ and triple products (``volumes'') $v_{i}\cdot (v_{j}\w v_{k})$. A polynomial of even power can be written entirely  in terms of scalar products, while a polynomial of odd power necessarily involves one triple product in each  of its product terms. Here, there is no triple product since we have only two vectors, so rotational-invariant polynomials are necessarily of even degree.
}

\medskip

We construct  the $|j,m,d\ra$ eigenstates by first constructing the states $|j,j,0\ra\propto (\alpha_{-}\dag)^{j}\,|0\ra$, which corresponds to the state with $(N_{-},N_{+},N_{3})=(j,0,0)$. Indeed, we check that:
\be
\begin{array}{lcl}
S_{3}\, (\alpha_{-}\dag)^{j}\,|0\ra
&=&
j\, (\alpha_{-}\dag)^{j}\,|0\ra
\vspace*{1mm}\\
S_{+}\, (\alpha_{-}\dag)^{j}\,|0\ra
&=&
0
\vspace*{1mm}\\
\hat{N}\, (\alpha_{-}\dag)^{j}\,|0\ra
&=&
j\, (\alpha_{-}\dag)^{j}\,|0\ra
\end{array}
\,.
\ee
Then we reach a state with $m<j$ by acting with the $\so(3)$ lowering operator $(S_{-})^{j-m}$ and with $d>0$ by acting with the $\sll_{2}(\R)$ raising operator $(f\dag)^d$,
\be
|j,m,d\ra\propto (f\dag)^d\,(S_{-})^{j-m}\,(\alpha_{-}\dag)^{j}\,|0\ra
\,,
\ee
where the latter two operators, $(f\dag)^d$ and $(S_{-})^{j-m}$, commute with each other by construction. These gives the wanted eigenvalues \eqref{jmd-eigenstates} of the three operators, $\mathfrak{C}$, $S_{3}$ and $N$.

We give the explicit normalization factors of the states $|j,m,d\ra$ below in Section \ref{appendix:a} using the explicit action of the $\so(3)$ and $\sll_{2}(\R)$ generators on those states.

\subsection{Counting basis states}

We would like to show that one can reconstruct the initial basis $|N_{-},N_{+},N_{3}\ra$ from the states $|j,m,d\ra$. Expanding the binomials $(S_{-})^{j-m}=({\alpha}^\dagger_{+}\alpha_{3}-{\alpha}^\dagger_{3}\alpha_{-})^{j-m}$ and $(f\dag)^d=(\alpha_{3}^2\dag+2\alpha_{-}\dag\alpha_{+}\dag)^d$, we see that the state  $|j,m,d\ra$ is a superposition of states $|N_{-},N_{+},N_{3}\ra$ with
\be
\left|\begin{array}{lcl}
N_{-}
&=&
m+a+b
\\
N_{+}
&=&
a+b
\\
N_{3}
&=&
(d-a)+(j-m-2b)
\\
N
&=&
j+2d
\end{array}\right.
\qquad\textrm{with}\quad
-j\le m \le +j
\,,\quad
0\le a \le d
\,,\quad
0\le b \le \f{j-m}2
\,.
\ee
Let us identify the sets of $|j,m,d\ra$ states that involve the same $|N_{-},N_{+},N_{3}\ra$ states. Working with $N=j+2d=N_{-}+N_{+}+N_{3}$ fixed  in $\N$, and choosing $K\in\N$ such that $0\le K\le 2N$, we check that:
\be
\textrm{the states }|j,m,d\ra\textrm{ with }
(j,d,j-m)=
\left|\begin{array}{l}
(N,0,K)\\
(N-2,1,K-2)\\
(N-4,2,K-4)\\
\dots
\end{array}\right.
\ee
\be
\textrm{involve states }
|N_{-},N_{+},N_{3}\ra
\textrm{ with }
(N_{-},N_{+},N_{3})=
\left|\begin{array}{l}
(N-K,0,K)\\
(N-K+1,1,K-2)\\
(N-K+2,2,K-4)\\
\dots
\end{array}\right.
\ee
Taking into account that, on the one hand $0\le (j-m)\le 2j$ and on the other hand $N_{3}\ge0$ and $N_{-}\ge 0$, it is fairly simple to check that these two sets at fixed $N$ and $K$ contain the same number of states, thus establishing an isomorphism. Moreover, any state $|N_{-},N_{+},N_{3}\ra$ with arbitrary $N_{a}\in\N$ belongs to one of those sets (the set with $N=N_{-}+N_{+}+N_{3}$ and $K=N_{3}+2N_{+}$), so that we get a one-to-one change of basis between the $|N_{-},N_{+},N_{3}\ra$ states and the $|j,m,d\ra$ states.

\subsection{Unitary representation of $\so(3)\times \sll_{2}(\R)$}
\la{appendix:a}

To summarize the algebraic structure, we have the algebra $\so(3)\times \sll_{2}(\R)\sim\su(2)\times \sll(2,\mathbb{R})$ with commutation relations,
\be
[S^3,S^\pm]=\pm S^\pm
\,,\quad
[S^+,S^-]=S^3,
\,,\qquad
[h,f_{\pm} ]=\pm 2f_{\pm}
\,,\quad
[f_{-},f_{+}]=4 h
\,,
\ee
with  reality conditions:
\be
h^\dagger=h,\quad f_-^\dagger = f_+,\qquad  (S^3)^\dagger= S^3,\quad (S^-)^\dagger= S^+.
\ee
We further require that the $\su(2)$ Casimir and the $\sll_{2}(\R)$ Casimir operators be equal:
\be
\mathfrak{C}
=S^3(S^3+1)+ 2S^-S^+= (h-\tfrac32)(h-\tfrac12) - f_+f_{-}. 
\ee
Working in the $\SU(2)$ representation of spin $j$, the Casimir is $\mathfrak{C}=\vS^2=j(j+1)$, this means that we are working in the corresponding unitary representation of $\SL(2,\R)$ with the same Casimir, i-e a unitary representation from the discrete series.

We define a orthogonal basis of normalized states $|j,m,d\rangle$  which diagonalizes the three operators $\vS^2$, $S^3$ and $h$, with $j$ and $d$ integers and $m$ running by integer steps from $-j$ to $+j$:
\be
\vS^2|j,m,d\rangle= j(j+1)|j,m,d\rangle
,\quad
S^3|j,m,d\rangle= m|j,m,d\rangle
,\quad
h|j,m,d\rangle = (j+2d+\tfrac32 ) |j,m,d\rangle.
\ee
Using the commutation relations defining the $\su(2)$ and $\sll_{2}(\R)$ algebras, $ [S^3, S^-]= - S^-$ and $[N, f^\dagger]= 2 f^\dagger $,
one obtains the action of the raising and lowering operators on those states:
\be
\begin{array}{clcll}
S_-&|j,m, d\rangle &=& \sqrt{(j-m +1)(j+m)/2 } &|j, m-1,d\rangle,\vspace*{1mm}\\
S_+&|j,m, d\rangle &=& \sqrt{(j-m )(j+m+1)/2 } & |j, m+1,d\rangle,\vspace*{1mm}\\
f_- &|j,m,d\rangle &=& \sqrt{(N-j) (N+j+1) } &|j,m,d-1\rangle,\vspace*{1mm}\\
f_+ &|j,m,d\rangle &=& \sqrt{2 (d +1)(2(j+d+1)+1) }& |j,m,d+1\rangle.
\end{array}
\ee
Iterating these relations, we can derive an arbitrary state $ |j,m, d\rangle$ from the state $|j,j, 0\rangle$, which is both a highest weight vector for the $\su(2)$ algebra and a lowest weight vector for the $\sll_{2}(\R)$ algebra:
\ba
 |j,m, d\rangle =  \sqrt{\f{(2j+1)(j+d)!(j+m)!}{ j! d!(2(j+d)+1)! (j -m)!}} (f^\dagger)^d( \sqrt{2} S^-)^{j-m} |j,j, 0\rangle .
\ea
This fixes all the normalization factors.

 
 \bibliographystyle{bib-style}
 \bibliography{StringFlux}

 \end{document}